\documentclass[aps,prd,showpacs,notitlepage,nofootinbib,preprintnumbers,amsmath,amssymb]{revtex4-1}

\usepackage{graphics,graphicx}
\usepackage{dcolumn}
\usepackage{bm}
\usepackage{epsfig}
\usepackage[usenames]{color}
\usepackage{hyperref} 
\usepackage{mathbbol}
\usepackage{epstopdf}
\usepackage{simplewick}
\usepackage[utf8]{inputenc} 
\usepackage{slashed}
\usepackage{stackrel}
\usepackage{mathrsfs}
\usepackage{xcolor}
\usepackage{subcaption}
\usepackage[most]{tcolorbox}
\usepackage{physics}

\allowdisplaybreaks

\usepackage{ragged2e} 
\DeclareCaptionJustification{justified}{\justifying}

\def\eq#1{{Eq.~(\ref{#1})}}
\def\eqs#1{{Eqs.~(\ref{#1})}}
\def\fig#1{{Fig.~\ref{#1}}}
\newcommand{\ben}{\begin{eqnarray*}}
\newcommand{\een}{\end{eqnarray*}}
\newcommand{\un}[1]{\underline{#1}}

\newcommand{\pd}{\partial}

\newcommand{\thalf}{\tfrac{1}{2}}
\newcommand{\llangle}{\Big\langle \!\! \Big\langle}
\newcommand{\rrangle}{\Big\rangle \!\! \Big\rangle}

\newcommand{\as}{\alpha_s}

\newcommand{\bas}{{\bar\alpha}_s}

\newcommand{\dhd}{{\textstyle d}
\lower.03ex\hbox{\kern-0.38em$^{\scriptstyle-}$}\kern-0.05em{}}
\newcommand{\dbar}{{\textstyle \delta}
\lower.03ex\hbox{\kern-0.38em$^{\scriptstyle-}$}\kern-0.05em{}}

\newcommand{\ul}[1]{\underline{#1}}

\newcommand{\tord}{\textrm{T} \:}
\newcommand{\atord}{\overline{\textrm{T}} \:}

\makeatletter
\DeclareRobustCommand{\cev}[1]{%
  {\mathpalette\do@cev{#1}}%
}
\newcommand{\do@cev}[2]{%
  \vbox{\offinterlineskip
    \sbox\z@{$\m@th#1 x$}%
    \ialign{##\cr
      \hidewidth\reflectbox{$\m@th#1\vec{}\mkern4mu$}\hidewidth\cr
      \noalign{\kern-\ht\z@}
      $\m@th#1#2$\cr
    }%
  }%
}
\makeatother

\newcommand{\dprime}{{\prime \prime}}
\newcommand{\vpol}[2]{V_{{#1}}^{\mathrm{pol}[#2]}}
\newcommand{\iUV}{
    \int \displaylimits^z_{\frac{1}{s x_{10}^2}} \frac{dz^\prime }{z^\prime} \int \displaylimits^{x_{10}^2}_{\frac{1}{z^\prime s}} \frac{dx_{21}^2}{x_{21}^2}
}
\newcommand{\iIR}{
\int \displaylimits^{z}_{\frac{\Lambda^2}{s}} \frac{dz^\prime}{z^\prime} \int \displaylimits^{\mathrm{min}[\frac{z}{z^\prime} x_{10}^2, \frac{1}{\Lambda^2}]}_{\mathrm{max}[x_{10}^2, \frac{1}{z^{\prime}s}]} \frac{d x_{21}^2}{x_{21}^2}
}

\newcommand{\inUV}{\int \displaylimits^{z^\prime}_{\frac{1}{sx_{10}^2}} \frac{dz^{\dprime}}{z^{\dprime}} \int \displaylimits^{\mathrm{min}[x_{10}^2, x_{21}^2 \frac{z^\prime}{z^{\dprime}}]}_{\frac{1}{z^{\dprime} s}} \frac{d x_{32}^2}{x_{32}^2}}

\newcommand{\inIR}{\int \displaylimits^{z^\prime \frac{x_{21}^2}{x_{10}^2}}_{\frac{\Lambda^2}{s}} \frac{dz^{\dprime}}{z^{\dprime}} 
    \int \displaylimits^{\mathrm{min}[\frac{z^\prime}{z^{\dprime}} x_{21}^2, \frac{1}{\Lambda^2}]}_{\mathrm{max}[x_{10}^2, \frac{1}{z^{\dprime}s}]} \frac{d x_{32}^2}{x_{32}^2} }

\newcommand{\wb}{\omega_b}

\begin{document}

\title{Orbital Angular Momentum at Small \texorpdfstring{$x$}{x} Revisited} 

\author{Yuri V. Kovchegov} 
         \email[Email: ]{kovchegov.1@osu.edu}
         \affiliation{Department of Physics, The Ohio State
           University, Columbus, OH 43210, USA}

\author{Brandon Manley}
  \email[Email: ]{manley.329@osu.edu}
	\affiliation{Department of Physics, The Ohio State
           University, Columbus, OH 43210, USA}

\begin{abstract}
We revisit the problem of the small Bjorken-$x$ asymptotics of the quark and gluon orbital angular momentum (OAM) distributions in the proton utilizing the revised small-$x$ helicity evolution derived recently in \cite{Cougoulic:2022gbk}. We relate the quark and gluon OAM distributions at a small $x$ to the polarized dipole amplitudes and their (first) impact-parameter moments. To obtain the OAM distributions, we derive novel small-$x$ evolution equations for the impact-parameter moments of the polarized dipole amplitudes in the double-logarithmic approximation (summing powers of $\as \ln^2(1/x)$ with $\as$ the strong coupling constant). We solve these evolution equations numerically and extract the leading large-$N_c$, small-$x$ asymptotics of the quark and gluon OAM distributions, which we determine to be
    \begin{align}
        L_{q+\bar{q}}(x, Q^2) \sim L_{G}(x,Q^2) \sim \Delta \Sigma(x, Q^2) \sim \Delta G(x,Q^2) \sim \left(\frac{1}{x}\right)^{3.66 \, \sqrt{\frac{\as N_c}{2\pi}}}, \notag
    \end{align}
in agreement with \cite{Boussarie:2019icw} within the precision of our numerical evaluation. (Here $N_c$ is the number of quark colors.) We also investigate the ratios of the quark and gluon OAM distributions to their helicity distribution counterparts in the small-$x$ region. 
\end{abstract}

~\\

\maketitle

\tableofcontents


\section{Introduction}

The orbital angular momentum (OAM) carried by the quarks and gluons inside the proton is the least explored but a potentially important component of the proton spin. The parton OAM contribution to the proton spin is quantified in terms of spin sum rules: the Jaffe-Manohar sum rule~\cite{Jaffe:1989jz} and Ji sum rule~\cite{Ji:1996ek}. The former separates gluon OAM from the gluon helicity. It reads
\begin{equation}
S_{q + \bar q} +L_{q + \bar q}+S_G+L_G=\frac{1}{2}\, ,  
\label{eqn:JM}
\end{equation}
where the contributions to the proton spin carried by the quark and gluon helicities are 
\begin{align}\label{eqn:SqSG}
    S_{q + \bar q} (Q^2) = \frac{1}{2} \int\limits_0^1 \dd{x} \, \Delta\Sigma(x,Q^2)\,, \qquad\qquad
    S_G(Q^2) = \int\limits_0^1 \dd{x} \, \Delta G(x,Q^2)\, . 
\end{align}
Here the quark flavor-singlet helicity parton distribution function (hPDF) is
\begin{equation}
\Delta\Sigma(x,Q^2) = \sum_{f=u,d,s,\ldots} 
\left[\Delta q_f (x,Q^2) + \Delta\bar{q}_f (x,Q^2)\right]
\label{eqn:DeltaSigma}
\end{equation}
with $\Delta q_f (x,Q^2)$, $\Delta\bar{q}_f (x,Q^2)$ the quark and anti-quark hPDFs, respectively, dependent on the momentum scale $Q^2$ and the longitudinal fraction $x$ of the proton's momentum carried by the parton. Subscript $f$ denotes the flavor. The gluon hPDF is denoted by $\Delta G(x,Q^2)$.

The OAM contributions $L_{q + \bar q} (Q^2)$ and $L_G (Q^2)$ can also be decomposed into integrals of the corresponding distributions $L_{q + \bar q} (x,Q^2)$ and $L_G (x,Q^2)$ over the momentum fraction $x$ \cite{Bashinsky:1998if, Hagler:1998kg, Harindranath:1998ve, Hatta:2012cs, Ji:2012ba}, 
\begin{align}\label{eqn:LqLG}
    L_{q + \bar q} (Q^2) = \int\limits_0^1 \dd{x} \, L_{q + \bar q} (x,Q^2)\,, \qquad\qquad
    L_G(Q^2) = \int\limits_0^1 \dd{x} \, L_G (x,Q^2)\,.
\end{align}
Note that, unlike $\Delta\Sigma(x,Q^2)$ and $\Delta G(x,Q^2)$, which are given by the matrix elements of twist-two operators, $L_{q + \bar q} (x,Q^2)$ and $L_G (x,Q^2)$ receive both twist-two and twist-three contributions \cite{Hatta:2012cs}.

The aim of research on the proton spin puzzle~\cite{Aidala:2012mv, Accardi:2012qut, Leader:2013jra, Aschenauer:2013woa, Aschenauer:2015eha, Boer:2011fh, Proceedings:2020eah, Ji:2020ena, AbdulKhalek:2021gbh} is to determine the distributions $\Delta\Sigma(x,Q^2)$, $\Delta G(x,Q^2)$, $L_{q + \bar q} (x, Q^2)$, and $L_G (x, Q^2)$ over as broad a range in $x$ and $Q^2$ as possible, in order to quantitatively understand the origin of the proton spin. Helicity PDFs $\Delta\Sigma(x,Q^2)$ and $\Delta G(x,Q^2)$ have received a lot of attention over the years: they have been extracted from the experimental data over the accessible range of $x$ and $Q^2$~\cite{Gluck:2000dy, Leader:2005ci, deFlorian:2009vb, Leader:2010rb, Jimenez-Delgado:2013boa, Ball:2013lla, Nocera:2014gqa, deFlorian:2014yva, Leader:2014uua, Sato:2016tuz, Ethier:2017zbq, DeFlorian:2019xxt, Borsa:2020lsz, Zhou:2022wzm, Cocuzza:2022jye}. At the same time, the OAM distributions are much less explored. For instance, in spite of the fact that their evolution in $Q^2$ is known \cite{Hagler:1998kg,Hoodbhoy:1998yb}, no extraction from the experimental data currently exists for the quark and gluon OAM distributions $L_{q + \bar q} (x, Q^2)$ and $L_G (x, Q^2)$ due to the lack of observables known to directly couple to these distributions (see \cite{Bhattacharya:2022vvo} for a promising recent proposal on how to measure the OAM distributions). In order to solve the spin puzzle, it appears important to understand as much as possible about the quark and gluon OAM distributions, both theoretically and phenomenologically.

The goal of this work is to understand the OAM distributions, $L_{q + \bar q} (x, Q^2)$, and $L_G (x, Q^2)$ at small $x$ by revisiting and revising the earlier attempt to determine the small-$x$ asymptotics of $L_{q + \bar q} (x, Q^2)$ and $L_G (x, Q^2)$ made in \cite{Kovchegov:2019rrz}. Determining the small-$x$ asymptotics of all four distributions $\Delta\Sigma(x,Q^2)$, $\Delta G(x,Q^2)$, $L_{q + \bar q} (x, Q^2)$, and $L_G (x, Q^2)$ is important, in part because no present or future experimental data would be able to probe these distributions down to $x=0$: the acceptance of any given experiment is limited by $x > x_{\rm min}$, with the minimum $x$-value, $x_{\rm min}$, scaling as the inverse of the center-of-mass energy squared, but always remaining above zero. A theoretical extrapolation to lower values of $x$, $x < x_{\rm min}$, appears necessary in order to fully map out the small-$x$ region of the helicity and OAM distributions. 

A pioneering work in this direction, done for the hPDFs $\Delta\Sigma(x,Q^2)$ and $\Delta G(x,Q^2)$, was carried out by Bartels, Ermolaev and Ryskin (BER) \cite{Bartels:1995iu,Bartels:1996wc}, employing the infrared evolution equations (IREE) formalism from \cite{Gorshkov:1966ht,Kirschner:1983di,Kirschner:1994rq,Kirschner:1994vc,Blumlein:1995jp, Griffiths:1999dj}. The IREE of \cite{Bartels:1996wc} led to a version of small-$x$ asymptotics of $\Delta\Sigma(x,Q^2)$ and $\Delta G(x,Q^2)$ in the leading double-logarithmic approximation (DLA), i.e., resumming the powers of the parameter $\as \, \ln^2 (1/x)$ with $\as$ the strong coupling constant.

In recent years, a novel approach to small-$x$ evolution for helicity distributions has been developed in \cite{Kovchegov:2015pbl, Hatta:2016aoc, Kovchegov:2016zex, Kovchegov:2016weo, Kovchegov:2017jxc, Kovchegov:2017lsr, Kovchegov:2018znm, Cougoulic:2019aja, Kovchegov:2020hgb, Cougoulic:2020tbc, Chirilli:2021lif, Kovchegov:2021lvz, Cougoulic:2022gbk}, based on the $s$-channel evolution/shock wave formalism previously constructed in \cite{Mueller:1994rr,Mueller:1994jq,Mueller:1995gb,Balitsky:1995ub,Balitsky:1998ya,Kovchegov:1999yj,Kovchegov:1999ua,Jalilian-Marian:1997dw,Jalilian-Marian:1997gr,Weigert:2000gi,Iancu:2001ad,Iancu:2000hn,Ferreiro:2001qy} for unpolarized high-energy scattering. The unpolarized scattering dominates at high energy, with the evolution from \cite{Mueller:1994rr,Mueller:1994jq,Mueller:1995gb,Balitsky:1995ub,Balitsky:1998ya,Kovchegov:1999yj,Kovchegov:1999ua,Jalilian-Marian:1997dw,Jalilian-Marian:1997gr,Weigert:2000gi,Iancu:2001ad,Iancu:2000hn,Ferreiro:2001qy} formulated in terms of infinite light-cone Wilson lines: this is the eikonal approximation. Helicity-dependent scattering is suppressed by one inverse power of the center-of-mass energy squared $s$ (or, equivalently, by one positive power of Bjorken $x \sim 1/s$): this is referred to as the sub-eikonal approximation. The small-$x$ shock-wave formalism for the sub-eikonal approximation and beyond (including sub-sub-eikonal corrections) have been developed over the recent decade in \cite{Altinoluk:2014oxa,Balitsky:2015qba,Balitsky:2016dgz, Kovchegov:2017lsr, Kovchegov:2018znm, Kovchegov:2018zeq, Chirilli:2018kkw, Jalilian-Marian:2018iui, Jalilian-Marian:2019kaf, Altinoluk:2020oyd, Kovchegov:2021iyc, Altinoluk:2021lvu, Kovchegov:2022kyy, Altinoluk:2022jkk, Altinoluk:2023qfr,Altinoluk:2023dww, Li:2023tlw, Santiago:2023rfl}. The operators appearing in sub-eikonal scattering consist of semi-infinite and finite light-cone Wilson lines with the sub-eikonal operators inserted between them. This emerging formalism was proposed to be called light-cone operator treatment (LCOT) in \cite{Cougoulic:2022gbk}.

In the $s$-channel/shock wave formalism at the DLA, novel small-$x$ evolution equations for the so-called ``polarized dipole amplitudes", related to helicity PDFs, were derived in \cite{Kovchegov:2015pbl, Kovchegov:2016zex, Kovchegov:2017lsr, Kovchegov:2018znm} (KPS). More recently, a new sub-eikonal operator was identified in \cite{Cougoulic:2022gbk} (see also \cite{Hatta:2016aoc}) whose evolution mixes with that of the operators considered in the original KPS papers, but which was omitted in those papers. Including this new operator into the sub-eikonal formalism resulted in a revised version of helicity evolution equations derived in \cite{Cougoulic:2022gbk} (KPS-CTT). These equations were solved numerically \cite{Cougoulic:2022gbk} and analytically \cite{Borden:2023ugd} in the large-$N_c$ limit, and numerically in the large-$N_c \& N_f$ limit \cite{Adamiak:2023okq} (with $N_f$ the number of quark flavors). The latter solution allowed for a successful analysis of the polarized deep inelastic scattering (DIS) and semi-inclusive DIS (SIDIS) data at small-$x$ performed in \cite{Adamiak:2023yhz}. The small-$x$ asymptotics of hPDFs found by BER \cite{Bartels:1996wc} and by solving the KPS-CTT evolution \cite{Cougoulic:2022gbk,Borden:2023ugd,Adamiak:2023okq} appear to be very close to each other, with the intercept (the power of $1/x$) different by less than $1 \%$ at large $N_c$ and by less than $3 \%$ at large $N_c \& N_f$. The origin of this small discrepancy is not entirely clear at the moment, with its potential origin detailed in the appendices of \cite{Kovchegov:2016zex,Borden:2023ugd}.

The BER IREE formalism was applied to the OAM distributions $L_{q + \bar q} (x, Q^2)$ and $L_G (x, Q^2)$ at small $x$ in \cite{Boussarie:2019icw}. The first analysis of the problem in the shock wave picture had been performed in \cite{Hatta:2016aoc}, and the large-$N_c$ version of the KPS evolution was applied to OAM distributions at small $x$ in \cite{Kovchegov:2019rrz}.  (See also \cite{Hatta:2018itc} for an OAM distributions analysis at small $x$ based on the Dokshitzer-Gribov-Lipatov-Altarelli-Parisi (DGLAP) evolution equations \cite{Gribov:1972ri, Altarelli:1977zs, Dokshitzer:1977sg}.) The small-$x$ asymptotics of $L_{q + \bar q} (x, Q^2)$ and $L_G (x, Q^2)$ obtained in \cite{Boussarie:2019icw} and in \cite{Kovchegov:2019rrz} were significantly different from each other already at large $N_c$, with more than $50 \%$ difference in the intercepts. This was consistent with the difference between the intercepts for hPDFs found by BER \cite{Bartels:1996wc} and, at the time, by KPS \cite{Kovchegov:2016zex}. Since the latter (KPS) result has now been superseded by the KPS-CTT evolution, resulting in a much smaller difference between the corresponding hPDFs intercepts and the ones obtained by BER \cite{Cougoulic:2022gbk,Borden:2023ugd,Adamiak:2023okq}, it appears necessary to revise the OAM calculation from  \cite{Kovchegov:2019rrz} in order to include the full KPS-CTT evolution into it. This is exactly the goal of the present work.  

Below, we build on the calculations from \cite{Kovchegov:2019rrz} to simplify $L_{q + \bar q} (x, Q^2)$ and $L_G (x, Q^2)$ at small $x$, rewriting them in terms of the ``polarized dipole amplitudes" from \cite{Kovchegov:2015pbl, Kovchegov:2016zex, Kovchegov:2017lsr, Kovchegov:2018znm,Cougoulic:2022gbk} in Sec.~\ref{sec:OAM_distrs}. The calculation is performed for the quark OAM distribution in Sec.~\ref{sec:quark_OAM} and for the gluon OAM distribution in Sec.~\ref{sec:gluon_OAM}. The end results, given by Eqs.~\eqref{final_quark_OAM} and \eqref{final_gluonOAM}, express $L_{q + \bar q}$ and $L_G$ in terms of the impact-parameter integrated polarized dipole amplitudes and, in addition, in terms of their impact-parameter moments. The moments are obtained by multiplying the dipole amplitudes by an impact parameter and then integrating over all impact parameters (see Eqs.~\eqref{moment_defs} below). 

Since the DLA is applied to the evolution equations for the impact-parameter integrated polarized dipole amplitudes \cite{Cougoulic:2022gbk}, we cannot simply reuse the solutions constructed in \cite{Cougoulic:2022gbk,Borden:2023ugd,Adamiak:2023okq} for the latter to obtain the small-$x$ asymptotics of the OAM distributions, since we also need moments of the dipole amplitudes. Instead, we need to go back to the equations constructed in \cite{Cougoulic:2022gbk} for the polarized dipole amplitudes without the impact parameter integration, multiply those by the impact parameters, integrate over them as well, and extract the DLA part of the resulting evolution. This is done in Sec.~\ref{sec:evolution} for the large-$N_c$ evolution, resulting in the coupled system of equations \eqref{all_oam_eqns} for the impact-parameter moments of polarized dipole amplitudes. These equations are discretized and solved numerically in Sec.~\ref{sec:num_sol}, resulting in the small-$x$ asymptotics of $L_{q + \bar q} (x, Q^2)$ and $L_G (x, Q^2)$ given by \eq{LLasympt} (and above in the Abstract), which is the main result of this work. Within the accuracy of our numerical solution, these asymptotics are consistent with the results of \cite{Boussarie:2019icw} based on BER IREE. Note, however, that the discrepancy in the hPDF intercepts between the large-$N_c$ BER results and the KPS-CTT evolution found analytically in \cite{Borden:2023ugd} is smaller than the numerical precision used in our present work: we anticipate that the same discrepancy will be present in the intercepts for the OAM distributions at hand. (See also \cite{Moch:2014sna, Blumlein:2021ryt} for a discrepancy between the IREE and the small-$x$ limit of the exact 3-loop calculations of spin-dependent DGLAP anomalous dimensions: however, unlike \cite{Borden:2023ugd}, the discrepancy found in \cite{Moch:2014sna, Blumlein:2021ryt} can be attributed to scheme dependence.)

An important quantity to consider is the ratio of the OAM distributions to the flavor-singlet helicity PDFs, calculated separately for quarks and for gluons \cite{Hatta:2016aoc,Hatta:2018itc,Boussarie:2019icw,More:2017zqp}. We analyse such ratios at small-$x$ in Sec.~\ref{sec:ratios}, extracting their continuum limits in two different ways, shown in Eqs.~\eqref{app1_res} and \eqref{app2_res}. We compare the resulting values for these ratios to those in the existing literature \cite{Hatta:2018itc,Boussarie:2019icw}. We summarize and conclude in Sec.~\ref{sec:conc}.


\section{Evaluation of the OAM distributions at small $x$}

\label{sec:OAM_distrs}


\subsection{Quark OAM}

\label{sec:quark_OAM}

Our notation for the light-cone components of four-vectors is $x^{\pm} = (x^0 \pm x^3)/\sqrt{2}$, while transverse vectors are defined as $\un x = (x^1, x^2)$ with $\un x_{ij} = \un x_i - \un x_j$ and $|\un x_{ij} | = x_{ij}$ for $i,j$ labeling the partons. For the transverse momenta we have $k_\perp = |{\un k}|$ with ${\un k} = (k^1, k^2)$. We use the following notation for fundamental Wilson lines on the $x^-$ light cone
\begin{align}\label{Vline}
    V_{\un x}[x_f^-, x_i^-] = \mathcal{P} \exp \left[\ ig \int \displaylimits^{x_f^-}_{x_i^-} dx^- A^+(0^+, x^-, \un x)  \right],
\end{align}
with the abbreviation $V_{\un x} = V_{\un x}[\infty, - \infty]$ for infinite lines. Here $\mathcal{P}$ is the path-ordering operator, $g$ the strong coupling constant, and $A^\mu = \sum_a A^{a\, \mu} \, t^a$ is the background gluon field with $t^a$ the $\mathrm{SU}(N_c)$ generators. Our proton is moving along the $x^+$ direction.

We base our derivation on that from \cite{Kovchegov:2019rrz}, which starts with the quark OAM defined in terms of the quark Wigner function $W (k,b)$ as 
\begin{align}\label{OAM0}
L_z = \int \frac{d^2 b_\perp d b^- \, d^2 k_\perp \, d k^+}{(2 \pi)^3} \, ({\un b} \times {\un k})_z \, W (k,b)
\end{align} 
and simplifies it at small $x$ using the SIDIS Wigner function with the future-pointing Wilson line staple. 

Starting from Eq.~(16) in \cite{Kovchegov:2019rrz}, we modify it by replacing \cite{Cougoulic:2022gbk, Kovchegov:2021iyc}
\begin{align}\label{repl}
    \bar{v}_{\sigma_1}(k_1)
    \left(\hat{V}_{\un{w}}^\dagger \right)^{ji}
    v_{\sigma_2}(k_2) \to 2 \sqrt{k_1^- k_2^-} \int d^2 z \,  \left( V^\dagger_{\un{z}, \un{w}; -\sigma_2, -\sigma_1}\right)^{ji}
\end{align}
in it, where $V_{\un{z}, \un{w}; \sigma_2, \sigma_1}$ is the quark $S$-matrix for scattering on background quark and gluon fields defined in \cite{Cougoulic:2022gbk}.
For our small-$x$ calculation, one needs to expand the $S$-matrix in ``eikonality", that is in the inverse powers of energy. Such an expansion gives \cite{Cougoulic:2022gbk, Altinoluk:2014oxa,Balitsky:2015qba,Balitsky:2016dgz, Kovchegov:2017lsr, Kovchegov:2018znm, Chirilli:2018kkw, Altinoluk:2020oyd, Kovchegov:2021iyc, Altinoluk:2021lvu}
\begin{align} 
    \label{polWL}
    V_{\un{x}, \un{y}; \sigma^\prime \sigma} = \delta_{\sigma \sigma^\prime} \, V_{\un{x}} \, \delta^2 (\un{x} - \un{y})  + \sigma \delta_{\sigma \sigma^\prime} \, V_{\un{x}}^{\mathrm{pol}[1]} \, \delta^2(\un{x}-\un{y}) + \delta_{\sigma \sigma^\prime} \, V_{\un{x}, \un{y}}^{\mathrm{pol}[2]} + \cdots .
\end{align}
The first term on the right-hand side of \eq{polWL} is the eikonal contribution, while the remaining shown terms are sub-eikonal (suppressed by exactly one power of energy). The ellipsis denote sub-sub-eikonal \cite{Kovchegov:2021iyc} terms and beyond, that is, everything suppressed by two or more powers of energy. 

The sub-eikonal operators, referred to as the ``polarized Wilson lines" in \cite{Kovchegov:2015pbl, Kovchegov:2016zex, Kovchegov:2016weo, Kovchegov:2017jxc, Kovchegov:2017lsr, Kovchegov:2018znm, Cougoulic:2019aja, Kovchegov:2020hgb, Cougoulic:2020tbc, Kovchegov:2021lvz, Cougoulic:2022gbk}, contain insertions of one or two gluon or quark operators into the light-cone Wilson lines. Specifically, we have \cite{Cougoulic:2022gbk}
\begin{align}\label{VqG_decomp}
V_{\un x}^{\textrm{pol} [1]} = V_{\un x}^{\textrm{G} [1]} + V_{\un x}^{\textrm{q} [1]}, \ \ \  V_{{\ul x}, {\un y}}^{\textrm{pol} [2]} = V_{{\ul x}, {\un y}}^{\textrm{G} [2]} + V_{{\ul x}}^{\textrm{q} [2]} \, \delta^2 ({\un x} - {\un y}) ,
\end{align}
where
\begin{subequations}\label{VqG}
\begin{align}
& V_{\un x}^{\textrm{G} [1]}  = \frac{i \, g \, P^+}{s} \int\limits_{-\infty}^{\infty} d{x}^- V_{\un{x}} [ \infty, x^-] \, F^{12} (x^-, {\un x}) \, \, V_{\un{x}} [ x^-, -\infty]  , \label{VG1} \\
& V_{\un x}^{\textrm{q} [1]}  = \frac{g^2 P^+}{2 \, s} \int\limits_{-\infty}^{\infty} \!\! d{x}_1^- \! \int\limits_{x_1^-}^\infty d x_2^- V_{\un{x}} [ \infty, x_2^-] \, t^b \, \psi_{\beta} (x_2^-,\un{x}) \, U_{\un{x}}^{ba} [x_2^-, x_1^-] \, \left[ \gamma^+ \gamma^5 \right]_{\alpha \beta} \, \bar{\psi}_\alpha (x_1^-,\un{x}) \, t^a \, V_{\un{x}} [ x_1^-, -\infty] , \label{Vq1} \\
& V_{{\ul x}, {\un y}}^{\textrm{G} [2]}  = - \frac{i \, P^+}{s} \int\limits_{-\infty}^{\infty} d{z}^- d^2 z \ V_{\un{x}} [ \infty, z^-] \, \delta^2 (\un{x} - \un{z}) \, \cev{D}^i (z^-, {\un z}) \, D^i  (z^-, {\un z}) \, V_{\un{y}} [ z^-, -\infty] \, \delta^2 (\un{y} - \un{z}) , \label{VxyG2} \\
& V_{{\ul x}}^{\textrm{q} [2]} = - \frac{g^2 P^+}{2 \, s} \int\limits_{-\infty}^{\infty} \!\! d{x}_1^- \! \int\limits_{x_1^-}^\infty d x_2^- V_{\un{x}} [ \infty, x_2^-] \, t^b \, \psi_{\beta} (x_2^-,\un{x}) \, U_{\un{x}}^{ba} [x_2^-, x_1^-] \, \left[ \gamma^+ \right]_{\alpha \beta} \, \bar{\psi}_\alpha (x_1^-,\un{x}) \, t^a \, V_{\un{x}} [ x_1^-, -\infty] \label{Vq2}.
\end{align}
\end{subequations}
Here $\psi$ and $\bar \psi$ are the quark and anti-quark field operators, $D^i = \pd^i - i g A^i$ and $\cev{D}^i = \cev{\pd}^i + i g A^i$ are the right- and left-acting covariant derivatives, respectively, with $i=1,2$ the transverse Lorentz index, $U_{\un{x}}^{ba} [x_2^-, x_1^-]$ is the adjoint light-cone Wilson line defined by analogy to the fundamental one in \eq{Vline}, $P^+$ is the large component of the proton's momentum, and $s$ is the center of mass energy squared for the quark--proton scattering. 

The contribution of $V_{\un{x}, \un{y}}^{\mathrm{pol}[2]}$ in \eq{polWL} was neglected in the analysis of \cite{Kovchegov:2019rrz} (and in the original KPS papers). Reinstating this term is our goal here: to this end, the replacement \eqref{repl} is handy, since it allows for parton propagation which is non-local in the transverse plane. 

Replacing $e^{i \un{k}_2 \cdot (\un{\xi} - \un{w})} \to e^{i \un{k}_2 \cdot (\un{\xi} - \un{z})}$ in addition to \eq{repl}, we rewrite Eq.~(16) from \cite{Kovchegov:2019rrz} as
\begin{align}
    \label{qOAMir1}
    L_q(x,Q^2) = +
    \frac{2 P^+}{(2\pi)^3} \int &d^2 k_\perp d^2 \zeta d^2 \xi \, e^{-i \un{k} \cdot \left( \un{\zeta} - \un{\xi} \right)} \left(
     \frac{\un{\zeta} + \un{\xi}}{2} \times \un{k}
    \right)  \\ \notag
    &\times \int d^2 w \;d^2 z \frac{d^2 k_1 dk_1^-}{(2\pi)^3} \frac{d^2 k_2}{(2\pi)^2} 
    e^{i \un{k}_1 \cdot (\un{w} - \un{\zeta}) + i \un{k}_2 \cdot (\un{\xi} - \un{z}) 
    } 
    \frac{\theta(k_1^-)}{\un{k}_1^2 \un{k}_2^2} \sum_{\sigma_1, \sigma_2} \bar{v}_{\sigma_2}(k_2)
    \frac{1}{2} \gamma^+ v_{\sigma_1}(k_1)
     \\\notag
    &\times \left\langle 
    \mathrm{T} \;V_{\un{\zeta}}^{ij}[\infty, -\infty] 
    2 \sqrt{k_1^- k_2^-} \left( V^\dagger_{\un{z}, \un{w}; -\sigma_2, -\sigma_1}\right)^{ji} 
    \right \rangle + \mathrm{c.c.}
\end{align}
with the quark OAM distribution $L_q(x,Q^2)$. (Since we are interested in $L_{q +{\bar q}} = L_q + L_{\bar q}$, the anti-quark OAM distribution $L_{\bar q}$ will be added later.) Here $x = k^+ / P^+$ is the light-cone momentum fraction of the proton's large momentum $P^+$ carried by the quark in question, while $\mathrm{T}$ is the time-ordering sign needed to separate the amplitude from the complex conjugate amplitude (which enters with the anti-time ordering sign $\overline{\mathrm{T}}$). We have used the polarization-dependent shock-wave/color glass condensate (CGC) \cite{Gribov:1984tu, Iancu:2003xm, Weigert:2005us, JalilianMarian:2005jf, Gelis:2010nm, Albacete:2014fwa, Kovchegov:2012mbw, Morreale:2021pnn} averaging 
\begin{align}\label{cgc_ave}
    \Big\langle \ldots \Big\rangle \equiv \frac{1}{2} \sum_{S_L} S_L\frac{1}{2 P^+ V^-} \langle P, S_L | \ldots | P, S_L \rangle
\end{align}
with the volume factor $V^- = \int d^2 x_\perp \, d x^-$ and the proton helicity $S_L$. This averaging is discussed in more detail in Appendix~A of \cite{Kovchegov:2019rrz}. Note that we are working in the same ``polarized DIS" scheme \cite{Adamiak:2023okq} as the helicity evolution calculation  \cite{Cougoulic:2022gbk}.

We employ the same $+ \leftrightarrow -$ interchanged Brodsky-Lepage spinors \cite{Lepage:1980fj} as in \cite{Kovchegov:2019rrz} (see also \cite{Kovchegov:2018znm}). They yield the following Dirac matrix element 
\begin{align}
    \label{diracME}
     \bar{v}_{\sigma_2}(k_2) \frac{1}{2} \gamma^+ v_{\sigma_1}(k_1) = \frac{1}{2} \delta_{\sigma_1 \sigma_2} \frac{\un k_2 \cdot \un k_1 + i \sigma_1(\un k_1 \times \un k_2)}{\sqrt{k_1^- k_2^-}},
\end{align}
where the cross-product of two-dimensional vectors is defined by ${\un u} \times {\un v} = u^1 v^2 - u^2 v^1 = \epsilon^{ij} \, u^i \, v^j$ with $\epsilon^{ij}$ the two-dimensional Levi-Civita symbol ($i,j = 1,2$ are the transverse indices). Substituting Eqs.~(\ref{polWL}) and (\ref{diracME}) into \eq{qOAMir1}, summing over $\sigma_1, \sigma_2$ and integrating over $\un k_1$ and $\un k_2$, we get 
\begin{align}\label{OAM3}
        L_q(x,Q^2) = - \frac{4 P^+}{(2\pi)^5} \int &d^2 k_\perp d^2 \zeta d^2 \xi \, e^{-i \un{k} \cdot \left( \un{\zeta} - \un{\xi} \right)} \left(
     \frac{\un{\zeta} + \un{\xi}}{2} \times \un{k}
    \right) 
    \int d^2 w \;d^2 z \int\limits^{p^-_2}_0 \frac{dk_1^-}{2\pi}
     \\\notag
    &\times \Bigg\langle -
    \frac{\un{\xi} - \un{z} }{|\un{\xi} - \un{z}|^2} \cdot  \frac{\un{\zeta} - \un{w} }{|\un{\zeta} - \un{w}|^2}
    \left\{ 
    \mathrm{T} \; \tr \left[V_{\un{\zeta}} \, V_{\un{w}}^\dagger \right] \delta^2 (\un{z}-\un{w})  + \mathrm{T \; tr} \left[ 
    V_{\un{\zeta}} \left( \vpol{\un{z}, \un{w}}{2} \right)^\dagger 
    \right]
    \right\}  \\ \notag
    & \hspace{1cm} + i \, 
    \frac{\un{\zeta} - \un{w} }{|\un{\zeta} - \un{w}|^2} \times  \frac{\un{\xi} - \un{z}}{|\un{\xi} - \un{z}|^2} 
    \,
    \mathrm{T \; tr} \left[ V_{\un{\zeta}} \left( \vpol{\un{w}}{1}\right)^\dagger \right] \delta^2 (\un{w}-\un{z}) 
    \Bigg\rangle + \mathrm{c.c.}.
\end{align}

Let us evaluate each term separately in the angle brackets of \eq{OAM3}. The eikonal term (the first trace in the angle brackets) gives, after we include the complex conjugate term, replace ${\un k} \to - {\un k}$ in it, and integrate over $\un z$ in the entire expression,
\begin{align}\label{eik1}
  + \frac{4 P^+}{(2\pi)^5} \int d^2 k_\perp d^2 \zeta d^2 \xi d^2 w \, e^{-i \un{k} \cdot \left( \un{\zeta} - \un{\xi} \right)} \left(
     \frac{\un{\zeta} + \un{\xi}}{2} \times \un{k}
    \right) 
      \int\limits^{p^-_2}_0 \frac{dk_1^-}{2\pi} \, \frac{\un{\xi} - \un{w} }{|\un{\xi} - \un{w}|^2} \cdot  \frac{\un{\zeta} - \un{w} }{|\un{\zeta} - \un{w}|^2} \left\langle \mathrm{T} \; \tr \left[V_{\un{\zeta}} \, V_{\un{w}}^\dagger \right] - \overline{\mathrm{T}} \; \tr \left[ V_{\un{w}} \, V_{\un{\zeta}}^\dagger \right] \right\rangle ,
\end{align}
where $\overline{\mathrm{T}}$ is the anti-time ordering sign. Similar to \cite{Kovchegov:2019rrz}, we can use  reflection symmetry with respect to the final-state cut for the Wilson lines \cite{Mueller:2012bn} to ``flip" the Wilson lines in the second trace of \eq{eik1} from the complex conjugate amplitude into the amplitude, effectively replacing $\atord \to \tord$ in that term. Defining the odderon amplitude by  \cite{Hatta:2005as,Kovchegov:2003dm}
\begin{align}\label{Odderon}
    \mathcal{O}_{\un{\zeta} \un{w}} = \frac{1}{2 i N_c} \Big{\langle}  \tord \tr [ V_{\underline{\zeta}}  V_{\underline{w}}^{\dagger }] - \tord \tr [ V_{\underline{w}}  V_{\underline{\zeta}}^{\dagger }] \Big{\rangle} 
\end{align}
we rewrite \eq{eik1} as
\begin{align}\label{eik2}
  + \frac{8 i P^+  \, N_c}{(2\pi)^5} \int d^2 k_\perp d^2 \zeta d^2 \xi d^2 w \, e^{-i \un{k} \cdot \left( \un{\zeta} - \un{\xi} \right)} \left(
     \frac{\un{\zeta} + \un{\xi}}{2} \times \un{k}
    \right) 
      \int\limits^{p^-_2}_0 \frac{dk_1^-}{2\pi} \, \frac{\un{\xi} - \un{w} }{|\un{\xi} - \un{w}|^2} \cdot  \frac{\un{\zeta} - \un{w} }{|\un{\zeta} - \un{w}|^2} \, \mathcal{O}_{\un{\zeta} \un{w}} = 0 .
\end{align}
This contribution is zero, because the integrand contains exactly one two-dimensional Levi-Civita symbol $\epsilon^{ij}$, coming from the cross-product. The odderon amplitude for the interaction with the longitudinally polarized proton, does not contain an $\epsilon^{ij}$: we know that the operator definition \eqref{Odderon} does not include an $\epsilon^{ij}$, the gluon fields entering the operator definition are the eikonal $A^+$ gluon fields, which also do not carry an $\epsilon^{ij}$ \cite{Kovchegov:1996ty}, and the odderon's small-$x$ evolution also has no $\epsilon^{ij}$ in it \cite{Bartels:1999yt,Kovchegov:2003dm,Hatta:2005as,Kovchegov:2012rz}. Therefore, after all the integrals are done, the expression \eqref{eik2} is a scalar quantity (it has no transverse indices) which does not depend on any transverse momenta, but contains an $\epsilon^{ij}$: therefore, it must be zero. Hence, the odderon does not contribute to the quark OAM distribution, unlike the spin-dependent odderon \cite{Boer:2015pni,Szymanowski:2016mbq,Dong:2018wsp,Kovchegov:2021iyc, Kovchegov:2022kyy} contributing to the Sivers function \cite{Sivers:1989cc}.

Dropping the eikonal term in \eq{OAM3} we are left with 
\begin{align}
    \label{qOAMint2}
        & L_q(x,Q^2) = - \frac{4 P^+}{(2\pi)^5} \int d^2 k_\perp d^2 \zeta d^2 \xi \, e^{-i \un{k} \cdot \left( \un{\zeta} - \un{\xi} \right)} \left(
     \frac{\un{\zeta} + \un{\xi}}{2} \times \un{k}
    \right) 
    \int d^2 w \;d^2 z \int\limits^{p^-_2}_0 \frac{dk_1^-}{2\pi}
     \\\notag
    &\times \Bigg\langle -
    \frac{\un{\xi} - \un{z} }{|\un{\xi} - \un{z}|^2} \cdot  \frac{\un{\zeta} - \un{w} }{|\un{\zeta} - \un{w}|^2} \,
    \mathrm{T \; tr} \left[ 
    V_{\un{\zeta}} \left( \vpol{\un{z}, \un{w}}{2} \right)^\dagger 
    \right]
       + i \, 
    \frac{\un{\zeta} - \un{w} }{|\un{\zeta} - \un{w}|^2} \times  \frac{\un{\xi} - \un{z} }{|\un{\xi} - \un{z}|^2} 
    \,
    \mathrm{T \; tr} \left[ V_{\un{\zeta}} \left( \vpol{\un{w}}{1}\right)^\dagger \right] \delta^2 (\un{w}-\un{z}) 
    \Bigg\rangle + \mathrm{c.c.}.
\end{align}
We need to add the anti-quark OAM distribution as well since $L_{q +{\bar q}} = L_q + L_{\bar q}$. The anti-quark contribution is calculated similarly, giving
\begin{align}\label{qbarOAM}
    & L_{\bar q} (x,Q^2) = -\frac{4 P^+}{(2\pi)^5} \int d^2 k_\perp d^2 \zeta d^2 \xi \, e^{-i \un{k} \cdot \left( \un{\zeta} - \un{\xi} \right)} \left(
     \frac{\un{\zeta} + \un{\xi}}{2} \times \un{k}
    \right) 
    \int d^2 w \;d^2 z \int\limits^{p^-_2}_0 \frac{dk_1^-}{2\pi}
     \\\notag
    &\times \Bigg\langle 
    \frac{\un{\xi} - \un{z} }{|\un{\xi} - \un{z}|^2} \cdot  \frac{\un{\zeta} - \un{w} }{|\un{\zeta} - \un{w}|^2}
    \, \mathrm{T \; tr} \left[ 
    V_{\un{\zeta}}^\dagger  \, \vpol{\un{z}, \un{w}}{2}
    \right]
       + i \,
    \frac{\un{\zeta} - \un{w} }{|\un{\zeta} - \un{w}|^2} \times  \frac{\un{\xi} - \un{z} }{|\un{\xi} - \un{z}|^2} 
    \,
    \mathrm{T \; tr} \left[ V_{\un{\zeta}}^\dagger \, \vpol{\un{w}}{1} \right] \delta^2 (\un{w}-\un{z}) 
    \Bigg\rangle + \mathrm{c.c.}.
\end{align}
Adding Eqs.~\eqref{qOAMint2} and \eqref{qbarOAM} and including the complex conjugate terms explicitly, while replacing ${\un k} \to - {\un k}$ in those terms, yields the OAM distribution for the quarks
\begin{align}\label{OAM8}
L_{q + \bar q} (x, Q^2)  \, & = - \frac{8 i P^+}{(2\pi)^5}    \int d^2 k_\perp \, d^{2} \zeta \, d^{2} \xi \, d^2 w \, d^2 z \, e^{- i {\un k} \cdot ({\un \zeta} - {\un \xi})} \left( \frac{{\un \zeta} + {\un \xi}}{2} \times {\un k}\right) \int\limits_0^{p_2^-} \frac{d k_1^-}{2\pi}   \\ 
& \times  \, \left\{ \frac{\ul{\zeta} - {\un w}}{|\ul{\zeta} - {\un w}|^2} \cdot \frac{\ul{\xi} - {\un z}}{|\ul{\xi} - {\un z}|^2} \,
\mbox{Im} \left\langle -\mbox{T} \, \tr \left[ V_{\ul \zeta} \, \left( V_{{\un z}, {\ul w}}^{\textrm{pol} [2]} \right)^\dagger \right] +  \mbox{T} \, \tr \left[ V_{\ul \zeta}^\dagger \, V_{{\un z}, {\ul w}}^{\textrm{pol} [2]} \right] \right\rangle \right. \notag \\
& \left. + \frac{\ul{\zeta} - {\un w}}{|\ul{\zeta} - {\un w}|^2} \times \frac{\ul{\xi} - {\un z}}{|\ul{\xi} - {\un z}|^2} \,
\mbox{Re} \left\langle \mbox{T} \, \tr \left[ V_{\ul \zeta} \, \left( V^{\textrm{pol} [1]}_{\ul w} \right)^\dagger \right] + \mbox{T} \, \tr \left[ V_{\ul \zeta}^\dagger \, V^{\textrm{pol} [1]}_{\ul w} \right] \right\rangle \, \delta^2 ({\un z} - {\un w})  \right\} . \notag
\end{align}

The second term in the curly brackets of \eq{OAM8} is the result obtained in \cite{Kovchegov:2019rrz}\footnote{There is a relative minus sign between the result here and in \cite{Kovchegov:2019rrz}. This is because one has to correct Eq.~(17) of \cite{Kovchegov:2019rrz} by replacing $\sigma \to -\sigma$ in agreement with \eq{polWL} above and with Eq.~(9) of \cite{Cougoulic:2022gbk}.}. Defining the polarized dipole amplitude of the first type \cite{Kovchegov:2018znm} 
\begin{align}\label{Qdef}
    Q_{\un w, \un \zeta}(zs) = \frac{1}{2 N_c} \mathrm{Re} \, \llangle 
    \mathrm{T\;tr}\left[ V_{\un \zeta} \left(V_{\un w}^{\mathrm{pol[1]}}\right)^\dagger \right] + \mathrm{T\;tr}\left[ V_{\un w}^{\mathrm{pol[1]}} V_{\un \zeta}^\dagger \right]
     \rrangle
\end{align}
with the energy-rescaled averaging of sub-eikonal operators defined by \cite{Kovchegov:2015pbl,Kovchegov:2018znm, Cougoulic:2022gbk}
\begin{align}\label{resc}
     \llangle \ldots \rrangle \equiv z \, s \, \Big \langle \ldots \Big \rangle ,
\end{align}
we rewrite this term as 
\begin{align}\label{Qterm}
- \frac{8iN_c}{(2\pi)^6} \int d^2 k_\perp \, d^{2} \zeta \, d^{2} \xi \, d^2 w \, e^{- i {\un k} \cdot ({\un \zeta} - {\un \xi})} \, \left( \frac{{\un \zeta} + {\un \xi}}{2} \times {\un k}\right) \, \int\limits^1_{\frac{\Lambda^2}{s}} \frac{dz}{z} \, \frac{\ul{\zeta} - {\un w}}{|\ul{\zeta} - {\un w}|^2} \times \frac{\ul{\xi} - {\un w}}{|\ul{\xi} - {\un w}|^2} \, Q_{\un w, \un \zeta}(zs).
\end{align}
We have defined the longitudinal momentum fraction $z = k^-_1 / p_2^-$ of some (parent) projectile's momentum $p_2^-$ carried by the soft (anti)-quark in the dipole \cite{Kovchegov:2018znm}. For the OAM distribution in question, $p_2^-$ can be thought of simply as the upper cutoff on the $k_1^-$ integral. The center of mass energy squared is $s = 2 P^+ p_2^-$, such that $z \, s = 2 P^+ k_1^-$.\footnote{There is a subtlety here \cite{Kovchegov:2015pbl, Kovchegov:2018znm, Kovchegov:2021lvz}: $z \, s$ in the argument of $Q$ in \eqref{Qdef} contains the $z$ of the softest parton in the dipole (or of the softest virtual parton in the previous evolution steps), while $z \, s$ in \eq{resc} contains the $z$ variable of the polarized parton. These two are not always the same.} We have also introduced the infrared (IR) cutoff $\Lambda$ for the transverse momenta.

The expression in \eq{Qterm} is simplified in Appendix~\ref{sec:simplifying} below. There, we also simplify the contribution of the first term in the curly brackets of \eq{OAM8}, omitted from the analysis in \cite{Kovchegov:2019rrz}. In doing so we define the following new polarized Wilson line \cite{Kovchegov:2018znm, Cougoulic:2022gbk}
\begin{align}\label{Vi0}
V_{\un{z}}^{i \, \textrm{G} [2]} \equiv \frac{P^+}{2 s} \, \int\limits_{-\infty}^{\infty} d {z}^- \, V_{\un{z}} [ \infty, z^-] \, \left[ {D}^i (z^-, \un{z}) - \cev{D}^i (z^-, \un{z}) \right] \, V_{\un{z}} [ z^-, -\infty]  
\end{align}
along with the polarized dipole amplitude of the second kind \cite{Kovchegov:2018znm, Cougoulic:2022gbk}
\begin{align}
    \label{Gi_def0}
    G^i_{\un w, \un \zeta}(zs) \equiv \frac{1}{2N_c} \mathrm{Re} \llangle \mathrm{T\,tr} \left[V_{\un \zeta}^\dagger V_{\un w}^{i\, G[2]} +  \left(V_{\un w}^{i\, G[2]}\right)^\dagger V_{\un \zeta}\right] \rrangle.
\end{align}
We arrive at
\begin{align}
    \label{qOAMint77}
    L_{q+\bar{q}}(x,Q^2) = & -\frac{8 iN_c}{(2\pi)^5}  \int d^2 k_\perp \, d^{2} x_{10} \, d^2 x_1 \, e^{i {\un k} \cdot \un{x}_{10}} \int\limits_\frac{\Lambda^2}{s}^{1} \frac{d z}{z} \,  \left\{ 
    \left[ - i\,  \frac{\un{x}_{10} \times \un k}{x_{10}^2 \, k^2_\perp} \, \left(\frac{2 \, {\un x}_1 - {\un x}_{10}}{2} \times \un k  \right) + \frac{1}{2} \frac{\un{x}_{10}}{x_{10}^2} \cdot \frac{\un k}{k^2_\perp} \right] Q_{10} (zs) \right. \\ \notag
    & \left.
   - \left[ \left( {\un x}_1 - {\un x}_{10} \right)\times {\un k}\right]
\frac{- k^i \, x_{10}^2 \, ( {\un k} \cdot {\un x}_{10} + i) + 2 i \, {\un k} \cdot {\un x}_{10} \, x_{10}^i}{k_\perp^2 \, x_{10}^4}
G^i_{10} (zs) 
\right\},
\end{align}
where we employ an abbreviated notation for the dipole amplitudes, $Q_{10} = Q_{{\un x}_1, {\un x}_0}$ and $G^i_{10} = G^i_{{\un x}_1, {\un x}_0}$. The integration over $\un x_1$ is performed while keeping ${\un x}_{10}$ fixed. 

At this point it appears desirable to integrate over the impact parameter $\un x_1$. To this end, we define the impact-parameter integrated polarized dipole amplitudes \cite{Cougoulic:2022gbk}
\begin{subequations}
    \label{int_defs}
\begin{align}
   & \int d^2 x_1 \, Q_{10} (zs) = Q(x_{10}^2, zs), \label{Qint_def} \\ 
   &  \int d^2 x_1 \, G^i_{10}(zs) = x_{10}^i \, G_1(x_{10}^2, zs) + \epsilon^{ij} x_{10}^j \, G_2(x_{10}^2, zs).
\end{align}
\end{subequations}
The small-$x$ quark and gluon helicity PDFs in the DLA can all be expressed in terms of the amplitudes $Q(x_{10}^2, zs)$ and $G_2(x_{10}^2, zs)$ from Eqs.~\eqref{int_defs} (the amplitude $G_1(x_{10}^2, zs)$ does not contribute to those quantities) \cite{Cougoulic:2022gbk}. However, a quick inspection of \eq{qOAMint77} reveals that the amplitudes defined in Eqs.~\eqref{int_defs} are not sufficient to describe $L_{q+\bar{q}}(x,Q^2)$ due to the presence of explicit factors of $\un x_1$ in the integrand. We, therefore, need to define the first impact-parameter moments of the dipole amplitudes $I_3, I_4, I_5$, and $I_6$ (hereafter denoted ``moment amplitudes") by
\begin{subequations}
    \label{moment_defs}
    \begin{align}
        & \int d^2 x_1 \, x_1^i \, Q_{10}(zs) = {x}_{10}^i \, I_3(x_{10}^2, zs) + \ldots, \label{moment_defs_Q} \\ 
         & \int d^2 x_1 \, x_1^i G^j_{10} (zs) = \epsilon^{ij} \, x_{10}^2 \, I_{4}(x_{10}^2,zs) + \epsilon^{ik} \, x_{10}^k \, x_{10}^j \, I_{5}(x_{10}^2 , zs) + \epsilon^{jk} \, x_{10}^k \, x_{10}^i \, I_{6}(x_{10}^2, zs) + \ldots. \label{456def}
    \end{align}
\end{subequations}
The ellipsis denote other possible tensor structures, with (in \eq{moment_defs_Q}) and without (in \eq{456def}) $\epsilon^{ij}$. 

Employing Eqs.~\eqref{int_defs} and \eqref{moment_defs} in \eq{qOAMint77} we see that $G_1(x_{10}^2 ,zs)$ does not contribute to $L_{q+\bar{q}}(x, Q^2)$. This is because, as we argued above, only terms with an even number of Levi-Civita symbols $\epsilon^{ij}$ in the integrand survive after all the integrations are carried out. Similarly, other possible tensor structures, denoted by ellipsis in Eqs.~\eqref{moment_defs}, do not contribute to $L_{q+\bar{q}}(x, Q^2)$ because they do not contain the number of $\epsilon^{ij}$ necessary for a non-zero contribution: we omit them here. We emphasize here that the moment amplitudes $I_3, I_4, I_5,$ and $I_6$ are new and do not appear in the helicity evolution of \cite{Cougoulic:2022gbk}, though, as we will shortly see, their small-$x$ DLA evolution can be derived starting from the evolution in \cite{Cougoulic:2022gbk}.

Substituting Eqs.~\eqref{int_defs} and \eqref{moment_defs} into \eq{qOAMint77}, we obtain 
\begin{align}   
    L_{q+\bar{q}}(x,Q^2) &= \frac{8 iN_c}{(2\pi)^5}  \int d^2 k_\perp \, d^2 x_{10} \, e^{i {\un k} \cdot \un x_{10}} \int\limits_\frac{\Lambda^2}{s}^{1} \frac{d z}{z} \\ \notag
    & \times \Bigg\{ 
      -(\un k \cdot \un x_{10} + i) \left[I_4(x_{10}^2, zs) - I_6(x_{10}^2, zs) \right] + \frac{\un k \cdot \un x_{10} }{k_\perp^2 \, x_{10}^2} \left[I_3(x_{10}^2, zs) - Q(x_{10}^2, zs) - 3\, G_2(x_{10}^2, zs) \right] 
      \\ \notag
    & + i\, \frac{(\un k \cdot \un x_{10} )^2 }{k_\perp^2 \, x_{10}^2} \left[I_5(x_{10}^2, zs) - I_6(x_{10}^2, zs) + 2 \, I_4(x_{10}^2, zs)\right] -  \frac{(\un k \cdot \un x_{10} )^3 }{k_\perp^2 \, x_{10}^2} \left[I_5(x_{10}^2, zs) + I_6(x_{10}^2, zs)\right] 
    \Bigg \}.
 \end{align}

Now we integrate over $\un k$ (with an upper limit of $Q^2$ on $k_\perp^2$), average over the angles of $\un x_{10}$, and impose lifetime ordering  $\frac{1}{x} \gg zs x_{10}^2 \gg 1$ \cite{Kovchegov:2015pbl, Cougoulic:2019aja}. In doing so, we discard terms that contribute a $\delta^2(\un x_{10})$. Such terms are suppressed by one logarithm of energy compared to the terms we keep below, and are, therefore, outside of our DLA approximation. We get
\begin{align}
    \label{final_quark_OAM}
    L_{q+\bar{q}}(x, Q^2) = \frac{N_c N_f}{2\pi^3} \int \displaylimits^1_{\Lambda^2/s} \frac{dz}{z} \int \displaylimits^{\mathrm{min}\left\{ \frac{1}{z Q^2}, \frac{1}{\Lambda^2}\right\}}_{\mathrm{max} \left\{ \frac{1}{zs}, \frac{1}{Q^2} \right\}} \frac{d x_{10}^2}{x_{10}^2} \Big[&Q(x_{10}^2,zs) +  3\, G_2(x_{10}^2,zs) - I_3(x_{10}^2,zs)  \\ \notag 
    & + 2\, I_4(x_{10}^2,zs) - I_5(x_{10}^2,zs) - 3\, I_6(x_{10}^2,zs) \Big],
\end{align}
where we have picked up a factor of $N_f$ from summing over quark flavors, while assuming, for simplicity, that all the dipole and moment amplitudes in the integrand are flavor-independent. This assumption needs to be revised in phenomenology (see \cite{Adamiak:2023yhz}), where $Q$ and, hence, $I_3$ would depend on the quark flavor.  

We conclude from \eq{final_quark_OAM} that in order to determine the quark OAM distribution at small $x$, we would need to know not only the impact-parameter integrated dipole amplitudes $Q$ and $G_2$, but also the moment amplitudes $I_3, I_4, I_5$, and $I_6$. To determine the latter we will need to construct small-$x$ evolution equations for the moment amplitudes starting from the equations obtained in \cite{Cougoulic:2022gbk}. This will be done below in the large-$N_c$ limit.

The result \eqref{final_quark_OAM} can be compared to the flavor-singlet quark helicity PDF at small-$x$ \cite{Cougoulic:2022gbk} (see \cite{Borden:2024bxa} for a correction to one of the integration limits)
\begin{align}
    \label{quark_hel}
    \Delta \Sigma(x, Q^2) = -\frac{N_c N_f}{2\pi^3} \int \displaylimits^1_{\Lambda^2/s} \frac{dz}{z} \int \displaylimits^{\mathrm{min}\left\{ \frac{1}{z Q^2}, \frac{1}{\Lambda^2}\right\}}_{\mathrm{max} \left\{ \frac{1}{zs}, \frac{1}{Q^2} \right\}} \frac{d x_{10}^2}{x_{10}^2} \Big[Q(x_{10}^2,zs) + 2\, G_2(x_{10}^2,zs) \Big].
\end{align}
We see that while quark helicity PDF is expressible in terms of the polarized dipole amplitudes, $Q(x_{10}^2, zs)$ and $G_2(x_{10}^2, zs)$, the quark OAM distribution involves not only these amplitudes but the moment amplitudes $I_3(x_{10}^2, zs), I_4(x_{10}^2, zs), I_5(x_{10}^2, zs),$ and $I_6(x_{10}^2, zs)$. Because of this dependence on the moment amplitudes and the lack of a linear combination $Q(x_{10}^2,zs) + 2\, G_2(x_{10}^2,zs)$ appearing in the quark OAM, there is no clear relation between the quark helicity PDF and the quark OAM at the level exhibited here. This is to be compared with Eqs.~(31) and (53) of \cite{Kovchegov:2019rrz} where such a relation was suggested, based on, as we now find, an incomplete analysis and the original (uncorrected) KPS evolution. To make any analytic statements about the ratio between the quark OAM and the quark helicity PDF will require analytic expressions for the polarized dipole and moment amplitudes. We leave this for future work \cite{Manley:2023}. However, we can still determine the ratio of these two distributions numerically in the large-$N_c$ limit. This will be done in Sec.~\ref{sec:ratios}.


\subsection{Gluon OAM}

\label{sec:gluon_OAM}

Let us now consider the gluon OAM distribution. We start with the expression for the gluon OAM distribution given by Eq.~(61) in \cite{Kovchegov:2019rrz}, 
\begin{align}
    \label{gOAMint1}
    L_G(x,Q^2) = \frac{4}{(2\pi)^3 x} &\int d\xi^- d^2 \xi_\perp d\zeta^- d^2 \zeta_\perp d^2 k_\perp \left(\frac{\un{\zeta} + \un{\xi}}{2} \times \un{k} \right) e^{i x P^+ (\xi^- - \zeta^-) - i \un{k} \cdot (\un{\xi} - \un{\zeta})} \notag \\
    & \times \left\langle 
    \mathrm{tr} \left[ 
    V_{\un{\zeta}}[-\infty, \zeta^-] F^{+i}(0^+, \zeta^-, \un{\zeta}) V_{\un{\zeta}}[\zeta^-, +\infty] V_{\un{\xi}}[+\infty, \xi^-] F^{+i}(0^+, \xi^-, \un{\xi}) V_{\un{\xi}}[\xi^-, -\infty]
    \right]
    \right \rangle .
\end{align}
As in \cite{Kovchegov:2019rrz}, we note this is the gluon {\sl dipole} OAM distribution: in Appendix~C of \cite{Kovchegov:2019rrz}, it was shown to be consistent with the canonical Jaffe-Manohar gluon OAM definition \cite{Jaffe:1989jz}.

We would like to simplify \eq{gOAMint1} at small $x$ following \cite{Hatta:2016aoc, Kovchegov:2017lsr, Cougoulic:2022gbk}. In any gauge where $A_\perp \to 0$ at $x^- \to \pm \infty$, we have 
\begin{align}\label{FtoA}
    \int^\infty_{-\infty} d\xi^- e^{ix P^+ \xi^-} V_{\un{\xi}}[\infty, \xi^-] F^{+j}(\xi) V_{\un{\xi}}[\xi^-, -\infty] = -\int^\infty_{-\infty} d\xi^- e^{ixP^+ \xi^-} V_{\un{\xi}}[\infty, \xi^-] \left( 
    \partial^j A^+ + i x P^+ A^j 
    \right) V_{\un{\xi}}[\xi^-, -\infty].
\end{align}
Further, expanding in $x$ following \cite{Cougoulic:2022gbk}, we can show that 
\begin{align}\label{simplification}
\int\limits_{-\infty}^\infty d \xi^- \, e^{i x P^+ \, \xi^-} \, V_{{\un \xi}} [+\infty, \xi^-] \, (\pd^i A^+ + i x P^+ A^i) (\xi) \,  V_{{\un \xi}} [\xi^-, -\infty] = \frac{1}{i g} \, \pd^i V_{\un \xi} - \frac{x \, s}{g} \, V_{\un \xi}^{i \, \textrm{G} [2]} + {\cal O} (x^2), 
\end{align}
where $V_{\un \xi}^{i \, \textrm{G} [2]}$ is defined in \eq{Vi0} above. Substituting Eqs.~\eqref{FtoA} and \eqref{simplification} into \eq{gOAMint1}, after some algebra one obtains 
\begin{align}\label{gOAMint4}
L_G (x, Q^2) = \frac{8 N_c}{g^2 \, (2\pi)^3} \,  \int d^2 x_1 \ d^2 x_0 \, d^2 k_\perp \, 
  e^{i \un{k} \cdot {\un x}_{10}} 
\left[ k^i \, \left( \frac{{\un x}_1 + {\un x}_0}{2} \times {\un k} \right)  
+ \frac{i}{2} \,  \epsilon^{ij} \, k^j \right] \, G^i_{10} \left( z s = \frac{Q^2}{x} \right).
\end{align} 

This is exactly what was found in \cite{Kovchegov:2019rrz}, up to the difference in definitions of the polarized dipole amplitude of the second type $G^i_{10}(zs)$, which also accounts for the sign difference between our \eq{gOAMint4} and  Eq.~(72) of \cite{Kovchegov:2019rrz}. Consequently, the discussion following Eq. (72) of \cite{Kovchegov:2019rrz} is still valid here since the revised helicity evolution for the polarized dipole amplitudes (and the revised moment-amplitude evolution below) does not alter the initial conditions of the polarized dipole amplitudes. Thus, at Born level we have $L_G(x,Q^2) = - \Delta G(x,Q^2)$ (cf. \cite{Hatta:2016aoc}). Note that, as we will see below, this relation between the gluon OAM and helicity distributions does not survive DLA small-$x$ evolution. 

As in the quark sector, we will work here with the first impact-parameter moments defined in \eqs{moment_defs}. Utilizing those definitions in \eq{gOAMint4} and integrating over $\un k$ (with $k_\perp < Q$) yields 
\begin{align}
    \label{final_gluonOAM}
    L_G(x,Q^2) &= - \frac{2 \, N_c}{\as \pi^2} \Bigg\{ 
    \left[2+ 6\, x_{10}^2 \frac{\partial}{\partial x_{10}^2} + 2 \, x_{10}^4 \frac{\partial^2}{\partial (x_{10}^2)^2} \right] \left[ I_4(x_{10}^2, zs) + I_5(x_{10}^2, zs) \right] 
    \\ \notag
    & \hspace{2cm} + \left[1 + x_{10}^2 \frac{\partial}{\partial x_{10}^2} \right] \left[ I_5(x_{10}^2, zs) + I_6(x_{10}^2, zs) \right]
    \Bigg\}_{x_{10}^2 = 1/Q^2, \,zs = Q^2/x}.
\end{align}
Somewhat surprisingly, the gluon OAM distribution is given entirely by the moment amplitudes from \eqs{moment_defs}, without the explicit dependence on the dipole amplitudes $Q$ and $G_2$. This conclusion is, however, consistent with Eq.~(79) in \cite{Kovchegov:2019rrz}. Again, the evolution equations for the moment amplitudes and their numerical solutions are constructed below.

Equation \eqref{final_gluonOAM} can be compared to the corresponding expression for the gluon helicity distribution at small $x$ \cite{Kovchegov:2017lsr,  Cougoulic:2022gbk} 
\begin{align}
    \label{gluon_hel}
    \Delta G(x,Q^2) = \frac{2 N_c}{\as \pi^2} \left[\left( 1+ x_{10}^2 \frac{\partial}{\partial x_{10}^2}\right) G_2 \left(x_{10}^2, zs = \frac{Q^2}{x} \right)\right]_{x^2_{10} = 1/Q^2}. 
\end{align}
Again, and this time similar to \cite{Kovchegov:2019rrz}, there is no clear connection between $L_G(x,Q^2)$ and $\Delta G(x, Q^2)$ here. Below, in Sec.~\ref{sec:ratios} we investigate the ratio of the gluon OAM to the gluon helicity PDF numerically. 

Before we derive evolution equations for the moment amplitudes, we briefly comment on the non-uniqueness of the moment definitions in \eqs{moment_defs}.


\subsection{Alternative definition of the moment amplitudes}

\label{sec:alt}

Here we comment on an alternative to \eqs{moment_defs} definition of the moment amplitudes. Instead of using \eqs{moment_defs}, we could define the first impact-parameter moments of the polarized dipole amplitudes in the following way,
\begin{subequations}
    \label{alt_moment_defs}
\begin{align}
    \int d^2 x_1 \, x_1^i \,  x^j_{10} \, G^j_{10}(zs) &= \epsilon^{ij} \, x_{10}^j \, x_{10}^2 \, J_1 (x_{10}^2, zs) + \ldots , \\
    \int d^2 x_1 \, x_1^i \, \nabla_{10}^j \, G^j_{10}(zs) &= \epsilon^{ij} \, x_{10}^j  \, J_2 (x_{10}^2, zs) + \ldots , 
\end{align}
\end{subequations}
where, similar to \eq{456def}, we have denoted by ellipsis the tensor structures that do not contain an $\epsilon^{ij}$ needed to contribute to the OAM distributions. Note we have not altered the definition of $I_3(x_{10}^2, zs)$ in \eq{moment_defs_Q}. The definitions \eqref{alt_moment_defs} are closer to the original moment definition used in \cite{Kovchegov:2019rrz} (see Eq.~(82) there). 

Using \eqs{alt_moment_defs} in Eqs.~(\ref{qOAMint77}) and (\ref{gOAMint4}) above, and performing steps similar to the above to simplify the OAM distributions at small $x$, we get 
\begin{align}
    L_{q+\bar{q}}(x, Q^2) & = \frac{N_c N_f}{2\pi^3} \int\displaylimits^1_{\frac{\Lambda^2}{s}} \frac{dz}{z} \int \displaylimits^{\mathrm{min}\left\{\frac{1}{z Q^2} , \frac{1}{\Lambda^2} \right\}}_{\mathrm{max} \left\{ \frac{1}{zs}, \frac{1}{Q^2} \right\}} \frac{d x_{10}^2}{x_{10}^2} \Big[
        Q(x_{10}^2, zs) + 3\, G_2(x_{10}^2, zs) \\ \notag 
        & \hspace{6cm} 
        - I_3(x_{10}^2, zs) - 3\, J_2(x_{10}^2, zs) + 8\, J_1(x_{10}^2, zs) 
    \Big]
\end{align}
in the quark sector and 
\begin{align}
    L_G(x,Q^2) = - \frac{2 N_c}{\as \pi^2} \left[\left( 1+ x_{10}^2 \frac{\partial}{\partial x_{10}^2}\right) J_2 \left(x_{10}^2, zs = \frac{Q^2}{x} \right)\right]_{x^2_{10} = 1/Q^2}
\end{align}
in the gluon sector. Now, it is clear that the original moment definitions were redundant - we can describe the same OAM distributions using only three degrees of freedom ($I_3, J_1, J_2$) instead of four ($I_3, I_4, I_5, I_6$). Furthermore, with the moment definitions in \eq{alt_moment_defs}, the gluon OAM assumes a compact form that mirrors the gluon helicity distribution, \eq{gluon_hel}. Below, we will see that, in the DLA, $I_6 = \frac{1}{2} G_2$ (up to initial conditions). This is likely due to the redundancy in the moments definitions we use above in \eqs{moment_defs}. 

In the DLA, the moments defined by \eqs{alt_moment_defs} constitute linear combinations of the original moments from \eqs{moment_defs}. Indeed, acting on \eqs{456def} with $x_{10}^j$ and $\nabla_{10}^j$ and employing \eqs{moment_defs}, we get the following relations between the two types of moments 
\begin{subequations}
\begin{align}
    J_1(x_{10}^2 ,zs) &= I_4(x_{10}^2, zs) + I_5(x_{10}^2,zs), \\ \label{alt_rel2}
    J_2(x_{10}^2, zs) &= 2\, I_4(x_{10}^2, zs) + 3\, I_5(x_{10}^2, zs) + I_6(x_{10}^2, zs) + 
    2 \, x_{10}^2 \, \frac{\pd}{\pd x_{10}^2} \left[I_4(x_{10}^2, zs) + I_5(x_{10}^2, zs) \right].
\end{align}
\end{subequations}
The last term in \eq{alt_rel2} contains a logarithmic derivative with respect to $x_{10}^2$: this removes one logarithm from $I_4$ and $I_5$ and should be discarded in the DLA as sub-leading. Thus, for the DLA evolution derived below, the moments from \eqs{alt_moment_defs} are nothing but linear combinations of the moments from \eqs{moment_defs}. 

Therefore, alternative moment definitions exist and we mention them for completeness.\footnote{We also acknowledge that there are other choices besides \eqs{alt_moment_defs} and \eqs{moment_defs}. For example, instead of taking the $\un x_1$-moment, one could extract a general moment with $\un x_1 - \beta \un x_{10}$ as the weight, where $\beta$ is an arbitrary constant (e.g., $\beta = \frac{1}{2}$ would correspond to a dipole impact-parameter moment). However, as one can show, the asymptotics of the OAM distributions and the conclusions of Section~\ref{sec:num_sol} are independent of $\beta$ or of any other alternative definition of the moment amplitudes. Therefore, we choose $\beta=0$ here for simplicity.} However, despite the apparent simplicity they induce in the OAM distributions, they do not appear to give any more physical insight than the original moment definitions do. For this reason, above and throughout the rest of this paper, we choose to work with the definitions in \eqs{moment_defs}. Now, let us turn our attention to the evolution of the moment amplitudes.


\section{Derivation of evolution equations for the moment amplitudes at large-$N_c$}

\label{sec:evolution}

Our next goal is to derive small-$x$ evolution equations for the moment amplitudes $I_3, I_4, I_5$, and  $I_6$ in the DLA. We will work in the large-$N_c$ limit. We will use the large-$N_c$ (pure-glue) analogue of $Q_{10}(zs)$, obtained from the latter by dropping the quark operator from it, namely \cite{Cougoulic:2022gbk} (cf. \eq{Qdef})
\begin{align}
    \label{G10_def}
    G_{10}(zs) \equiv \frac{1}{2 N_c} \mathrm{Re} \, \llangle \mathrm{T\, tr} \left[V_{\un 0 } \, \left( V_{\un 1}^{\mathrm{G[1]}}\right)^\dagger \right] + \mathrm{T\, tr} \left[V_{\un 1 }^{\mathrm{G[1]}} \, V_{\un 0}^{\dagger}\right] \rrangle
\end{align}
with
\begin{align}
    \int d^2 x_1 \, G_{10}(zs) \equiv G(x_{10}^2, z s) .
\end{align}

To determine the evolution of $I_3(x_{10}^2,zs)$, let us start with the large-$N_c$ evolution equation for $G_{10}(zs)$. From Eq.~(118) of \cite{Cougoulic:2022gbk}, we have 
\begin{align}
    \label{118}
    G_{10}(zs) &= G^{(0)}_{10}(zs) + \frac{\as N_c}{2\pi^2} \int \displaylimits^z_{\frac{\Lambda^2}{s}} \frac{dz^\prime}{z^\prime} \int d^2 x_2 \Bigg\{ 
    2 \left[ \frac{1}{x_{21}^2} - \frac{\un{x}_{21}}{x^2_{21}} \cdot \frac{\un{x}_{20}}{x^2_{20}} \right]\left[S_{20}(z^\prime s) \,G_{21}(z^\prime s) + S_{21}(z^\prime s) \, \Gamma^{\mathrm{gen}}_{20,21}(z^\prime s) \right]  \\ \notag
    & +\left[
     2 \frac{\epsilon^{ij} x_{21}^j}{x_{21}^4} - \frac{\epsilon^{ij} \left( x_{20}^j + x_{21}^j \right)}{x_{21}^2 x_{20}^2} - \frac{2 \, \un{x}_{20} \times \un{x}_{21}}{x_{20}^2 x_{21}^2} \left(
     \frac{x_{21}^i}{x_{21}^2} - \frac{x_{20}^i}{x_{20}^2} 
     \right)
    \right] \left[S_{20}(z^\prime s)\,  G_{21}^i(z^\prime s) + S_{21}(z^\prime s) \, \Gamma^{i \, \mathrm{gen}}_{20,21}(z^\prime s) \right]  \\ \notag
    &+ \frac{x_{10}^2}{x_{21}^2 x_{20}^2} \left[S_{20}(z^\prime s) \,  G_{12}(z^\prime s) - \Gamma^{\mathrm{gen}}_{10,21}(z^\prime s) \right]
    \Bigg\} , 
\end{align}
where the generalized polarized dipole amplitudes are \cite{Kovchegov:2017lsr}
\begin{subequations}\label{generals}
\begin{align}
    \Gamma^\mathrm{gen}_{10,21}(z s) &\equiv G_{10}(zs) \, \theta(x_{21} - x_{10}) + \Gamma_{10,21}(zs) \, \theta(x_{10} - x_{21}), \\ 
    \Gamma^{i\, \mathrm{gen}}_{10, 21}(zs) &\equiv G^i_{10}(zs) \, \theta(x_{21} - x_{10}) + \Gamma^i_{10,21} \, \theta(x_{10} - x_{21}).
\end{align}
\end{subequations}
Here, $\Gamma_{10,21}(zs)$ and $\Gamma^i_{10,21}(zs)$ are the ``neighbor" dipole amplitudes --- auxiliary functions necessary to enforce lifetime ordering in the evolution \cite{Kovchegov:2015pbl}. Their operator definitions are the same as for $G_{10}$ and $G^i_{10}$ in Eqs.~\eqref{G10_def} and \eqref{Gi_def0}, respectively, but the lifetime cutoff on their evolution, which is not shown explicitly in those definitions, is different, dependent on the transverse size of the adjacent dipole \cite{Kovchegov:2018znm, Cougoulic:2019aja, Cougoulic:2022gbk}. In \eq{118} and below, the inhomogeneous term in the integral equations is given by the initial condition, and is denoted by the superscript $(0)$. 

Since we are working purely in the DLA, we set $S_{21} = S_{20} = 1$ as these unpolarized $S$-matrices deviate from $1$ only at the single logarithmic level. Next, we multiply each side of \eq{118} by a factor of $x_1^m$ and integrate over $x_1$ (while keeping $\un x_{10}$ and other inter-parton distances fixed). We get 
\begin{align}
    \label{nonsimp_i3evoeqn}
   & x_{10}^m I_3(x_{10}^2,zs)= x_{10}^m I_3^{(0)} (x_{10}^2,zs) + \frac{\as N_c}{2\pi} \int\displaylimits^z_{\frac{\Lambda^2}{s}} \frac{dz^\prime}{z^\prime} \int d^2 x_2 \\ \notag 
    & \times 
    \Bigg\{ 
        2 \left[ \frac{1}{x_{21}^2} - \frac{\un{x}_{21}}{x^2_{21}} \cdot \frac{\un{x}_{20}}{x^2_{20}} \right]\left[x_{21}^m \left( I_3(x_{21}^2,z^\prime s) - G(x_{21}^2,z^\prime s) - \Gamma^{\mathrm{gen}}(x_{20}^2, x_{21}^2, z^\prime s)\right) - x_{20}^m \, \Gamma^{\mathrm{gen}}_3 (x_{20}^2, x_{21}^2, z^\prime s)\right] \\ \notag 
    & \hspace{1cm}+ \left[
     2 \frac{\epsilon^{ij} x_{21}^j}{x_{21}^4} - \frac{\epsilon^{ij} \left( x_{20}^j + x_{21}^j \right)}{x_{21}^2 x_{20}^2} - \frac{2 \, \un{x}_{20} \times \un{x}_{21}}{x_{20}^2 x_{21}^2} \left(
     \frac{x_{21}^i}{x_{21}^2} - \frac{x_{20}^i}{x_{20}^2} 
     \right)
    \right] 
    \\ \notag
       & \hspace{1.5cm}
    \times \Big[
        \epsilon^{mi} x_{21}^2 I_4(x_{21}^2, z^\prime s) + \epsilon^{mk} x_{21}^k \, x_{21}^i \, I_5(x_{21}^2, z^\prime s) 
        + \epsilon^{ik} \, x_{21}^k \, x_{21}^m \left[ I_6(x_{21}^2, z^\prime s) - G_2(x_{21}^2,z^\prime s) \right] 
        \\ \notag 
        & \hspace{3cm}
        + \epsilon^{mi} x_{20}^2 \, \Gamma^{\mathrm{gen}}_4(x_{20}^2,x_{21}^2, z^\prime s) + \epsilon^{mk} x_{20}^k \, x_{20}^i \, \Gamma^{\mathrm{gen}}_5(x_{20}^2,x_{21}^2, z^\prime s) 
        \\ \notag 
        & \hspace{3cm} 
        + \epsilon^{ik} x_{20}^k \, x_{20}^m \, \Gamma^{\mathrm{gen}}_6(x_{20}^2, x_{21}^2, z^\prime s) - \epsilon^{ik} x_{20}^k \, x_{21}^m \, \Gamma^{\mathrm{gen}}_2 (x_{20}^2, x_{21}^2,z^\prime s) 
    \Big] \\ \notag
    & \hspace{2cm} + \frac{x_{10}^2}{x_{21}^2 x_{20}^2} \left[-x_{21}^m I_3(x_{21}^2, z^\prime s) - x_{10}^m \Gamma^{\mathrm{gen}}_3 (x_{10}^2, x_{21}^2, z^\prime s) \right]
    \Bigg\} ,
\end{align}
where we have, analogously to Eqs.~\eqref{int_defs} and \eqref{moment_defs}, defined 
\begin{subequations}
    \label{genNeigh_defs}
\begin{align}
    & \int d^2 x_1 \, \Gamma^{\mathrm{gen}}_{10,21}(zs) = \Gamma^{\mathrm{gen}}(x_{10}^2,x_{21}^2, z^\prime s), \\
    & \int d^2 x_1 \, \Gamma^{i\, \mathrm{gen}}_{10, 21}(zs) = \epsilon^{ik} x_{10}^k \, \Gamma^{\mathrm{gen}}_2(x_{10}^2, x_{21}^2, z^\prime s) + \ldots , \\ 
    & \int d^2 x_1 \, x_1^m \, \Gamma^{\mathrm{gen}}_{10,21} (zs) = x_{10}^m \, \Gamma_3^\mathrm{gen} (x_{10}^2, x_{21}^2, zs) + \ldots , \\
    & \int d^2 x_1 \, x^m_1 \, \Gamma^{i\, \mathrm{gen}}_{10, 21}(zs) \\ 
    & \hspace*{1cm} = \epsilon^{mi} x_{10}^2 \, \Gamma^{\mathrm{gen}}_4(x_{10}^2, x_{21}^2, z^\prime s)  + \epsilon^{mk} x_{10}^k \, x_{10}^i \, \Gamma^{\mathrm{gen}}_5(x_{10}^2, x_{21}^2, z^\prime s) + 
    \epsilon^{ik} x_{10}^k \, x_{10}^m  \, \Gamma^{\mathrm{gen}}_6(x_{10}^2, x_{21}^2, z^\prime s) + \ldots \, . \notag
\end{align}
\end{subequations}
Here every generalized polarized dipole amplitude can be written by analogy to \eqs{generals},
\begin{subequations}
    \begin{align}
        & \Gamma^{\mathrm{gen}} (x_{10}^2, x_{21}^2, z^\prime s) = G (x_{10}^2, z^\prime s) \, \theta (x_{21} - x_{10}) + \Gamma (x_{10}^2, x_{21}^2, z^\prime s) \, \theta (x_{10} - x_{21} ) , \\
        & \Gamma^{\mathrm{gen}}_2(x_{10}^2, x_{21}^2, z^\prime s) = G_2 (x_{10}^2, z^\prime s) \, \theta (x_{21} - x_{10}) + \Gamma_2(x_{10}^2, x_{21}^2, z^\prime s) \, \theta (x_{10} - x_{21} ) , \\
        & \Gamma^{\mathrm{gen}}_p (x_{10}^2, x_{21}^2, z^\prime s) = I_p (x_{10}^2, z^\prime s) \, \theta (x_{21} - x_{10}) + \Gamma_p (x_{10}^2, x_{21}^2, z^\prime s) \, \theta (x_{10} - x_{21} ) ,
    \end{align}
\end{subequations}
for $p=3, 4,5,6$. As before, there are other possible tensor structures in \eqs{genNeigh_defs}, denoted by ellipsis, but, as one can check, they do not contribute to \eq{nonsimp_i3evoeqn}, and we omit them here. \eq{nonsimp_i3evoeqn} has divergences both in the IR ($ x_{21} \approx x_{20} \gg x_{10}$) and the ultraviolet (UV) ($x_{21} \ll x_{10} \approx x_{20}$) regions of the $x_{21}$ integral. Note that there are no divergences coming from the $x_{20} \ll x_{10} \approx x_{21}$ UV region. Also note that to single out the IR divergence, we rewrite ${\un x}_{20} = {\un x}_{21} + {\un x}_{10}$ and expand the kernel in the powers of $x_{10}/x_{21}$; for the UV divergence we expand in the powers of $x_{21}/x_{10}$.  Keeping only the divergent terms\footnote{In extracting the DLA parts of the evolution equations here and below, we assume that the impact-parameter integrated amplitudes, $G, G_2, I_3, I_4, I_5, I_6, \Gamma, \Gamma_2, \Gamma_3, \Gamma_4, \Gamma_5, \Gamma_6$, are not proportional to the non-zero integer powers of the dipole sizes. We further assume that dependence on the dipole sizes enters only as perturbatively small ($\sim \sqrt{\as}$ or $\sim \as$) powers or logarithms of the dipole sizes. These assumptions are the same as those used in \cite{Kovchegov:2018znm, Cougoulic:2022gbk} and are supported by explicit calculations of the Born-level initial conditions for the moment amplitudes (see \cite{Cougoulic:2022gbk, Kovchegov:2016zex, Kovchegov:2017lsr} for the initial conditions for the impact-parameter integrated polarized dipole amplitudes).}, after some algebra we arrive at the DLA evolution equation for $I_3$,
\begin{align}
    \label{I3_eqn}
    I_3(x_{10}^2, zs) &= I_3^{(0)}(x_{10}^2, zs)  \\ \notag 
    & 
    + \frac{\as N_c}{4\pi} \iUV \,\big[
    2\,\Gamma_3 (x_{10}^2, x_{21}^2, z^\prime s) - 4\,\Gamma_4(x_{10}^2, x_{21}^2, z^\prime s) + 2 \, \Gamma_5(x_{10}^2, x_{21}^2, z^\prime s) \\ \notag
    & \hspace{8cm} + 6\, \Gamma_6(x_{10}^2, x_{21}^2, z^\prime s) -2 \, \Gamma_2(x_{10}^2, x_{21}^2, z^\prime s) \big]\\ \notag &+ \frac{\as N_c}{4\pi}  \iIR \, 
        \big[ 4\, I_3(x_{21}^2, z^\prime s)  -4\, I_4(x_{21}^2, z^\prime s) + 2\, I_5(x_{21}^2, z^\prime s) \\ \notag
       & \hspace{8cm} + 6\, I_6(x_{21}^2, z^\prime s) 
        -4 \, G(x_{21}^2, z^\prime s)  - 6 \, G_2(x_{21}^2, z^\prime s) \big].
\end{align}
When imposing the integration limits in \eq{I3_eqn} we have employed the $x^-$-lifetime ordering \cite{Kovchegov:2015pbl, Cougoulic:2019aja, Cougoulic:2022gbk} and cut off all the dipole sizes by $1/\Lambda$ in the IR.

To derive evolution equations for $I_4, I_5,$ and $I_6$, we start with the evolution equation for $G^i_{10}(zs)$ (Eq.~(128) in \cite{Cougoulic:2022gbk}), which reads 
\begin{align}\label{Gi_eq}
    & G^i_{10}(zs) = G^{i(0)}_{10}(zs) + \frac{\as N_c}{4\pi^2} \int \displaylimits^z_{\frac{\Lambda^2}{s}} \frac{dz^\prime}{z^\prime}  \int d^2 x_2 \Bigg\{
    2 \, \frac{x_{10}^2}{x_{21}^2 x_{20}^2} \left[S_{20}(z^\prime s) G_{12}^i(z^\prime s) - \Gamma^{i \, \mathrm{gen}}_{10,21}(z^\prime s) \right] \\ \notag
    & + 2 \left[\frac{\epsilon^{ij} x_{21}^j}{x_{21}^2} - \frac{\epsilon^{ij} x_{20}^j}{x_{20}^2} + 2 \, x_{21}^i \frac{{\un x}_{21} \times {\un x}_{20}}{x_{21}^2 x_{20}^2}\right]\left[S_{20}(z^\prime s) \, G_{21}(z^\prime s) + S_{21}(z^\prime s) \, \Gamma^{\mathrm{gen}}_{20,21}(z^\prime s) \right]
    \\\notag
    & + \left[\delta^{ij}\left(\frac{3}{x_{21}^2} - 2 \frac{{\un x}_{20} \cdot {\un x}_{21}}{x_{20}^2 x_{21}^2} - \frac{1}{x_{20}^2} \right) - 2 \frac{x_{21}^i x_{20}^j}{x_{21}^2 x_{20}^2} \left(2 \frac{{\un x}_{20} \cdot {\un x}_{21}}{x_{20}^2} + 1 \right) + 2 \frac{x_{21}^i x_{21}^j}{x_{21}^2 x_{20}^2}\left(2 \frac{{\un x}_{20} \cdot {\un x}_{21}}{x_{21}^2} + 1 \right) + 2 \frac{x_{20}^i x_{20}^j}{x_{20}^4} - 2 \frac{x_{21}^i x_{21}^j}{x_{21}^4}\right]
    \\ \notag 
    & \times \left[S_{20}(z^\prime s) \, G_{21}^j(z^\prime s) + S_{21}(z^\prime s) \, \Gamma^{j \, \mathrm{gen}}_{20,21}(z^\prime s) \right]
    \Bigg\}.
\end{align}
Once again we set $S_{21} = S_{20} = 1$ since we are only interested in the DLA result. Multiplying \eq{Gi_eq} by $x_1^m$ and integrating over $x_1$ (while keeping $\un x_{10}$ and other inter-parton distances fixed) we arrive at 
\begin{align}
    \label{nosimp_i4}
   &\epsilon^{mi} x_{10}^2 I_4(x_{10}^2, zs) + \epsilon^{mk} x_{10}^k x_{10}^i I_5(x_{10}^2, zs) + \epsilon^{ik} x_{10}^k x_{10}^m I_6(x_{10}^2, zs) = \\ \notag
   &
   \epsilon^{mi} x_{10}^2 I^{(0)}_4(x_{10}^2, zs) + \epsilon^{mk} x_{10}^k x_{10}^i I^{(0)}_5(x_{10}^2, zs) + \epsilon^{ik} x_{10}^k x_{10}^m I^{(0)}_6(x_{10}^2, zs)  \\ \notag 
   & + \frac{\as N_c}{4\pi^2} \int \displaylimits^z_{\frac{\Lambda^2}{s}} \frac{dz^\prime}{z^\prime}  \int d^2 x_2 \Bigg\{
    2 \, \frac{x_{10}^2}{x_{21}^2 x_{20}^2} \Big[\epsilon^{mi} x_{21}^2 I_4(x_{21}^2, z^\prime s) + \epsilon^{mk} x_{21}^k x_{21}^i I_5(x_{21}^2 , z^\prime s) + \epsilon^{ik} x_{21}^k x_{21}^m I_6(x_{21}^2 , z^\prime s)
    \\ \notag 
    & \hspace{4cm}   - \epsilon^{mi} x_{10}^2 \Gamma^{\mathrm{gen}}_4(x_{10}^2, x_{21}^2, z^\prime s)
    - \epsilon^{mk} x_{10}^k x_{10}^i \Gamma^{\mathrm{gen}}_5(x_{10}^2, x_{21}^2, z^\prime s) - \epsilon^{ik} x_{10}^k x_{10}^m \Gamma^{\mathrm{gen}}_6(x_{10}^2, x_{21}^2, z^\prime s)
    \Big] \\ \notag
    & + 2 \, \left[\frac{\epsilon^{ij} x_{21}^j}{x_{21}^2} - \frac{\epsilon^{ij} x_{20}^j}{x_{20}^2} + 2 \, x_{21}^i \frac{{\un x}_{21} \times {\un x}_{20}}{x_{21}^2 x_{20}^2}\right] \Big[
        x_{21}^m (I_3(x_{21}^2, z^\prime s) - G(x_{21}^2, z^\prime s) - \Gamma^{\mathrm{gen}}(x_{20}^2, x_{21}^2, z^\prime s)) 
       + x_{20}^m  \Gamma^{\mathrm{gen}}_3(x_{20}^2, x_{21}^2, z^\prime s) 
    \Big] \\ \notag
   &  + \Bigg[\delta^{ij}\left(\frac{3}{x_{21}^2} - 2 \, \frac{{\un x}_{20} \cdot {\un x}_{21}}{x_{20}^2 x_{21}^2} - \frac{1}{x_{20}^2} \right) - 2 \, \frac{x_{21}^i x_{20}^j}{x_{21}^2 x_{20}^2} \left(2 \frac{{\un x}_{20} \cdot {\un x}_{21}}{x_{20}^2} + 1 \right) + 2 \frac{x_{21}^i x_{21}^j}{x_{21}^2 x_{20}^2}\left(2 \frac{{\un x}_{20} \cdot {\un x}_{21}}{x_{21}^2} + 1 \right) + 2 \frac{x_{20}^i x_{20}^j}{x_{20}^4} - 2 \frac{x_{21}^i x_{21}^j}{x_{21}^4}\Bigg]  \\ \notag
   & \times 
   \Big[
        \epsilon^{mj} x_{21}^2 I_4(x_{21}^2, z^\prime s) +  \epsilon^{mk} x_{21}^k x_{21}^j I_5(x_{21}^2, z^\prime s) + \epsilon^{jk} x_{21}^k x_{21}^m (I_6(x_{21}^2, z^\prime s) - G_2(x_{21}^2, z^\prime s)) + \epsilon^{mj} x_{20}^2 \Gamma^{\mathrm{gen}}_4(x_{20}^2, x_{21}^2, z^\prime s) \\ \notag 
        & \hspace{2cm} + \epsilon^{mk} x_{20}^k \, x_{20}^j \, \Gamma^{\mathrm{gen}}_5(x_{20}^2, x_{21}^2, z^\prime s) + \epsilon^{jk} x_{20}^k \, x_{20}^m  \, \Gamma^{\mathrm{gen}}_6(x_{20}^2, x_{21}^2, z^\prime s) -
        \epsilon^{jk} x_{20}^k \, x_{21}^m \, 
         \Gamma^{\mathrm{gen}}_2(x_{20}^2, x_{21}^2, z^\prime s)
   \Big]
    \Bigg\}.
\end{align}
Equation~\eqref{nosimp_i4} only has divergences in the $x_{21} \approx x_{20} \gg x_{10}$ IR region of the $x_{21}$ integral. Note that, when extracting the divergences, similar to \cite{Cougoulic:2022gbk} we neglect terms with logarithmic derivatives of the amplitudes with respect to dipole sizes, since such terms are not DLA. Keeping only the divergent terms and matching the coefficients multiplying the same tensor structures on both sides of the equation, after some extensive but straightforward algebra, we arrive at the DLA evolution equations for $I_4, I_5$ and $I_6$. We combine these with \eq{I3_eqn} by writing, in matrix form, 
\begin{align}
    \label{oam_eqns}
    \begin{pmatrix}
    I_3 \\ 
    I_4 \\
    I_5 \\
    I_6 
    \end{pmatrix}(x_{10}^2, zs) &= \begin{pmatrix}
    I_3^{(0)} \\ 
    I_4^{(0)} \\
    I_5^{(0)} \\
    I_6^{(0)} 
    \end{pmatrix} (x_{10}^2, zs)  
    + \frac{\as N_c}{4\pi} \iUV \,  \begin{pmatrix}
    2\,\Gamma_3 - 4\,\Gamma_4 + 2 \, \Gamma_5 + 6\, \Gamma_6 -2 \, \Gamma_2 \\ 
    0 \\
    0 \\
    0 
    \end{pmatrix}(x_{10}^2, x_{21}^2, z^\prime s) \\\notag &+ \frac{\as N_c}{4\pi} \iIR \, \begin{pmatrix}
        4 & -4 & 2 & 6 & -4 &  - 6 \\
        0 & 4 & 2 & -2 & 0 &  1 \\
        -2 & 2 & -1 & -3 & 2 &  3 \\
        0 & 0 & 0 & 0 & 2 &  4 \\
    \end{pmatrix} 
    \begin{pmatrix}
    I_3 \\ 
    I_4 \\
    I_5 \\
    I_6 \\ 
    G \\
    G_2
    \end{pmatrix} (x_{21}^2, z^\prime s).
\end{align}
The equations for the moment neighbor dipole amplitudes can be found by analogy, employing existing techniques \cite{Cougoulic:2022gbk, Kovchegov:2016zex, Kovchegov:2018znm, Kovchegov:2017lsr, Cougoulic:2019aja, Kovchegov:2021lvz}
\begin{align}
    &\begin{pmatrix} \label{oam_neigh_eqn}
    \Gamma_3 \\ 
    \Gamma_4 \\
    \Gamma_5 \\
    \Gamma_6 
    \end{pmatrix}(x_{10}^2, x_{21}^2, z^\prime s)  = \begin{pmatrix}
    I_3^{(0)} \\ 
    I_4^{(0)} \\
    I_5^{(0)} \\
    I_6^{(0)} 
    \end{pmatrix} (x_{10}^2, z^\prime s)  \\ \notag 
    & 
    + \frac{\as N_c}{4\pi} \inUV \, \begin{pmatrix}
    2\,\Gamma_3 - 4\,\Gamma_4 + 2 \, \Gamma_5 + 6\, \Gamma_6 -2 \, \Gamma_2 \\ 
    0 \\
    0 \\
    0 
    \end{pmatrix}(x_{10}^2, x_{32}^2, z^{\prime \prime} s)
    \\\notag &
    + \frac{\as N_c}{4\pi} \inIR \, \begin{pmatrix}
        4 & -4 & 2 & 6 & -4 &  - 6 \\
        0 & 4 & 2 & -2 & 0 &  1 \\
        -2 & 2 & -1 & -3 & 2 &  3 \\
        0 & 0 & 0 & 0 & 2 &  4 \\
    \end{pmatrix} 
    \begin{pmatrix}
    I_3 \\ 
    I_4 \\
    I_5 \\
    I_6 \\
    G \\ 
    G_2
    \end{pmatrix} (x_{32}^2, z^{\prime \prime} s).
\end{align}

We notice that \eqs{oam_eqns} and (\ref{oam_neigh_eqn}) do not close on their own as they mix with the dipole amplitudes $G, G_2$ and the neighbor dipole amplitude $\Gamma_2$. However, with the large-$N_c$ helicity evolution equations for $G, G_2, \Gamma,$ and $\Gamma_2$ derived in \cite{Cougoulic:2022gbk}, we have a closed set of equations which we list here for convenience,
\begin{tcolorbox}[colback=blue!10!white]
\begin{subequations}
\label{all_oam_eqns}
\begin{align}
    \label{helEvo_eqn1}
    & G(x_{10}^2,zs) = G^{(0)}(x_{10}^2, zs) + \frac{\as N_c}{2\pi} \iUV \Bigg[ 
    \Gamma(x_{10}^2, x_{21}^2, z^\prime s) + 3\, G(x_{21}^2, z^\prime s) \\ \notag 
    & \hspace{7cm} + 2\, G_2(x_{21}^2, z^\prime s) + 2\, \Gamma_2(x_{10}^2, x_{21}^2, z^\prime s) 
    \Bigg] ,
    \\ 
    & \Gamma(x_{10}^2, x_{21}^2, z^\prime s) = G^{(0)}(x_{10}^2, z^\prime s) + \frac{\as N_c}{2\pi} \inUV \Bigg[ 
    \Gamma(x_{10}^2, x_{32}^2, z^{\dprime} s) + 3\, G(x_{32}^2, z^{\dprime} s) \label{Gamma_eq} \\ 
    \notag 
    & \hspace{7cm} + 2\, G_2(x_{32}^2, z^{\dprime} s) + 2\, \Gamma_2(x_{10}^2, x_{32}^2, z^{\dprime} s) 
    \Bigg] , 
    \\
    & G_2(x_{10}^2, zs) = G_2^{(0)}(x_{10}^2, zs) + \frac{\as N_c}{\pi} \iIR \Big[ 
      G(x_{21}^2, z^\prime s) + 2\, G_2(x_{21}^2, z^\prime s) 
    \Big] , \label{G2_eq}
    \\
    \label{helEvo_eqn4}
     & \Gamma_2(x_{10}^2, x_{21}^2, z^\prime s) = G_2^{(0)}(x_{10}^2, z^\prime s) + \frac{\as N_c}{\pi} \inIR \Big[ 
      G(x_{32}^2, z^{\dprime} s) + 2\, G_2(x_{32}^2, z^{\dprime} s) 
    \Big] , 
    \\
    \label{final_oam_eqns}
    & 
    \begin{pmatrix}
    I_3 \\ 
    I_4 \\
    I_5 \\
    I_6 
    \end{pmatrix}(x_{10}^2, zs) = \begin{pmatrix}
    I_3^{(0)} \\ 
    I_4^{(0)} \\
    I_5^{(0)} \\
    I_6^{(0)} 
    \end{pmatrix} (x_{10}^2, zs)  
    + \frac{\as N_c}{4\pi} \iUV \,  
    \begin{pmatrix}
    2\,\Gamma_3 - 4\,\Gamma_4 + 2 \, \Gamma_5 + 6\, \Gamma_6 -2 \, \Gamma_2  \\ 
    0 \\
    0 \\
    0 
    \end{pmatrix}(x_{10}^2, x_{21}^2, z^\prime s) \notag \\
    & \hspace*{2.5cm} + \frac{\as N_c}{4\pi} \iIR \, \begin{pmatrix}
        4 & -4 & 2 & 6 & -4 &  - 6 \\
        0 & 4 & 2 & -2 & 0 &  1 \\
        -2 & 2 & -1 & -3 & 2 &  3 \\
        0 & 0 & 0 & 0 & 2 &  4 \\
    \end{pmatrix} 
    \begin{pmatrix}
    I_3 \\ 
    I_4 \\
    I_5 \\
    I_6 \\ 
    G \\
    G_2
    \end{pmatrix} (x_{21}^2, z^\prime s) , \\ 
& 
  \begin{pmatrix} \label{oam_neigh_eqns}
    \Gamma_3 \\ 
    \Gamma_4 \\
    \Gamma_5 \\
    \Gamma_6 
    \end{pmatrix}(x_{10}^2, x_{21}^2, z^\prime s) = \begin{pmatrix}
    I_3^{(0)} \\ 
    I_4^{(0)} \\
    I_5^{(0)} \\
    I_6^{(0)} 
    \end{pmatrix} (x_{10}^2, z^\prime s)  \\ \notag 
    & \hspace*{3cm}
    + \frac{\as N_c}{4\pi} \inUV \, \begin{pmatrix}
    2\,\Gamma_3 - 4\,\Gamma_4 + 2 \, \Gamma_5 + 6\, \Gamma_6 -2 \, \Gamma_2 \\ 
    0 \\
    0 \\
    0 
    \end{pmatrix}(x_{10}^2, x_{32}^2, z^{\prime \prime} s)
    \\ \notag 
    & \hspace*{3cm}
    + \frac{\as N_c}{4\pi} \inIR \, \begin{pmatrix}
        4 & -4 & 2 & 6 & -4 &  - 6 \\
        0 & 4 & 2 & -2 & 0 &  1 \\
        -2 & 2 & -1 & -3 & 2 &  3 \\
        0 & 0 & 0 & 0 & 2 &  4 \\
    \end{pmatrix} 
    \begin{pmatrix}
    I_3 \\ 
    I_4 \\
    I_5 \\
    I_6 \\
    G \\ 
    G_2
    \end{pmatrix} (x_{32}^2, z^{\prime \prime} s).
\end{align}
\end{subequations}
\end{tcolorbox}
Let us reiterate that, as in \cite{Cougoulic:2022gbk}, $1/\Lambda$ here is an IR cutoff on all dipole sizes and \eqs{all_oam_eqns} are valid only for $x_{10} < 1/\Lambda$. Furthermore, the neighbor dipole amplitudes are defined only for $x_{10} \geq x_{21}$. 

Equations \eqref{all_oam_eqns} need to be solved with the appropriate initial conditions (inhomogeneous terms) $G^{(0)}$, $G_2^{(0)}$, $I_3^{(0)}$, $I_4^{(0)}$, $I_5^{(0)}$, and $I_6^{(0)}$. The solution of \eqs{all_oam_eqns} would give us the flavor-singlet quark and gluon OAM distributions and the flavor-singlet quark and gluon helicity PDFs defined in Eqs.~(\ref{final_quark_OAM}), (\ref{final_gluonOAM}), (\ref{quark_hel}),  and (\ref{gluon_hel}), in the DLA and the large-$N_c$ limit. We emphasize that while \eqs{helEvo_eqn1}-(\ref{helEvo_eqn4}) were derived in \cite{Cougoulic:2022gbk}, \eqs{final_oam_eqns} and (\ref{oam_neigh_eqns}) are new.

Before concluding this Section, let us note that comparing the equations for $I_6$ and $G_2$ in \eqref{all_oam_eqns} one readily observes that 
\begin{align}\label{I6sol}
I_{6}(x_{10}^2, zs) = I^{(0)}_{6}(x_{10}^2, zs)  - \thalf \, G^{(0)}_{2}(x_{10}^2, zs) + \thalf \, G_2 (x_{10}^2, zs) \approx \thalf \, G_2 (x_{10}^2, zs),
\end{align}
where the approximation in the last step is valid for sufficiently high energies, when the initial conditions become negligibly small compared to $I_6$ and $G_2$. Hence, we see that $I_6 \approx G_2 /2$, along with $\Gamma_6 \approx \Gamma_2 /2$, and, as mentioned above in Sec.~\ref{sec:alt}, the number of our independent dipole amplitudes is reduced.


\section{Numerical solution} 

\label{sec:num_sol}

Due to the complexity of \eqs{all_oam_eqns}, we solve them numerically in this Section. We closely follow the procedure outlined in \cite{Cougoulic:2022gbk}. We first extract the hgh energy asymptotics of the dipole and moment amplitudes in order to determine the small-$x$ asymptotics of the OAM distributions. We then study the ratio of the OAM distributions to helicity PDFs for quarks and for gluons.


\subsection{Discretization of the evolution equations for the moment amplitudes}

Let us start by rewriting \eqs{final_oam_eqns} and (\ref{oam_neigh_eqns}) in terms of the following variables \cite{Kovchegov:2016weo, Kovchegov:2020hgb, Cougoulic:2022gbk}:
\begin{subequations}
    \label{rescaled_vars}
\begin{align}
    \eta & = \sqrt{\frac{\as N_c}{2\pi}} \ln \frac{zs}{\Lambda^2}, \hspace{2cm} \eta^\prime = \sqrt{\frac{\as N_c}{2\pi}} \ln \frac{z^\prime s}{\Lambda^2}, \hspace{2cm}  \eta^{\dprime} = \sqrt{\frac{\as N_c}{2\pi}} \ln \frac{z^{\dprime}s}{\Lambda^2}, \\ 
    s_{10} & = \sqrt{\frac{\as N_c}{2\pi}} \ln \frac{1}{x_{10}^2 \Lambda^2}, \hspace{1.35cm} s_{21} = \sqrt{\frac{\as N_c}{2\pi}} \ln \frac{1}{x_{21}^2 \Lambda^2}, \hspace{1.45cm} s_{32} = \sqrt{\frac{\as N_c}{2\pi}} \ln \frac{1}{x_{32}^2 \Lambda^2}. 
\end{align}
\end{subequations}

Equations \eqref{final_oam_eqns} and (\ref{oam_neigh_eqns})
become
\begin{subequations}
\label{rescaled_oam_eqns}
\begin{align}
\label{rescaled_oam_eqn}
    \begin{pmatrix}
    I_3 \\ 
    I_4 \\
    I_5 \\
    I_6 
    \end{pmatrix}(s_{10}, \eta) &= \begin{pmatrix}
    I_3^{(0)} \\ 
    I_4^{(0)} \\
    I_5^{(0)} \\
    I_6^{(0)} 
    \end{pmatrix} (s_{10}, \eta)  
    + \frac{1}{2} 
    \int \displaylimits^{\eta}_{s_{10}} d \eta^\prime \int \displaylimits^{\eta^\prime}_{s_{10}} d s_{21} 
    \,  \begin{pmatrix}
    2\,\Gamma_3 - 4\,\Gamma_4 + 2 \, \Gamma_5 + 6\, \Gamma_6 -2 \, \Gamma_2 \\ 
    0 \\
    0 \\
    0 
    \end{pmatrix}(s_{10}, s_{21}, \eta^\prime) \\\notag &+ \frac{1}{2} 
    \int \displaylimits^{s_{10}}_{0} d s_{21} 
    \int \displaylimits^{\eta - s_{10} + s_{21}}_{s_{21}} d \eta^\prime
    \, \begin{pmatrix}
        4 & -4 & 2 & 6 & -4 &  - 6 \\
        0 & 4 & 2 & -2 & 0 &  1 \\
        -2 & 2 & -1 & -3 & 2 &  3 \\
        0 & 0 & 0 & 0 & 2 &  4 \\
    \end{pmatrix} 
    \begin{pmatrix}
    I_3 \\ 
    I_4 \\
    I_5 \\
    I_6 \\ 
    G \\
    G_2
    \end{pmatrix} (s_{21}, \eta^\prime), \\ 
    \label{rescaled_oam_neigh_eqn}
\begin{pmatrix}
    \Gamma_3 \\ 
    \Gamma_4 \\
    \Gamma_5 \\
    \Gamma_6 
    \end{pmatrix}(s_{10}, s_{21}, \eta^\prime)  &= \begin{pmatrix}
    I_3^{(0)} \\ 
    I_4^{(0)} \\
    I_5^{(0)} \\
    I_6^{(0)} 
    \end{pmatrix} (s_{10}, \eta^\prime)  \\ \notag 
    & 
    + \frac{1}{2} 
    \int \displaylimits^{\eta^\prime}_{s_{10}} d \eta^{\dprime} \int \displaylimits^{\eta^{\dprime}}_{\mathrm{max}[s_{10}, s_{21} + \eta^{\dprime} - \eta^\prime]} d s_{32}  
    \, \begin{pmatrix}
    2\,\Gamma_3 - 4\,\Gamma_4 + 2 \, \Gamma_5 + 6\, \Gamma_6 -2 \, \Gamma_2 \\ 
    0 \\
    0 \\
    0 
    \end{pmatrix}(s_{10}, s_{32}, \eta^{\dprime})
    \\\notag &
    + \frac{1}{2} 
    \int \displaylimits^{s_{10}}_{0} d s_{32} 
    \int \displaylimits^{\eta^\prime - s_{21} + s_{32}}_{s_{32}} d \eta^{\dprime}
    \, \begin{pmatrix}
        4 & -4 & 2 & 6 & -4 &  - 6 \\
        0 & 4 & 2 & -2 & 0 &  1 \\
        -2 & 2 & -1 & -3 & 2 &  3 \\
        0 & 0 & 0 & 0 & 2 &  4 \\
    \end{pmatrix} 
    \begin{pmatrix}
    I_3 \\ 
    I_4 \\
    I_5 \\
    I_6 \\
    G \\ 
    G_2
    \end{pmatrix} (s_{32}, \eta^{\dprime}) ,
\end{align}
\end{subequations}
where the ordering $0 \leq s_{10} \leq \eta$ is assumed in \eq{rescaled_oam_eqn} while $0 \leq s_{10} \leq s_{21} \leq \eta^\prime$ is assumed in \eq{rescaled_oam_neigh_eqn}. Note that we omit the helicity equations \eqref{helEvo_eqn1}-(\ref{helEvo_eqn4}) here for brevity, but since \eqref{final_oam_eqns} and (\ref{oam_neigh_eqns}) close only when the equations for $G, G_2, \Gamma$, and $\Gamma_2$ are included, we also solve \eqs{helEvo_eqn1}-(\ref{helEvo_eqn4}) numerically in exactly the same way as described in~\cite{Cougoulic:2022gbk}.\footnote{We thank Josh Tawabutr for providing us with his numerical code for solving \eqs{helEvo_eqn1}-(\ref{helEvo_eqn4}).} 

Now we discretize the integrals in \eqs{rescaled_oam_eqns} with step size $\delta$ in both $\eta$ and $s_{10}$ directions. We express the discretized form of the dipole and moment amplitudes along with their corresponding neighbor dipole amplitudes as 
\begin{align}
\begin{aligned}[c]
    & G_{ij} = G(i \delta, j \delta), \\
    & G_{2, ij} = G_2(i \delta, j \delta), \\
    & I_{3, ij} = I_3(i \delta, j \delta), \\
    & I_{4, ij} = I_4(i \delta, j \delta), \\
    & I_{5, ij} = I_5(i \delta, j \delta), \\
    & I_{6, ij} = I_6(i \delta, j \delta), 
\end{aligned}
     \hspace{1cm} 
\begin{aligned}[c]
     & \Gamma_{ikj} = \Gamma (i \delta, k \delta, j \delta), \\ 
     & \Gamma_{2, ikj} = \Gamma_2 (i \delta, k \delta, j \delta), \\
    & \Gamma_{3, ikj} = \Gamma_3 (i \delta, k \delta, j \delta), \\ 
     & \Gamma_{4, ikj} = \Gamma_4 (i \delta, k \delta, j \delta), \\ 
    & \Gamma_{5, ikj} = \Gamma_5 (i \delta, k \delta, j \delta), \\ 
    & \Gamma_{6, ikj} = \Gamma_6 (i \delta, k \delta, j \delta).
\end{aligned}          
\end{align}
We then rewrite \eqs{rescaled_oam_eqns} in the discretized form as 
\begin{subequations}
\label{disc_oam_eqns}
\begin{align}
    \begin{pmatrix}
    I_{3, ij} \\ 
    I_{4,ij} \\
    I_{5, ij} \\
    I_{6, ij} 
    \end{pmatrix} &= \begin{pmatrix}
   I_{3, ij}^{(0)} \\ 
    I_{4,ij}^{(0)} \\
    I_{5, ij}^{(0)} \\
    I_{6, ij}^{(0)} 
    \end{pmatrix} 
    + \frac{\delta^2}{2} 
        \sum_{j^\prime = i}^{j-1} \sum^{j^\prime}_{i^\prime=i}
    \,  \begin{pmatrix}
    2\,\Gamma_{3, i i^\prime j^\prime} - 4\,\Gamma_{4,i i^\prime j^\prime} + 2 \, \Gamma_{5, i i^\prime j^\prime} + 6\, \Gamma_{6,i i^\prime j^\prime} -2 \, \Gamma_{2,i i^\prime j^\prime} \\ 
    0 \\
    0 \\
    0 
    \end{pmatrix} \\\notag &
    \hspace{2cm}
    + \frac{\delta^2}{2} 
    \sum_{i^\prime = 0}^{i-1} \sum_{j^\prime = i^\prime}^{j-i+i^\prime} 
    \, \begin{pmatrix}
        4 & -4 & 2 & 6 & -4 &  - 6 \\
        0 & 4 & 2 & -2 & 0 &  1 \\
        -2 & 2 & -1 & -3 & 2 &  3 \\
        0 & 0 & 0 & 0 & 2 &  4 \\
    \end{pmatrix} 
    \begin{pmatrix}
    I_{3, i^\prime j^\prime} \\ 
    I_{4, i^\prime j^\prime} \\
    I_{5, i^\prime j^\prime} \\
    I_{6, i^\prime j^\prime} \\ 
    G_{i^\prime j^\prime} \\
    G_{2, i^\prime j^\prime}
    \end{pmatrix} , \\ 
\begin{pmatrix}
    \Gamma_{3, ikj} \\ 
    \Gamma_{4, ikj} \\
    \Gamma_{5, ikj} \\
    \Gamma_{6, ikj}
    \end{pmatrix}  &= \begin{pmatrix}
    I_{3, ij}^{(0)} \\ 
    I_{4, ij}^{(0)} \\
    I_{5, ij}^{(0)} \\
    I_{6, ij}^{(0)}
    \end{pmatrix}  
    + \frac{\delta^2}{2} 
    \sum_{j^\prime =i}^{j-1} \sum_{i^\prime = \mathrm{max}[i, k+j^\prime -j]}^{j^\prime} 
    \, \begin{pmatrix}
    2\,\Gamma_{3, i i^\prime j^\prime} - 4\,\Gamma_{4,i i^\prime j^\prime} + 2 \, \Gamma_{5, i i^\prime j^\prime} + 6\, \Gamma_{6,i i^\prime j^\prime} -2 \, \Gamma_{2,i i^\prime j^\prime} \\ 
    0 \\
    0 \\
    0 
    \end{pmatrix}
    \\ \notag &
    \hspace{2cm}
    + \frac{\delta^2}{2} 
    \sum_{i^\prime =0}^{i-1} \sum_{j^\prime = i^\prime}^{j-k+i^\prime} 
    \, \begin{pmatrix}
        4 & -4 & 2 & 6 & -4 &  - 6 \\
        0 & 4 & 2 & -2 & 0 &  1 \\
        -2 & 2 & -1 & -3 & 2 &  3 \\
        0 & 0 & 0 & 0 & 2 &  4 \\
    \end{pmatrix} 
    \begin{pmatrix}
    I_{3, i^\prime j^\prime} \\ 
    I_{4, i^\prime j^\prime} \\
    I_{5, i^\prime j^\prime} \\
    I_{6, i^\prime j^\prime} \\
    G_{i^\prime j^\prime} \\ 
    G_{2, i^\prime j^\prime}
    \end{pmatrix}.
\end{align}
\end{subequations}

In order to speed up the numerical computation, we follow the method of \cite{Kovchegov:2020hgb} and write recursive relations for each of the equations in \eqs{disc_oam_eqns}. By subtracting $I_{p,i, j-1}$ from $I_{p,i, j}$ and $\Gamma_{p, i, k-1, j-1}$ from $\Gamma_{p, i, k, j}$ for all $p=3,4,5,6$, we get the following recursion relations 
\begin{subequations}
    \label{recursive}
    \begin{align}
    \begin{pmatrix}
    I_{3, ij} \\ 
    I_{4,ij} \\
    I_{5, ij} \\
    I_{6, ij} 
    \end{pmatrix} &= 
    \begin{pmatrix}
    I_{3, ij}^{(0)} - I_{3, i(j-1)}^{(0)} + I_{3, i(j-1)} \\ 
    I_{4, ij}^{(0)} - I_{4, i(j-1)}^{(0)} + I_{4,i(j-1)} \\
    I_{5, ij}^{(0)} - I_{5, i(j-1)}^{(0)} + I_{5, i(j-1)} \\
    I_{6, ij}^{(0)} - I_{6, i(j-1)}^{(0)} + I_{6, i(j-1)} 
    \end{pmatrix} 
    + \frac{\delta^2}{2} 
    \sum_{i^\prime = 0}^{i-1}
    \, \begin{pmatrix}
        4 & -4 & 2 & 6 & -4 &  - 6 \\
        0 & 4 & 2 & -2 & 0 &  1 \\
        -2 & 2 & -1 & -3 & 2 &  3 \\
        0 & 0 & 0 & 0 & 2 &  4 \\
    \end{pmatrix} 
    \begin{pmatrix}
    I_{3, i^\prime (i^\prime + j - i)} \\ 
    I_{4, i^\prime (i^\prime + j - i)} \\
    I_{5, i^\prime (i^\prime + j - i)} \\
    I_{6, i^\prime (i^\prime + j - i)} \\ 
    G_{i^\prime (i^\prime + j - i)} \\
    G_{2, i^\prime (i^\prime + j - i)}
    \end{pmatrix} 
    \\ \notag &
    \hspace{2cm}
     + \frac{\delta^2}{2} 
       \sum_{i^\prime = i}^{j-1} 
    \,  \begin{pmatrix}
    2\,\Gamma_{3, i i^\prime(j-1)} - 4\,\Gamma_{4,i i^\prime(j-1)} + 2 \, \Gamma_{5, i i^\prime (j-1)} + 6\, \Gamma_{6,i i^\prime (j-1)} -2 \, \Gamma_{2,i i^\prime (j-1)} \\ 
    0 \\
    0 \\
    0 
    \end{pmatrix}
    , \\ 
\begin{pmatrix}
    \Gamma_{3, ikj} \\ 
    \Gamma_{4, ikj} \\
    \Gamma_{5, ikj} \\
    \Gamma_{6, ikj}
    \end{pmatrix}  &= 
    \begin{pmatrix}
    I_{3, ij}^{(0)} - I_{3, i(j-1)}^{(0)} + \Gamma_{3, i(k-1)(j-1)} \\ 
    I_{4, ij}^{(0)} - I_{4, i(j-1)}^{(0)} + \Gamma_{4, i(k-1)(j-1)} \\
    I_{5, ij}^{(0)} - I_{5, i(j-1)}^{(0)} + \Gamma_{5, i(k-1)(j-1)} \\
    I_{6, ij}^{(0)} - I_{6, i(j-1)}^{(0)} + \Gamma_{6, i(k-1)(j-1)}
    \end{pmatrix} 
     \\ \notag &
    \hspace{2cm}
    + \frac{\delta^2}{2} 
    \sum_{i^\prime = k-1}^{j-1}
    \, \begin{pmatrix}
    2\,\Gamma_{3, i i^\prime (j-1)} - 4\,\Gamma_{4,i i^\prime  (j-1)} + 2 \, \Gamma_{5, i i^\prime  (j-1)} + 6\, \Gamma_{6,i i^\prime  (j-1)} -2 \, \Gamma_{2,i i^\prime (j-1)} \\ 
    0 \\
    0 \\
    0 
    \end{pmatrix}, 
    \end{align}
\end{subequations}
where $0 \leq i < j$ and $0 \leq i < k \leq j$ for $i \leq i_{\mathrm{max}}$, $j \leq j_{\mathrm{max}}$ as can be seen from \eqs{disc_oam_eqns}. For $i = j$ and $i = k$, the amplitudes are 
\begin{align}\label{recursive2}
    & I_{3, ii} = I_{3, ii}^{(0)}, \  I_{4, ii} = I_{4, ii}^{(0)}, \ I_{5, ii} = I_{5, ii}^{(0)}, \ I_{6, ii} = I_{6, ii}^{(0)}, \\
    & \Gamma_{3, iij} = I_{3, ij}, \ \Gamma_{4, iij} = I_{4, ij}, \ \Gamma_{5, iij} = I_{5, ij}, \ \Gamma_{6, iij} = I_{6, ij}. \notag
\end{align}
Note that the physical regions where the evolution in \eqs{disc_oam_eqns} applies are $j \ge i \ge 0$ and $j \ge k \ge i \ge 0$, corresponding to the $0 \leq s_{10} \leq \eta$ and $0 \leq s_{10} \leq s_{21} \leq \eta^\prime$ conditions shown above. Following \cite{Kovchegov:2016weo, Kovchegov:2020hgb, Cougoulic:2022gbk, Adamiak:2023okq}, the dipole amplitudes and moment-amplitudes will be left equal to the inhomogeneous terms outside those regions.

\begin{figure}[ht!]
     \centering
     \begin{subfigure}[b]{0.48\textwidth}
         \centering
         \includegraphics[width=\textwidth]{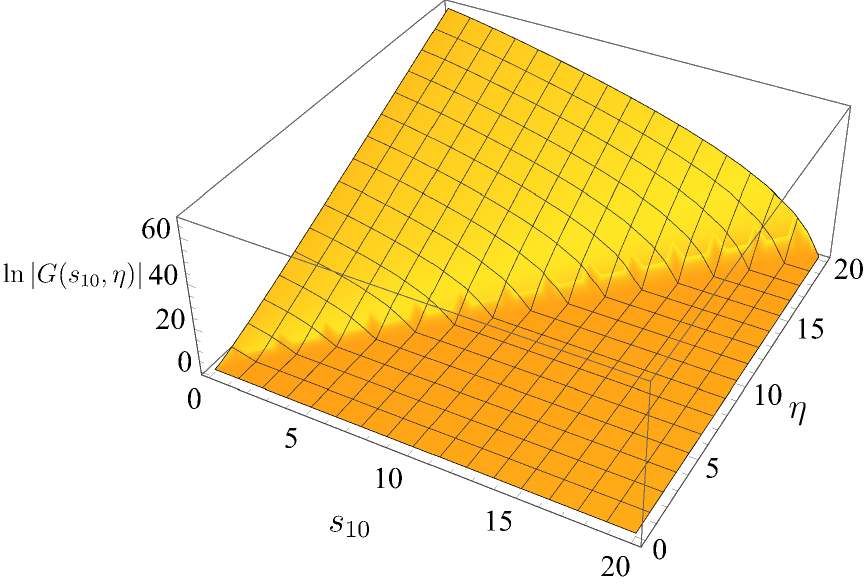}
         \caption{}
     \end{subfigure}
     \begin{subfigure}[b]{0.48\textwidth}
         \centering
         \includegraphics[width=\textwidth]{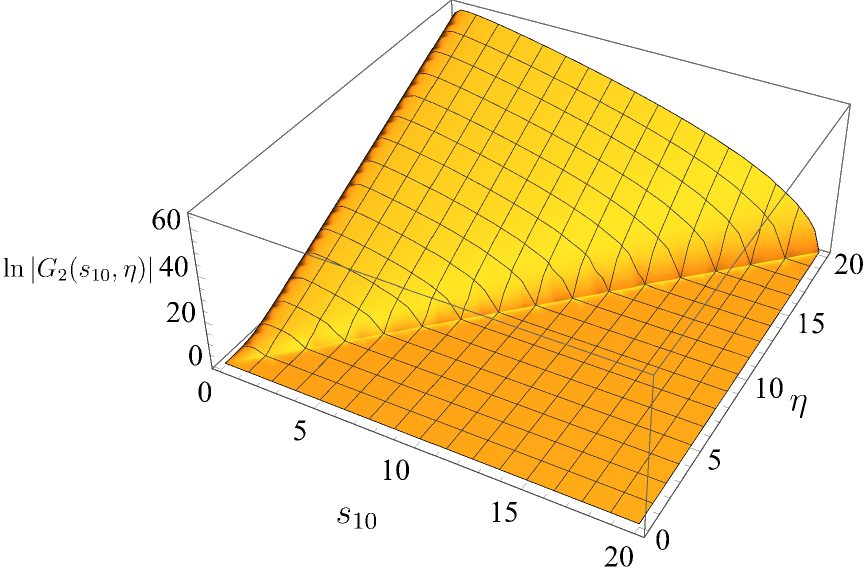}
         \caption{}
     \end{subfigure}
     \begin{subfigure}[b]{0.48\textwidth}
         \centering
         \includegraphics[width=\textwidth]{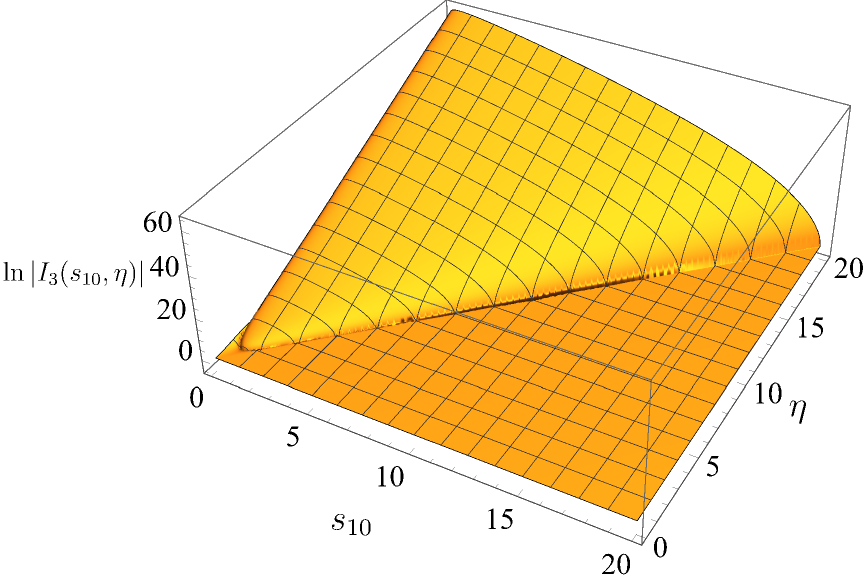}
         \caption{}
     \end{subfigure}
     \begin{subfigure}[b]{0.48\textwidth}
         \centering
         \includegraphics[width=\textwidth]{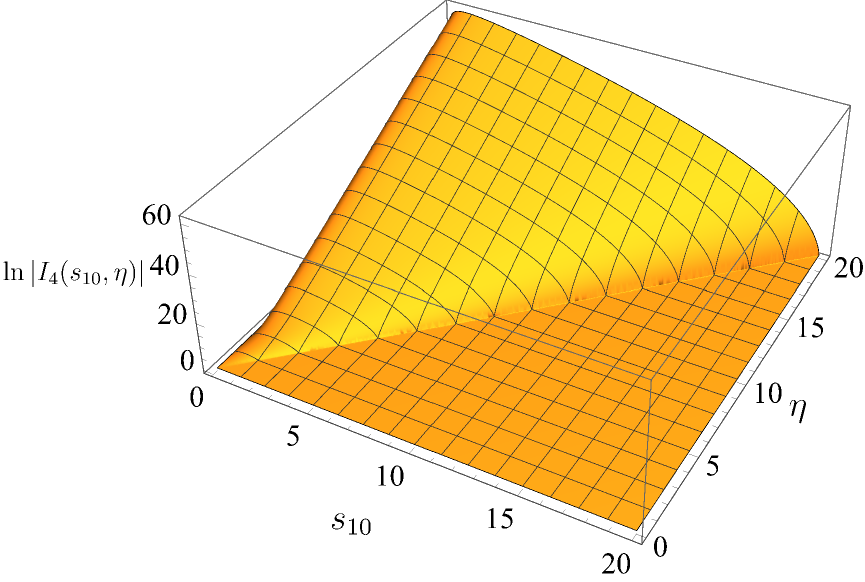}
         \caption{}
     \end{subfigure}
     \begin{subfigure}[b]{0.48\textwidth}
         \centering
         \includegraphics[width=\textwidth]{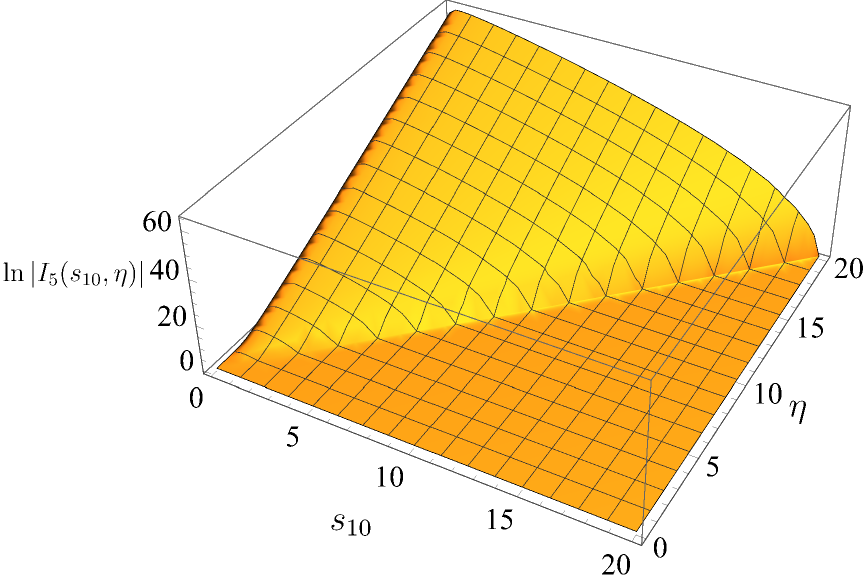}
         \caption{}
     \end{subfigure}
     \begin{subfigure}[b]{0.48\textwidth}
         \centering
         \includegraphics[width=\textwidth]{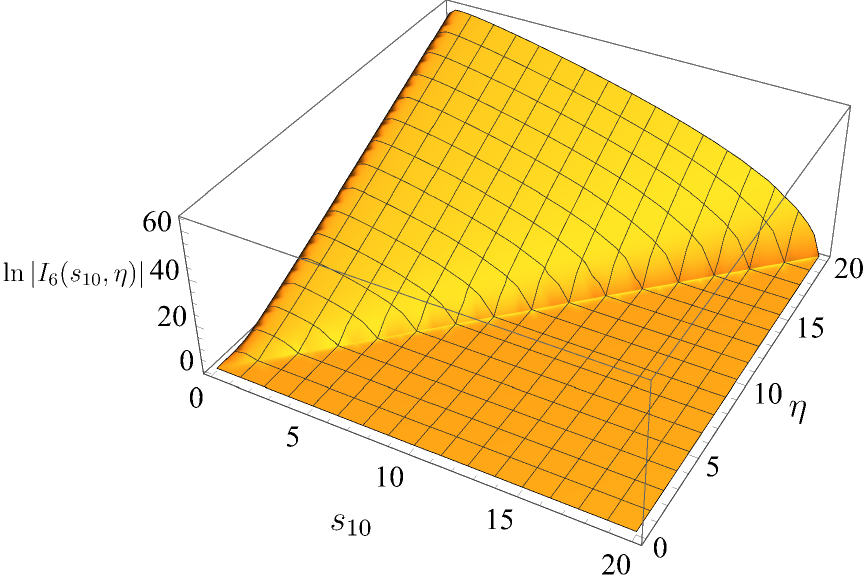}
         \caption{}
     \end{subfigure}
        \caption{Plots of the logarithms of the absolute values of the polarized dipole amplitudes $G$ and $G_2$ and moment amplitudes $I_p$ as functions of $s_{10}$ and $\eta$ in the $\eta, s_{10} \in [0,20]$ range resulting from our numerical solution of \eqs{recursive} and \eqref{recursive2} with a step size of $\delta = 0.025$. All the inhomogeneous terms are set to 1.}
        \label{fig:log_plots_moments}
\end{figure}

We have solved \eqs{recursive} and \eqref{recursive2}, along with their counterparts for $G$ and $G_2$ (see Eqs.~(258) in \cite{Cougoulic:2022gbk}), numerically for a variety of different initial conditions. Similar to the case of helicity distributions \cite{Kovchegov:2016weo, Kovchegov:2020hgb, Cougoulic:2022gbk, Adamiak:2023okq}, we found that the $x$-dependence of the resulting leading small-$x$ asymptotics is independent of the initial conditions (the inhomegenous terms). We plot our solution for the unit initial conditions $G^{(0)}_{ij} = G^{(0)}_{2, ij} = I_{3, ij}^{(0)} = I_{4, ij}^{(0)} = I_{5, ij}^{(0)} = I_{6, ij}^{(0)} =1$ in \fig{fig:log_plots_moments}. We employ a step size of $\delta = 0.025$. In \fig{fig:log_plots_moments} we plot $\ln |G|$, $\ln |G_2|$, $\ln |I_3|$, $\ln |I_4|$, $\ln |I_5|$, and $\ln |I_6|$ in panels (a), (b), (c), (d), (e), and (f) respectively, in the $0 \leq \eta, s_{10} \leq 20$ range.

\subsection{Small-$x$ asymptotics of the OAM distributions} 

\label{smallxasymp}

\begin{figure}[ht]
     \centering
     \begin{subfigure}[b]{0.48\textwidth}
         \centering
         \includegraphics[width=\textwidth]{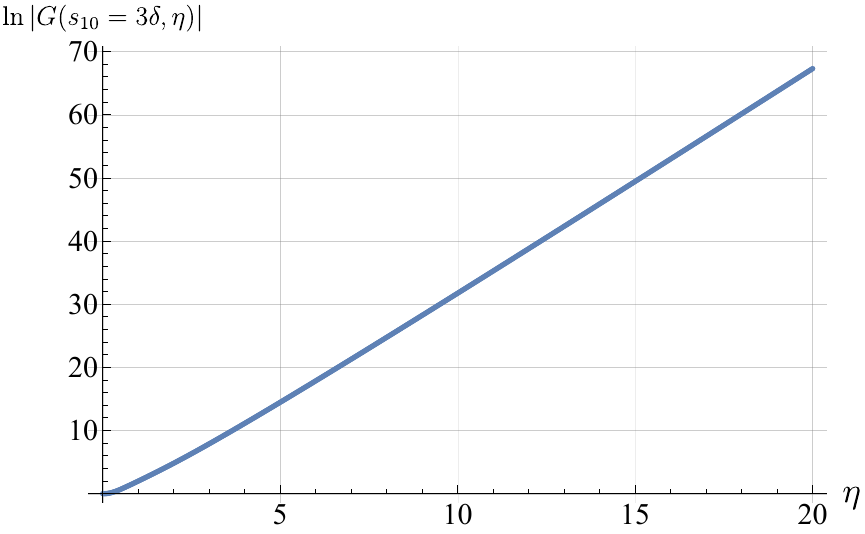}
         \caption{}
     \end{subfigure}
     \begin{subfigure}[b]{0.48\textwidth}
         \centering
         \includegraphics[width=\textwidth]{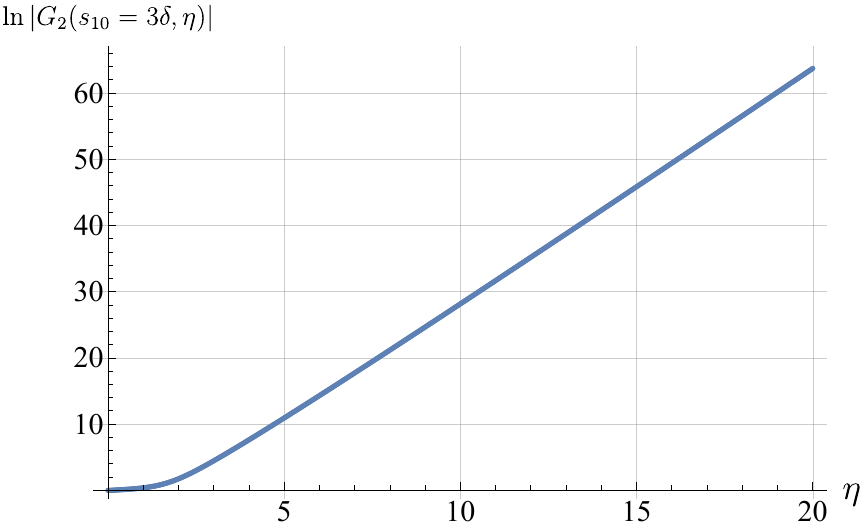}
         \caption{}
     \end{subfigure}
     \begin{subfigure}[b]{0.48\textwidth}
         \centering
         \includegraphics[width=\textwidth]{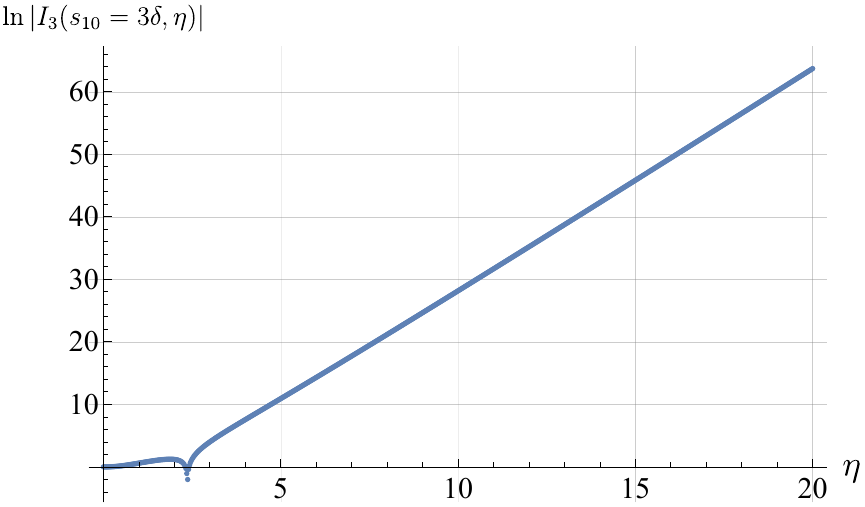}
         \caption{}
     \end{subfigure}
     \begin{subfigure}[b]{0.48\textwidth}
         \centering
         \includegraphics[width=\textwidth]{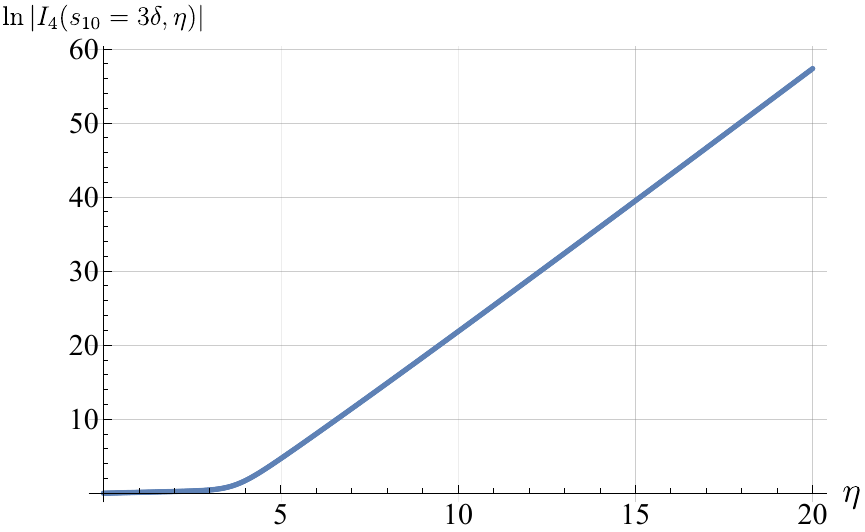}
         \caption{}
     \end{subfigure}
     \begin{subfigure}[b]{0.48\textwidth}
         \centering
         \includegraphics[width=\textwidth]{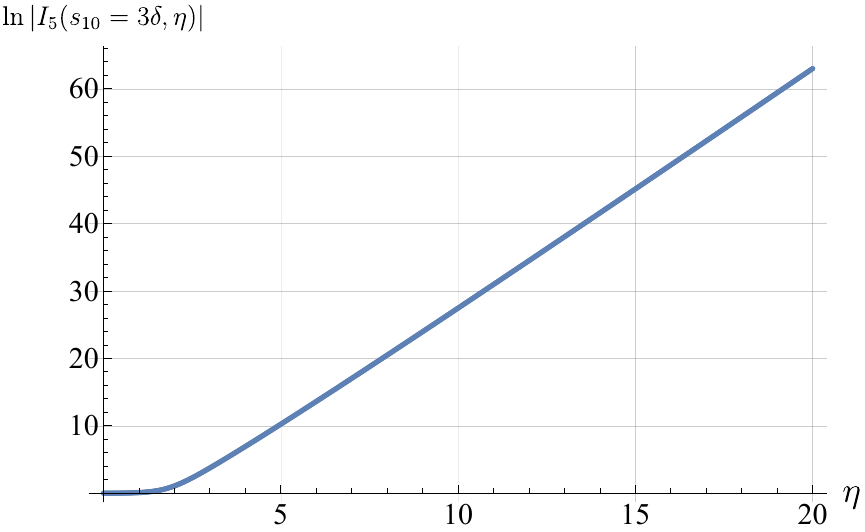}
         \caption{}
     \end{subfigure}
     \begin{subfigure}[b]{0.48\textwidth}
         \centering
         \includegraphics[width=\textwidth]{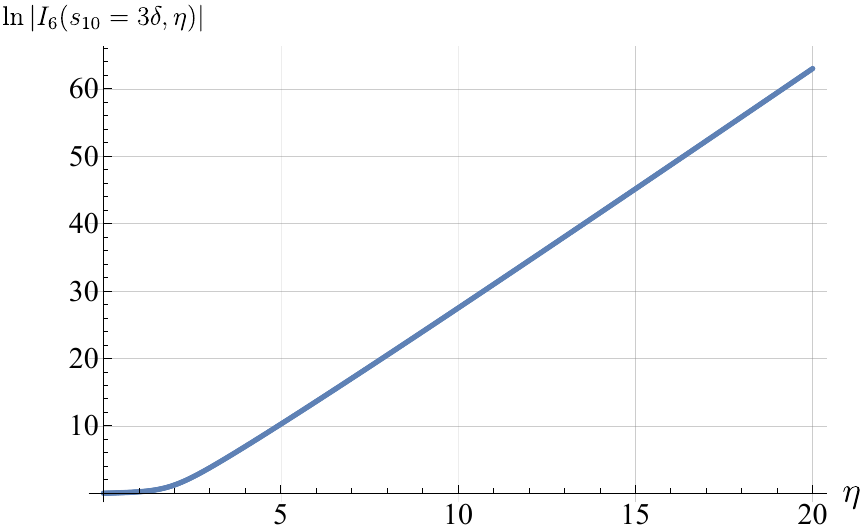}
         \caption{}
     \end{subfigure}
        \caption{Plots of the logarithms of the absolute values of the polarized dipole amplitudes $G$ and $G_2$ and moment amplitudes $I_p$ as functions of $\eta$ for $s_{10} = 3 \, \delta$, resulting from our numerical solution with a step size of $\delta = 0.025$. All the inhomogeneous terms are set to 1. The kink in the $\ln |I_3|$ plot (in panel (c)) corresponds to a sign reversal in $I_3$.}
        \label{fig:eta_plots_moments}
\end{figure}

From \fig{fig:log_plots_moments}, we see that the moment amplitudes $I_p$ grow exponentially with $\eta$, similar to the polarized dipole amplitudes $G$ and $G_2$.
This corresponds to a power-law growth in the center-of-mass energy squared $zs$. For the purposes of extracting the asymptotics of the OAM distributions, it is sufficient to determine the asymptotic form of the moment amplitudes as $\eta \to \infty$ at $s_{10} = 0$ \cite{Kovchegov:2016weo, Kovchegov:2020hgb, Cougoulic:2022gbk, Adamiak:2023okq}. To this end, we plot the logarithms of the absolute values of the polarized dipole and moment amplitudes as functions of $\eta$ for $s_{10} = 3 \, \delta$  in \fig{fig:eta_plots_moments} for $0 \leq \eta \leq 20$ and a step size of $\delta =0.025$. (We chose $s_{10} = 3 \, \delta$ and not $s_{10} = 0$ to avoid the feature of the evolution for $G_2$, $I_4$, $I_5$, and $I_6$ that the evolution is ``turned off" for $s_{10} = 0$ exactly, with the integral containing the kernel vanishing and those amplitudes given by their inhomogeneous terms; see, for instance, \eq{rescaled_oam_eqn}.)

All the functions in \fig{fig:eta_plots_moments} increase linearly with $\eta$ in the region sufficiently far away from $\eta=0$. (Near $\eta=0$, the initial conditions and the discretization errors likely play a dominant role.) Therefore, we have the following ansätze as $\eta \to \infty$ \cite{Kovchegov:2016weo, Kovchegov:2020hgb, Cougoulic:2022gbk, Adamiak:2023okq}
\begin{align}
    \label{moment_ansatz}
    I_p (s_{10} = 0, \eta) \sim e^{\alpha_p \eta \sqrt{\frac{2\pi}{\as N_c}}}
\end{align}
where $p \in \{3,4,5,6\}$ and $\alpha_p$'s are the intercepts. To cross-check the numerical calculation with \cite{Cougoulic:2022gbk}, we also use the ansatz \eq{moment_ansatz} for $G$ and $G_2$ as defined in Eq.~(261) of \cite{Cougoulic:2022gbk}. We then will compare our results for the intercepts $\alpha_h$ and $\alpha_{h,2}$ of $G$ and $G_2$ (defined by analogy to \eq{moment_ansatz}) with the calculations of \cite{Cougoulic:2022gbk}. 

Since this exponential growth is dominant for larger $\eta$, we deduce the $\alpha_p$'s by regressing the logarithms of the moment amplitudes over $0.75 \, \eta_{\mathrm{max}} \leq \eta \leq \eta_{\mathrm{max}}$ for a given step size $\delta$ \cite{Kovchegov:2016weo, Kovchegov:2020hgb, Cougoulic:2022gbk, Adamiak:2023okq}. Here $\eta_{\mathrm{max}}$ is the maximum value of $\eta$ in our simulation. We can take the continuum limit ($\delta \to 0, \eta_{\mathrm{max}} \to \infty$) by repeating this regression for other choices of $\delta$ and $\eta_{\mathrm{max}}$, fitting the results for $\alpha_p$ in the $(\delta, 1/\eta_{\mathrm{max}})$ space with a polynomial-model surface and extrapolating this model to $\delta = 0, 1/\eta_{\mathrm{max}} = 0$ \cite{Kovchegov:2016weo, Kovchegov:2020hgb, Cougoulic:2022gbk, Adamiak:2023okq}. For consistency with the numerical solution of the helicity evolution equations for $G$ and $G_2$, we choose the values of $\delta$ and $\eta_{\mathrm{max}}$ to be the same as in \cite{Cougoulic:2022gbk}. We detail each of these step sizes and the range of $\eta_{\mathrm{max}}$ at a particular step size given by $\eta_{\mathrm{max}} \in \{10, 20, \ldots, M(\delta) \}$ in Table~\ref{tab:max_eta_values}. 

\begin{table}[ht!]
    \centering
    \begin{tabular}{|c|c|c|c|c|c|c|c|c|c|c|}
        \hline
        $\delta$ &  \,0.0125\, & \,0.016\, & \,0.025 \,& \,0.032\, & \,0.0375 \,& \,0.05\, & \,0.0625\, & \,0.075\, &\, 0.08\, &\, 0.1\, \\ \hline
        $M(\delta)$ & 10 & 10 & 20 & 20 & 30 & 40 & 50 & 60 & 60 & 70 \\ 
        \hline
    \end{tabular}
    \caption{Maximum values of $\eta_{\mathrm{max}}$, $M(\delta)$, for each step size $\delta$. There are 32 ($\delta, 1/\eta_{\mathrm{max}}$)-data points in total.}
    \label{tab:max_eta_values}
\end{table}

\begin{figure}[ht!]
     \centering
     \begin{subfigure}[b]{0.48\textwidth}
         \centering
         \includegraphics[width=\textwidth]{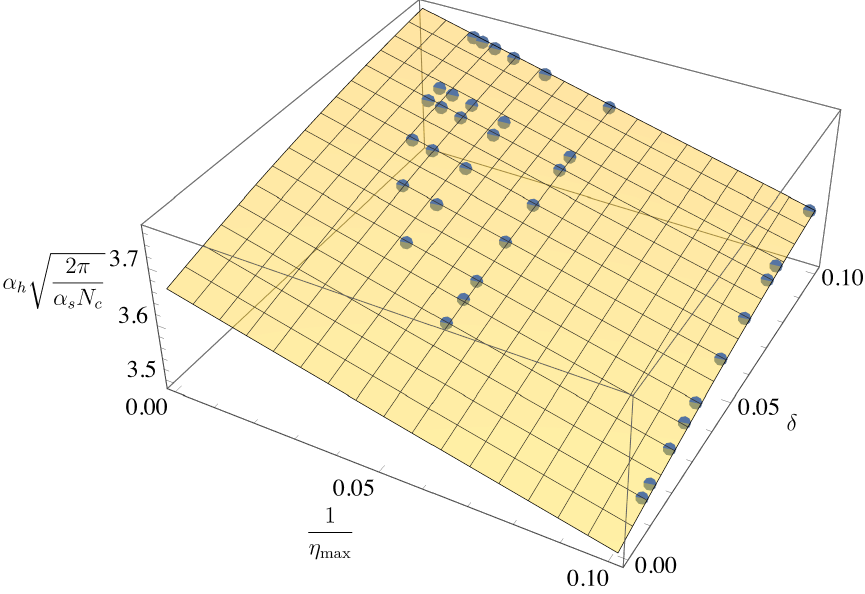}
         \caption{}
     \end{subfigure}
     \begin{subfigure}[b]{0.48\textwidth}
         \centering
         \includegraphics[width=\textwidth]{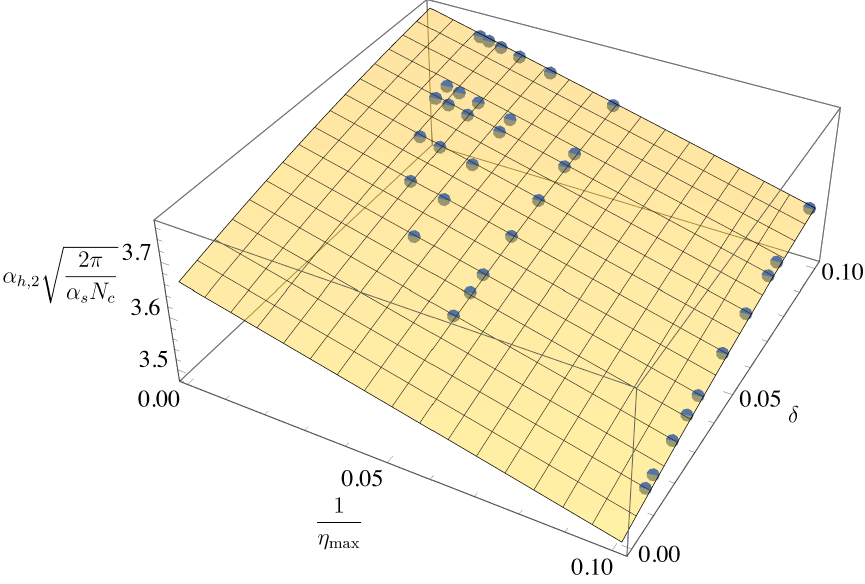}
         \caption{}
     \end{subfigure}
     \begin{subfigure}[b]{0.48\textwidth}
         \centering
         \includegraphics[width=\textwidth]{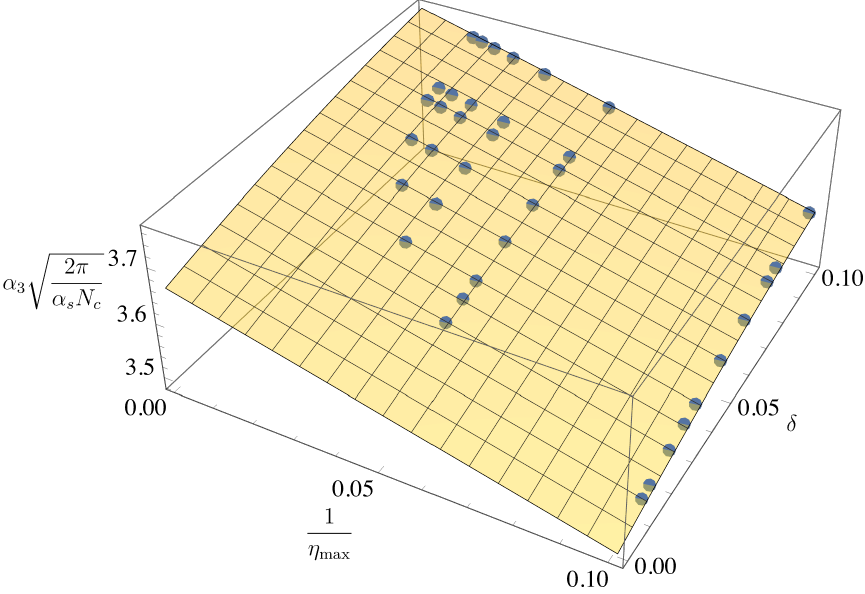}
         \caption{}
     \end{subfigure}
     \begin{subfigure}[b]{0.48\textwidth}
         \centering
         \includegraphics[width=\textwidth]{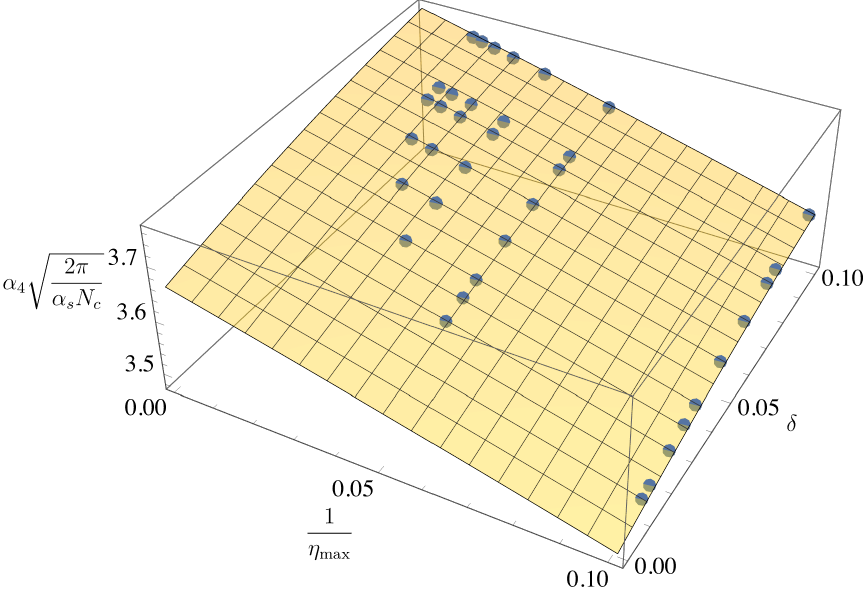}
         \caption{}
     \end{subfigure}
     \begin{subfigure}[b]{0.48\textwidth}
         \centering
         \includegraphics[width=\textwidth]{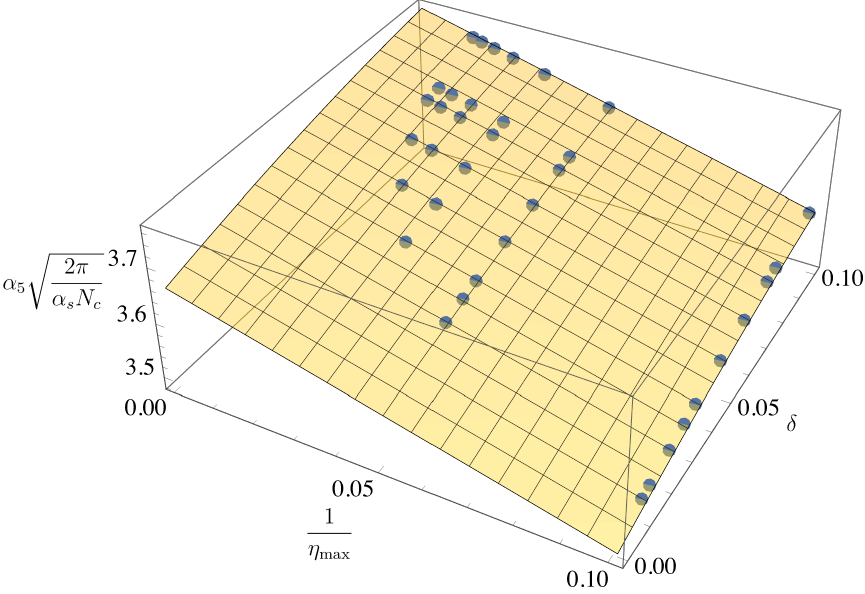}
         \caption{}
     \end{subfigure}
     \begin{subfigure}[b]{0.48\textwidth}
         \centering
         \includegraphics[width=\textwidth]{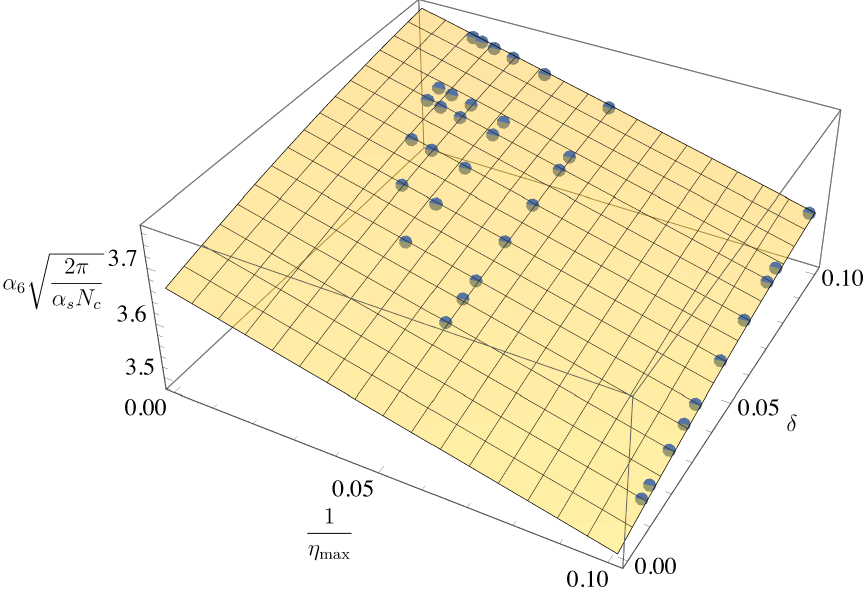}
         \caption{}
     \end{subfigure}
        \caption{Surfaces generated by model 3 for each intercept plotted versus $(\delta, 1/\eta_{\mathrm{max}})$ along with the corresponding data points at each $\delta$ and $1/\eta_{\mathrm{max}}$ where we performed a numerical simulation. The continuum limit corresponds to the value of the model 3 surface at $\delta =0$ and $1/\eta_{\mathrm{max}} =0$ (the left corner of each panel). }
        \label{fig:intercept_models}
\end{figure}

Similar to \cite{Kovchegov:2016weo, Kovchegov:2020hgb, Cougoulic:2022gbk, Adamiak:2023okq}, we model the intercepts $\alpha_p$ using $\delta$ and $1/\eta_{\mathrm{max}}$ as independent variables. We use polynomial regression models of various degrees, employing polynomials of different orders in $\delta$ and $1/\eta_{\mathrm{max}}$, weighted by the uncertainties of the intercepts, which we determine through regression at the 95\% confidence level. Once we have the best models, we can give an estimate for the intercepts in the continuum limit by extrapolating to $\delta=0, \, 1/\eta_{\mathrm{max}} = 0$. In particular we consider the following models with increasing polynomial degree:
\begin{itemize}
     \item Model 1: $\alpha_p = a_1$,
     \item Model 2: $\alpha_p = a_1 + a_2 \delta + \frac{a_3}{\eta_{\mathrm{max}}}$ ,
     \item Model 3: $\alpha_p = a_1 + a_2 \delta + \frac{a_3}{\eta_{\mathrm{max}}} + a_4 \delta^2 + \frac{a_5 \delta}{\eta_{\mathrm{max}}} + \frac{a_6}{\eta_{\mathrm{max}}^2}$ ,
     \item Model 4: $\alpha_p = a_1 + a_2 \delta + \frac{a_3}{\eta_{\mathrm{max}}} + a_4 \delta^2 + \frac{a_5 \delta}{\eta_{\mathrm{max}}} + \frac{a_6}{\eta_{\mathrm{max}}^2} + a_7 \delta^3 + \frac{a_8 \delta^2}{\eta_{\mathrm{max}}} + \frac{a_9 \delta}{\eta_{\mathrm{max}}^2} + \frac{a_{10}}{\eta_{\mathrm{max}}^3} $ .
\end{itemize}
Once we fit and evaluate all four models to our estimates for each $\alpha_p$ (including $\alpha_h$ and $\alpha_{h,2}$), we observe the Akaike information criterion (AIC) \cite{Akaike:1974} decrease significantly from model 1 to 2 and from model 2 to 3. The AIC has a minimum at model 3 for all intercepts. Indeed, examining model 4 reveals a higher AIC and insignificant parameters (via a 10\% level $t$-test). Therefore, we decide to use model 3, the quadratic model, for all of our intercepts. The estimated parameters for each best fit model are given in Table~\ref{tab:intercept table}. We plot these models along with the numerical data in \fig{fig:intercept_models}. 

\begin{table}[ht!]
    \centering
    \begin{tabular}{|c|c|c|c|c|c|c|}
        \hline 
         & $a_1$ & $a_2$ & $a_3$ & $a_4$ & $a_5$ & $a_6$ \\
         \hline
        $\alpha_h$ & $3.661$ & $1.502$ & $-1.732$ & $-4.41$ & $0.10$ & $0.34$ \\
        \hline
        $\alpha_{h,2}$ & $3.660$ & $1.514$ & $-1.709$ & $-4.45$ & $-0.16$ & $-0.43$ \\ 
        \hline
        $\alpha_{3}$ & $3.660$ & $1.511$ & $-1.714$ & $-4.44$ & $-0.11$ & $-0.37$ \\ 
        \hline
        $\alpha_4$ & $3.660$ & $1.519$ & $-1.691$ & $-4.47$ & $-0.31$ & $-1.21$ \\ 
        \hline
        $\alpha_5$ & $3.660$ & $1.512$ & $-1.712$ & $-4.44$ & $-0.13$ & $-0.41$ \\ 
        \hline
        $\alpha_6$ & $3.660$ & $1.514$ & $-1.709$ & $-4.45$ & $-0.16$ & $-0.43$ \\
        \hline
    \end{tabular}
    \caption{Best fit parameter values for model 3 for each intercept.}
    \label{tab:intercept table}
\end{table}

Taking into account the residuals of the quadratic model and the uncertainties at each data point, we arrive at the following estimates for the amplitude intercepts
\begin{subequations}
    \label{intercept_ests}
\begin{align}
    \label{intercept_estG}
        \alpha_h &= \left(3.661 \pm 0.002 \right) \sqrt{\frac{\as N_c}{2\pi}}, \\
    \label{intercept_estG2}
        \alpha_{h,2} &= \left(3.660 \pm 0.002 \right) \sqrt{\frac{\as N_c}{2\pi}},  \\ 
        \alpha_3 &= \left(3.660 \pm 0.002 \right) \sqrt{\frac{\as N_c}{2\pi}}, \\
        \alpha_4 &= \left(3.660 \pm 0.002 \right) \sqrt{\frac{\as N_c}{2\pi}}, \\ 
        \alpha_5 &= \left(3.660 \pm 0.002 \right) \sqrt{\frac{\as N_c}{2\pi}}, \\
        \alpha_6 &= \left(3.660 \pm 0.002 \right) \sqrt{\frac{\as N_c}{2\pi}}. 
\end{align}
\end{subequations}
Comparing \eqs{intercept_estG} and (\ref{intercept_estG2}) with Eq.~(263) of \cite{Cougoulic:2022gbk}, we see we are in agreement with the earlier numerical result for $\alpha_h$ and $\alpha_{h,2}$. Furthermore, we also find the estimates for $\alpha_h$ and $\alpha_{h,2}$ are in good agreement with the analytic result \cite{Borden:2023ugd}. 

Finally, employing \eqs{moment_ansatz} and (\ref{intercept_ests}) in \eqs{final_quark_OAM} and (\ref{final_gluonOAM}), we obtain the following large-$N_c$, small-$x$ asymptotics for the quark and gluon OAM distributions:
\begin{align}\label{LLasympt1}
    L_{q+\bar{q}}(x,Q^2) \sim L_G(x,Q^2) \sim \left( \frac{1}{x} \right)^{3.66 \sqrt{\frac{\as N_c}{2\pi}}}.
\end{align}
Comparing \eq{LLasympt1} to Eq.~(265) of \cite{Cougoulic:2022gbk}, we find the quark and gluon OAM distributions have the same small-$x$ asymptotics as the quark and gluon helicity PDF along with the $g_1$ structure function: 
\begin{align}\label{LLasympt}
    L_{q+\bar{q}}(x,Q^2) \sim L_G(x,Q^2) \sim 
    \Delta \Sigma(x, Q^2) \sim \Delta G (x,Q^2) \sim g_1(x,Q^2) \sim 
    \left( \frac{1}{x} \right)^{3.66 \sqrt{\frac{\as N_c}{2\pi}}}.
\end{align}

The result \eqref{LLasympt1} appears to be in agreement with the OAM distributions asymptotics found in \cite{Boussarie:2019icw}, at least within the precision of our numerical solution. However, the intercept found in \cite{Boussarie:2019icw} for OAM distributions is the same as the BER intercept for hPDFs \cite{Bartels:1996wc}. Moreover, for hPDFs, the analytic solution for the large-$N_c$ equations for $G$ and $G_2$ from \cite{Cougoulic:2022gbk} found in \cite{Borden:2023ugd} has a different intercept from BER, with the numerical difference appearing only in the third decimal point. Therefore, for OAM distributions, we also expect that an analytic solution of \eqs{all_oam_eqns}, when constructed,  would lead to a slightly different large-$N_c$ intercept than that found in \cite{Boussarie:2019icw}.


\subsection{OAM distribution to helicity PDF ratios}
\label{sec:ratios}

\begin{figure}[ht]
     \centering
     \begin{subfigure}[b]{0.48\textwidth}
         \centering
         \includegraphics[width=\textwidth]{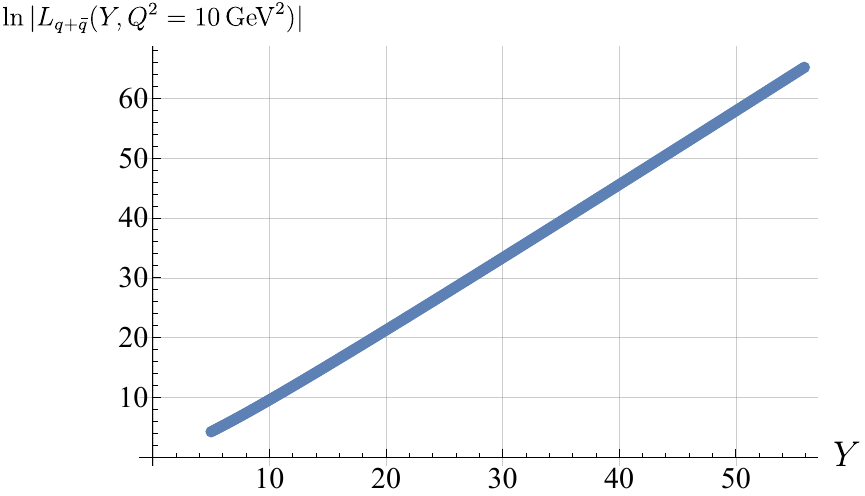}
         \caption{}
     \end{subfigure}
     \begin{subfigure}[b]{0.48\textwidth}
         \centering
         \includegraphics[width=\textwidth]{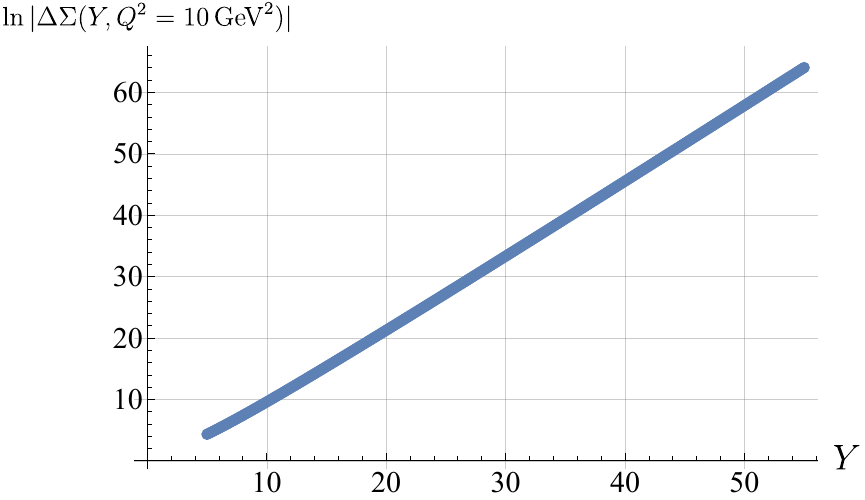}
         \caption{}
     \end{subfigure}
     \begin{subfigure}[b]{0.48\textwidth}
         \centering
         \includegraphics[width=\textwidth]{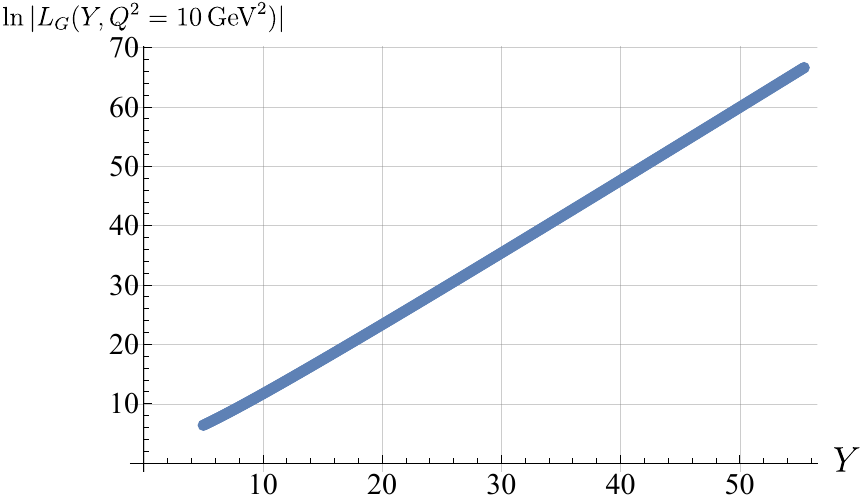}
         \caption{}
     \end{subfigure}
     \begin{subfigure}[b]{0.48\textwidth}
         \centering
         \includegraphics[width=\textwidth]{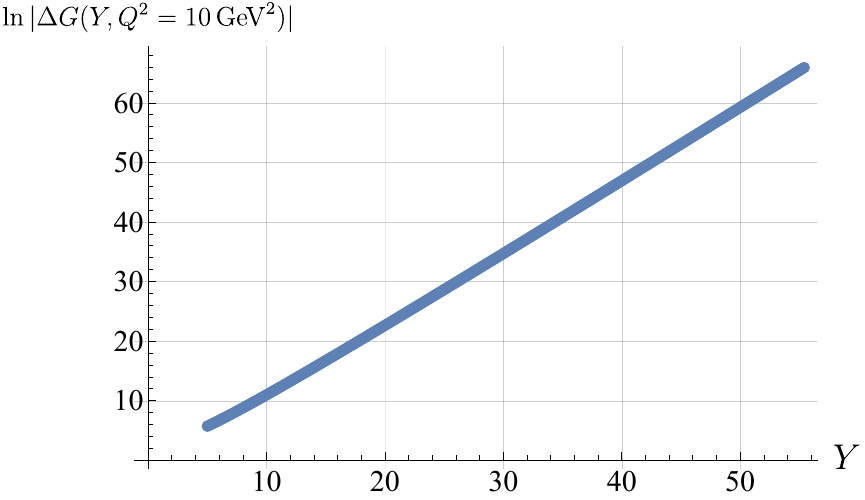}
         \caption{}
     \end{subfigure}
        \caption{Plots of the logarithms of the absolute values of OAM distributions and helicity PDFs as functions of rapidity $Y= \ln(1/x)$ at $Q^2=10 \,\mathrm{GeV^2}$ for $\delta = 0.025$.}
        \label{fig:dist_plots}
    \end{figure}

\begin{figure}[ht]
     \centering
     \begin{subfigure}[b]{0.48\textwidth}
         \centering
         \includegraphics[width=\textwidth]{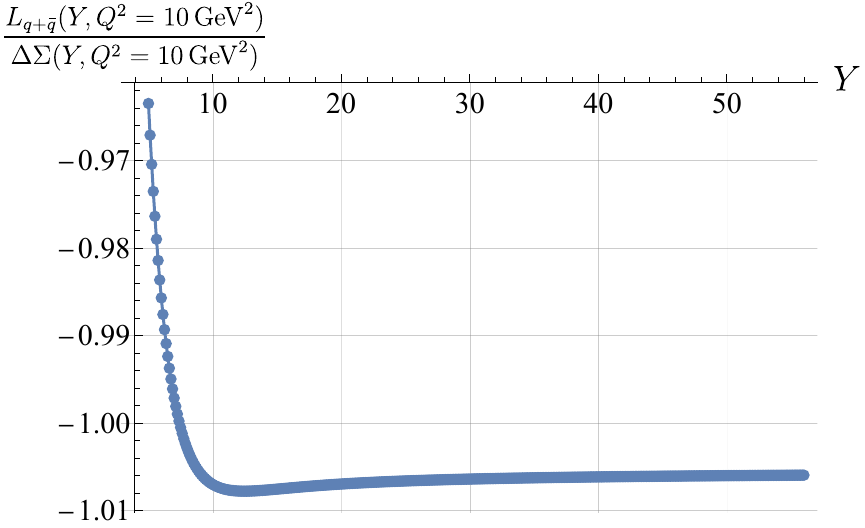}
         \caption{}
     \end{subfigure}
     \begin{subfigure}[b]{0.48\textwidth}
         \centering
         \includegraphics[width=\textwidth]{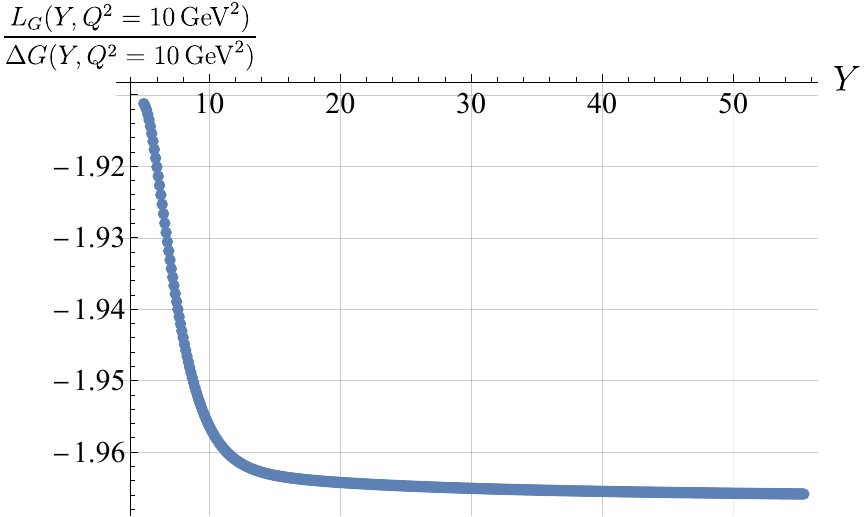}
         \caption{}
     \end{subfigure}
        \caption{Plots of the OAM distributions to helicity PDFs ratios for quarks (left panel) and gluons (right panel) as functions of rapidity $Y= \ln(1/x)$ at $Q^2=10 \,\mathrm{GeV^2}$ for $\delta = 0.025$.}
        \label{fig:ratio_plots}
\end{figure}

Another important quantity to investigate beyond the small-$x$ asymptotics of the OAM distributions is the ratios of the OAM distributions to the helicity PDFs, which were previously studied in \cite{Hatta:2016aoc,Hatta:2018itc,Boussarie:2019icw,More:2017zqp}. With the numerical solution of the moment amplitudes in hand, we can compute these ratios numerically. First, let us rewrite \eqs{final_quark_OAM}, (\ref{quark_hel}), (\ref{final_gluonOAM}), and (\ref{gluon_hel}) in terms of the rescaled variables of \eqs{rescaled_vars} 
\begin{subequations}
    \label{scaled_DFs}
\begin{align}
    \label{s_Lq}
    L_{q+\bar{q}}(Y, Q^2) &= \frac{N_f}{\as \pi^2} 
    \int \displaylimits^{\sqrt{\bas} \ln \frac{Q^2}{\Lambda^2}}_{0} d s_{10} 
    \int \displaylimits^{s_{10} + \sqrt{\bas} Y}_{s_{10}} d \eta 
    \left[Q + 3\, G_2 - I_3 + 2\, I_4 - I_5 - 3\, I_6 \right](s_{10}, \eta) ,
     \\ 
    \label{s_DE}
     \Delta \Sigma(Y, Q^2) &= -\frac{N_f}{\as \pi^2} 
    \int \displaylimits^{\sqrt{\bas} \ln \frac{Q^2}{\Lambda^2}}_{0} d s_{10} 
    \int \displaylimits^{s_{10} + \sqrt{\bas} Y}_{s_{10}} d \eta 
    \big[ Q(s_{10}, \eta)  + 2\, G_2(s_{10}, \eta)  \big],
    \\ 
    \label{s_LG}
    L_G(Y,Q^2) &= - \frac{2 N_c}{\as \pi^2} \Big[ 2 \, I_4 + 3\, I_5 + I_6\Big]\left(s_{10} = \sqrt{\bas} \ln \frac{Q^2}{\Lambda^2}, \, \eta = \sqrt{\bas}\ln \frac{Q^2}{\Lambda^2} + \sqrt{\bas} Y \right),
    \\
    \label{s_DG}
    \Delta G(Y,Q^2) &= \frac{2 N_c}{\as \pi^2}  G_2 \left(s_{10} = \sqrt{\bas} \ln \frac{Q^2}{\Lambda^2}, \, \eta = \sqrt{\bas}\ln \frac{Q^2}{\Lambda^2} + \sqrt{\bas} Y \right).
\end{align}
\end{subequations}
where we have used $s \approx Q^2/x$, and defined the rapidity variable $Y=\ln 1/x$ and  $\bas \equiv \frac{\as N_c}{2\pi}$. From here to the end of this Section we use a different argument of these distributions, $L_{q+\bar{q}}(x, Q^2) \to L_{q+\bar{q}}(Y, Q^2), \Delta \Sigma(x, Q^2) \to \Delta \Sigma(Y, Q^2), L_G(x,Q^2) \to L_G(Y,Q^2), \Delta G(x,Q^2) \to \Delta G(Y,Q^2)$, since $Y=\ln 1/x$ appears to be a more natural variable for our analysis (cf. \cite{Hatta:2018itc}). Using our numerical solution, we can compute the quantities in \eqs{scaled_DFs} numerically and then calculate the ratios of \eq{s_Lq} to \eq{s_DE} and \eq{s_LG} to \eq{s_DG} for the quark and gluon OAM to hPDF ratios respectively. The logarithms of the absolute values of these distributions as functions of $Y$ are plotted in \fig{fig:dist_plots}, and the OAM distribution to hPDF ratios in \fig{fig:ratio_plots} for $Q^2 = 10 \, \mathrm{GeV}^2$ and $\Lambda = 0.938 \, \mathrm{GeV}$. 

Based on Figs.~\ref{fig:dist_plots} and \ref{fig:ratio_plots}, we see that there are two distinct methods for estimating the ratios of the OAM distributions to the helicity PDFs. The first one involves fitting the distributions directly with some ansätze for their functional form then computing the ratios using the functions found, and the second one involves fitting the ratios directly with  similar ansätze. We then compare the two to ensure consistency.


\subsubsection{Method 1: Fitting the distributions directly}
\label{method1sec}

From Fig.~\ref{fig:dist_plots}, we observe a linear dependence on $Y$ which motivates the following ansätze (which are valid up to $\mathcal{O}(1/Y^2)$):
    \begin{subequations}
        \label{approach1_ansatz}
    \begin{align}
        \ln |L_{q+\bar{q}}(Y, Q^2)| &= \bar{\alpha}_{q} Y + \bar{\beta}_{q}  + \bar{\gamma}_q \ln Y  + \frac{\bar{\delta}_{q}}{Y}, \\ 
        \ln |\Delta \Sigma(Y, Q^2)| &= \alpha_{q} Y +\beta_{q}  + \gamma_q \ln Y  + \frac{\delta_{q}}{Y}, \\ 
        \ln |L_G(Y, Q^2)| &= \bar{\alpha}_G Y + \bar{\beta}_G  + \bar{\gamma}_G \ln Y  + \frac{\bar{\delta}_G} {Y}, \\ 
        \ln |\Delta G(Y, Q^2)| &=  \alpha_G Y +\beta_G  + \gamma_G \ln Y  + \frac{\delta_G}{Y}.
    \end{align}
    \end{subequations}
The $\ln Y$ and $1/Y$ terms are motivated by the analytic solution to the helicity evolution equations presented in Appendix~\ref{ansatz_app} below: we expect the structure of the asymptotic expressions for OAM distributions to be of the same type.

\begin{table}[ht!]
    \centering
    \begin{tabular}{|c|c|c|c|c|c|c|c|c|c|c|}
        \hline
        $\delta$ &  \,0.025 \,& \,0.032\, & \,0.0375 \,& \,0.05\, & \,0.0625\, & \,0.075\, &\, 0.08\, &\, 0.1\, \\ \hline
        $M_Y(\delta)$ & 55 & 55 & 85 & 114 & 143 & 172 & 172 & 201 \\ 
        \hline
    \end{tabular}
    \caption{Maximum values of $Y_{\mathrm{max}}$, $M_Y(\delta)$, for each step size $\delta$ as determined by requiring that the upper limits of the $\eta$ integrals in \eqs{s_Lq} and (\ref{s_DE}) are less than or equal to $\eta_{\mathrm{max}}$. Note that we exclude $\delta = 0.0125$ and $\delta = 0.016$ here as the numerically feasible range of $Y$ values is too small for those $\delta$'s to get a reliable fit.} 
    \label{tab:max_Y_values}
\end{table}

For a given $Q^2$, we fit the parameters in \eqs{approach1_ansatz} for each step size $\delta$ and $Y_{\mathrm{max}}$. In contrast to the intercept extraction presented above and in the earlier works \cite{Kovchegov:2016weo, Kovchegov:2020hgb, Cougoulic:2022gbk, Adamiak:2023okq}, to obtain the OAM distribution to hPDF ratios we want to capture the preasymptotic behavior in Figs.~\ref{fig:dist_plots} and \ref{fig:ratio_plots}. In other words, since all the intercepts are the same in \eqs{approach1_ansatz}, as we saw in \eqs{intercept_ests}, the parameters $\beta$ and $\gamma$ from \eqs{approach1_ansatz} (and, to a lesser degree, $\delta$) are the ones that determine the ratios. Consequently, instead of restricting ourselves to the largest 25\% of $Y$ values, we wish to include as wide a region in $Y$ as possible, $Y_{\mathrm{min}} \leq Y \leq Y_{\mathrm{max}}$, but where the distributions and ratios are independent of the initial conditions. One can check by explicitly computing the quantities in \eqs{scaled_DFs} for a range of different initial conditions that $Y_{\mathrm{min}} \approx 30$ makes for a good choice of rapidity above which the sensitivity of the solution to the initial conditions is lost (except for in the overall normalization). The fits below are therefore for the region $30 \leq Y \leq Y_{\mathrm{max}}$. Using the upper limit of the $\eta$ integrals in \eqs{s_Lq} and (\ref{s_DE}) to fix $Y_{\mathrm{max}}$ by requiring that $\sqrt{\bas} \left(Y_{\mathrm{max}} + \ln \frac{Q^2}{\Lambda^2} \right) = \eta_{\mathrm{max}}$, we can write an analog of Table~\ref{tab:max_eta_values} for $Y_{\mathrm{max}}$. This is shown in Table~\ref{tab:max_Y_values} for $Q^2 = 10 \, \mathrm{GeV}^2$. In a similar fashion to the way we took the continuum limit for the intercepts above, we can take the continuum limit for each parameter appearing in \eqs{approach1_ansatz} by extrapolating the best-fit model to $\delta \to 0$ and $1/Y_{\mathrm{max}} \to 0$. For all of the parameters, the best-fit model was always quadratic (as for the intercepts above). The results of this procedure are given in Table~\ref{tab:params_CL}. 

\begin{table}[ht!]
    \centering
    \begin{tabular}{|c|c|c|c|c|}
    \hline
         &  $\alpha$ & $\beta$ & $\gamma$ & $\delta$ \\
        \hline
      $\ln |L_{q+\bar{q}}(Y, Q^2=10~\mbox{GeV}^2)|$  & $1.2649 \pm 0.0001$ & $ 0.547 \pm 0.005 $ & $-1.500 \pm 0.002$ & $-1.64 \pm 0.03$ \\ 
      $\ln |\Delta \Sigma(Y, Q^2=10~\mbox{GeV}^2) |$ & $1.2649 \pm 0.0002$  & $0.542 \pm 0.005$ & $-1.499 \pm 0.002$ & $-1.67 \pm 0.03$ \\ 
       $\ln |L_G(Y, Q^2=10~\mbox{GeV}^2) |$   & $1.2649 \pm 0.0001$ & $2.57 \pm 0.01$  & $-1.500 \pm 0.007 $ & $-1.698 \pm 0.003$ \\ 
       $\ln |\Delta G(Y, Q^2=10~\mbox{GeV}^2) |$  & $1.2649 \pm 0.0001$ & $1.90 \pm 0.01$ & $-1.500 \pm 0.008 $ &  $-1.672 \pm 0.003$ \\
       \hline
    \end{tabular}
    \caption{Continuum limit results for the parameters in \eqs{approach1_ansatz}. Parameters are given for $\as =0.25, N_c =3, N_f = 3$ (in the prefactors of \eqs{s_Lq} and (\ref{s_DE})), $\Lambda = 938 \, \mathrm{MeV}, Q^2 = 10\, \mathrm{GeV}^2$.}
    \label{tab:params_CL}
\end{table}

The intercepts are consistent with the results from \cite{Cougoulic:2022gbk} and Section \ref{smallxasymp} for $\as = 0.25, N_c=3$. Instead of the usual $\gamma = -1/2$ from unpolarized evolution  \cite{Kuraev:1977fs, Balitsky:1978ic}, we see our results are consistent $\gamma = -3/2$ for all distributions. This result is in agreement with the analytic evaluation of hPDFs in Appendix~\ref{ansatz_app} below, based on the exact solution of the large-$N_c$ evolution in \cite{Borden:2023ugd}. Additionally, within the precision of our numerical approximation, we have $\alpha_{q/G} = \bar{\alpha}_{q/G}$ and $\gamma_{q/G} = \bar{\gamma}_{q/G}$. Therefore, the ratios of the distributions are coming solely from the $\beta$ and $\delta$ parameters and are given by (with the overall minus signs inferred from the numerical solution in \fig{fig:ratio_plots})
\begin{subequations}\label{app1_res}
\begin{align}
    \frac{L_{q + \bar{q}}(Y, Q^2= 10 \, \mathrm{GeV}^2)}{\Delta \Sigma(Y,Q^2= 10 \, \mathrm{GeV}^2) } &= -\exp(\bar{\beta}_q - \beta_{q})\left[1 + \frac{\bar{\delta}_q - \delta_q}{Y} + \ldots \right] = -(1.005 \pm 0.007) - \frac{(0.03 \pm 0.04)}{Y}+ \ldots , \label{q_ratio1} \\ 
      \frac{L_G(Y, Q^2= 10 \, \mathrm{GeV}^2)}{\Delta G (Y,Q^2= 10 \, \mathrm{GeV}^2)} &= -\exp(\bar{\beta}_G - \beta_{G})\left[1 + \frac{\bar{\delta}_G - \delta_G}{Y} + \ldots \right] = -(1.96 \pm 0.03) + \frac{(0.052 \pm 0.007)}{Y} + \ldots , 
\end{align}
\end{subequations}
where the ellipsis denote terms further suppressed by additional inverse powers of $Y$.

Before comparing these results to those in the literature, let us use another method for estimating the OAM to hPDF distribution ratios: we will fit the ratios directly. 


\subsubsection{Method 2: Fitting the ratios directly}

The plots in \fig{fig:ratio_plots} appear to approach a constant with growing $Y$. We therefore have the following alternative ansätze (as in method 1, the form of the ansätze is justified in Appendix~\ref{ansatz_app}):
    \begin{subequations}
        \label{ansatz2}
    \begin{align}
        \frac{L_{q+\bar{q}}(Y, Q^2)}{\Delta \Sigma(Y, Q^2)} &= A_q + \frac{B_q}{Y}, \\ 
         \frac{L_G(Y, Q^2)}{\Delta G(Y, Q^2)} &= A_G + \frac{B_G}{Y}.
    \end{align}
     \end{subequations}
Similar to \eqs{approach1_ansatz}, \eqs{ansatz2} are valid up to $\mathcal{O}(1/Y^2)$ terms. We can fit the ratios directly to extract $A$ and $B$ for each $\delta$ and $Y_{\mathrm{max}}$. Then we can repeat the procedure as described above to get an estimate for the parameters in the continuum limit ($\delta \to 0$ and $1/Y_{\mathrm{max}} \to 0$). As above for the intercepts and in method 1, quadratic models work best for all the parameters in \eqs{ansatz2}. The resulting continuum limit parameters we extract from our numerical solution give the following ratios:
\begin{subequations}\label{app2_res}
\begin{align}
        \frac{L_{q + \bar{q}}(Y, Q^2= 10 \, \mathrm{GeV}^2)}{\Delta \Sigma(Y,Q^2= 10 \, \mathrm{GeV}^2) } &= - (1.0005 \pm 0.0002) - \frac{(0.026 \pm 0.001)}{Y}, \label{q_ratio2} \\ 
        \frac{L_G(Y, Q^2= 10 \, \mathrm{GeV}^2)}{\Delta G (Y,Q^2= 10 \, \mathrm{GeV}^2)} &= - (1.96657 \pm 0.00003) + \frac{(0.0531 \pm 0.0002)}{Y}.
\end{align}
\end{subequations}
Comparing \eqs{app1_res} and \eqref{app2_res}, we see the two approaches are generally consistent within the precision of our numerical approximation. 

The results in \eqs{app1_res} and \eqref{app2_res} should be compared to the predictions from \cite{Boussarie:2019icw}. Taking Eqs.~(6) and (7) in \cite{Boussarie:2019icw}, and comparing them to our \eq{LLasympt}, we conclude that the parameter $\alpha$ from \cite{Boussarie:2019icw} is perturbatively small, $\alpha \sim \sqrt{\as} \ll 1$. Therefore, in the strict DLA, we need to neglect $\alpha$ compared to $1$, recasting Eqs.~(6) and (7) of \cite{Boussarie:2019icw} as 
\begin{subequations}\label{BHY}
\begin{align}
    \frac{L_{q+\bar{q}}(Y, Q^2)}{\Delta \Sigma(Y,Q^2)} &= -1, \label{BHYq} \\
    \frac{L_{G}(Y, Q^2)}{\Delta G(Y,Q^2)} &= -2.
\end{align}
\end{subequations}
Comparing \eqs{BHY} (see also Eq.~(24) in \cite{Boussarie:2019icw}) to \eqs{app1_res} and \eqref{app2_res}, we see that we are in good numerical agreement with \cite{Boussarie:2019icw} in both the quark and gluon sectors, though the precision of our Eqs.~\eqref{app2_res} appears to indicate a possible minor disagreement with \eqs{BHY}.

From \fig{fig:ratio_plots}~(a) we can see that the approach to the asymptotic value for the $L_{q+\bar{q}}/\Delta\Sigma$ ratio is rather slow, with the numbers for the ratios in Eqs.~\eqref{q_ratio1} and \eqref{q_ratio2} reached within a few percent level for extremely small $x$, $x \lesssim 4 \times 10^{-18}$ (corresponding to $Y \gtrsim 40$): saturation effects are very likely to become important in the proton well before such low values of $x$ are reached (even at $Q^2 = 10$~GeV$^2$), invalidating the linearized approximation we employed here for small-$x$ evolution (by putting $S=1$), along with most of our conclusions. Extending our plots in Figs.~\ref{fig:dist_plots} and \ref{fig:ratio_plots} to very high $Y$ values, corresponding to these extremely low $x$ values, is only done for the theoretical determination of the asymptotics outside of the saturation region. We also note that for $Y < Y_\mathrm{min} = 30$, the curves for the ratios plotted in \fig{fig:ratio_plots} are dependent on the initial conditions for our evolution \eqref{all_oam_eqns}.


\section{Conclusions and outlook}
\label{sec:conc}

In conclusion, let us reiterate our main results. We have revisited and revised the analysis of \cite{Kovchegov:2019rrz} for the quark and gluon OAM distributions. Our DLA expressions for the quark and gluon OAM distributions at small $x$ are given by \eqs{final_quark_OAM} and \eqref{final_gluonOAM}, respectively: the former contains new terms compared to the similar expression in  \cite{Kovchegov:2019rrz}. As in \cite{Kovchegov:2019rrz}, the OAM distributions depend on the impact parameter moments of the dipole amplitudes, which we refer to as moment amplitudes and define them in \eqs{moment_defs} somewhat differently from \cite{Kovchegov:2019rrz}.

To determine the OAM distributions at small $x$ we employ the results of \cite{Cougoulic:2022gbk} to construct the large-$N_c$ DLA evolution equations \eqref{all_oam_eqns} mixing the moment amplitudes $I_p$ for $p=3,4,5,6$ with the impact-parameter integrated polarized dipole amplitudes $G$ and $G_2$. Since the evolution equations in \cite{Cougoulic:2022gbk} augment and revise the KPS helicity evolution from \cite{Kovchegov:2015pbl, Kovchegov:2016zex, Kovchegov:2017lsr, Kovchegov:2018znm}, our evolution \eqref{all_oam_eqns} is also different from that in \cite{Kovchegov:2019rrz}. Solving the equations \eqref{all_oam_eqns} numerically we obtained the small-$x$ asymptotics of the OAM distributions shown above in \eq{LLasympt}. Within the precision of our numerical solution, we find agreement with the earlier results in \cite{Boussarie:2019icw} based on BER IREE \cite{Bartels:1996wc}, while anticipating a disagreement at the next significant digit \cite{Borden:2023ugd}.

We have also calculated the  $L_{q+\bar{q}}/\Delta\Sigma$ and $L_G/\Delta G$ ratios at very small $x$ using two different methods, with the results shown above in  \eqs{app1_res} and \eqref{app2_res}. In both the quark and gluon sectors, our results for the ratios appear to be in good numerical agreement with those found in \cite{Boussarie:2019icw}.

Further work on the subject may parallel the ongoing studies of helicity PDFs at small $x$. It appears possible \cite{Manley:2023} to construct an analytic solution of \eqs{all_oam_eqns} following the approach presented in \cite{Borden:2023ugd}: this would verify and provide an additional insight into our conclusions here. The full large-$N_c \& N_f$ helicity evolution equations exist \cite{Kovchegov:2015pbl, Kovchegov:2016zex, Kovchegov:2017lsr, Kovchegov:2018znm, Cougoulic:2022gbk} and can be utilized to expand our results, both numerically and analytically, beyond the large-$N_c$ limit we employed here (and the $N_f/N_c \to 0$ limit of the large-$N_c \& N_f$ approximation we have briefly explored as well). The large-$N_c \& N_f$ limit for OAM distributions (and hPDFs) includes quarks as well, in addition to gluons, and is, therefore, more realistic: it may eventually allow one to do phenomenology of OAM distributions at small $x$ similar to the study of hPDFs and the $g_1$ structure function carried out in \cite{Adamiak:2023yhz}. This may produce the much-needed numerical predictions for OAM distributions at small $x$, beginning to quantitatively address this important component of the proton spin puzzle.



\section*{Acknowledgments}

We are grateful to Yossathorn (Josh) Tawabutr for his extensive advice on the numerical solution performed here and to Yoshitaka Hatta and M. Gabriel Santiago for helpful and encouraging discussions. 

This material is based upon work supported by the U.S. Department of Energy, Office of Science, Office of Nuclear Physics under Award Number DE-SC0004286.

The work is performed within the framework of the Saturated Glue (SURGE) Topical Theory Collaboration.


\appendix


\section{Simplifying the quark OAM distribution}

\label{sec:simplifying}

Let us begin by simplifying the contribution of the first term in the curly brackets of \eq{OAM8}, omitted from the analysis in \cite{Kovchegov:2019rrz} 
\begin{align}
    \label{qOAMint4}
     & - \frac{8 i P^+}{(2\pi)^5} \int d^2 k_\perp d^2 \zeta d^2 \xi \, e^{-i \un{k} \cdot \left( \un{\zeta} - \un{\xi} \right)} \left(
     \frac{\un{\zeta} + \un{\xi}}{2} \times \un{k}
    \right) 
    \int d^2 w \;d^2 z \int\limits^{p^-_2}_0 \frac{dk_1^-}{2\pi}  \\\notag
    &\hspace{6cm} \times 
    \frac{\un{\xi} - \un{z}}{|\un{\xi} - \un{z}|^2} \cdot  \frac{\un{\zeta} - \un{w}}{|\un{\zeta} - \un{w}|^2} \, 
    \mbox{Im} \left\langle - \mbox{T} \, \tr \left[ V_{\ul \zeta} \, \left( V_{{\un z}, {\ul w}}^{\textrm{pol} [2]} \right)^\dagger \right] + \mbox{T} \, \tr \left[ V_{\ul \zeta}^\dagger \, V_{{\un z}, {\ul w}}^{\textrm{pol} [2]} \right] \right\rangle.
\end{align}
Employing $\vpol{\un{x}, \un{y}}{2} = V^{G [2]}_{\un{x}, \un{y}} + V^{q [2]}_{\un{x}} \delta^2(\un{x}- \un{y})$ (see \eq{VqG_decomp}), we rewrite it as
\begin{align} 
    \label{qOAMint5}
      - \frac{8 i P^+}{(2\pi)^5} &\int d^2 k_\perp d^2 \zeta d^2 \xi \, e^{-i \un{k} \cdot \left( \un{\zeta} - \un{\xi} \right)} \left(
     \frac{\un{\zeta} + \un{\xi}}{2} \times \un{k}
    \right) 
    \int d^2 w \;d^2 z \int\limits^{p^-_2}_0 \frac{dk_1^-}{2\pi}  \, 
    \frac{\ul{\zeta} - {\un w}}{|\ul{\zeta} - {\un w}|^2} \cdot \frac{\ul{\xi} - {\un z}}{|\ul{\xi} - {\un z}|^2}
     \\\notag
    &\times \Im \Bigg\langle -
    \mathrm{T \; tr} \left[ 
    V_{\un{\zeta}} \left( V^{q [2]}_{\un{z}} \right)^\dagger 
    \right] \delta^2(\un{z}-\un{w}) 
    + \tord \; \tr \left[ V_{\un{\zeta}}^\dagger \, 
     V^{q [2]}_{\un{z}}  
    \right] \delta^2(\un{z}-\un{w}) 
     - \mathrm{T \; tr} \left[ 
    V_{\un{\zeta}} \left( V^{G [2]}_{\un{z}, \un{w}} \right)^\dagger 
    \right] + \tord \tr \left[ V_{\un{\zeta}}^\dagger \, 
     V^{G [2]}_{\un{z}, \un{w}} 
    \right]
    \Bigg\rangle.
\end{align}

First consider the contribution of the $V^{q [2]}$-containing terms to \eq{qOAMint5},
\begin{align}
    \label{qOAMint6}
& - \frac{8 i P^+}{(2\pi)^5}    \int d^2 k_\perp \, d^{2} \zeta \, d^{2} \xi \, d^2 w \,  e^{- i {\un k} \cdot ({\un \zeta} - {\un \xi})} \left( \frac{{\un \zeta} + {\un \xi}}{2} \times {\un k}\right) \int\limits_0^{p_2^-} \frac{d k_1^-}{2\pi}  \\ 
& \times  \, \frac{\ul{\zeta} - {\un w}}{|\ul{\zeta} - {\un w}|^2} \cdot \frac{\ul{\xi} - {\un w}}{|\ul{\xi} - {\un w}|^2} \,
\mbox{Im} \left\langle - \mbox{T} \, \tr \left[ V_{\ul \zeta} \, V_{{\ul w}}^{q [2] \dagger} \right] + \mbox{T} \, \tr \left[ V_{\ul \zeta}^\dagger \, V_{{\ul w}}^{q [2]} \right] \right\rangle  .  \notag
\end{align}
Noticing that, under passive PT-symmetry transformation, 
\begin{align}
V_{\ul \zeta} \ \xrightarrow[]{\textrm{PT}} \ V_{- \ul \zeta}^\dagger , \ \ \ V_{{\un w}}^{q \, [2]}  \ \xrightarrow[]{\textrm{PT}} \ V_{-{\un w}}^{q \, [2] \, \dagger} , 
\end{align}
along with the fact that under time reversal the time-ordering operation $\tord$ becomes $\atord$, one can show that \eq{qOAMint6} is PT-odd. Therefore, it does not contribute to $L_{q + \bar q}(x,Q^2)$, which is PT-even. (In the definition of OAM distributions, the proton polarization $S_L$ is either factored out or, equivalently, projected out per \eq{cgc_ave}; while the orbital angular momentum vector $\vec L$ is PT-odd, the OAM distribution $L(x, Q^2)$ is PT-even both for quarks and for gluons.)

Next we consider the contribution of $V^{G[2]}$ to \eq{qOAMint5},
\begin{align}
\label{LG2}
&  - \frac{8 i P^+}{(2\pi)^5}    \int d^2 k_\perp \, d^{2} \zeta \, d^{2} \xi \, d^2 w \, d^2 z \, e^{- i {\un k} \cdot ({\un \zeta} - {\un \xi})} \left( \frac{{\un \zeta} + {\un \xi}}{2} \times {\un k}\right) \int\limits_0^{p_2^-} \frac{d k_1^-}{2\pi}   \\ 
& \times  \, \frac{\ul{\zeta} - {\un w}}{|\ul{\zeta} - {\un w}|^2} \cdot \frac{\ul{\xi} - {\un z}}{|\ul{\xi} - {\un z}|^2} \,
\mbox{Im} \left\langle - \mbox{T} \, \tr \left[ V_{\ul \zeta} \, \left( V_{{\un z}, {\ul w}}^{\textrm{G} [2]} \right)^\dagger \right] + \mbox{T} \, \tr \left[ V_{\ul \zeta}^\dagger \, V_{{\un z}, {\ul w}}^{\textrm{G} [2]} \right] \right\rangle . \notag
\end{align}
Employing \eq{VxyG2} we rewrite the term containing the first trace in \eq{LG2} as
\begin{align}\label{LG2_1}
&  + \frac{8 i P^+}{(2\pi)^5}    \int d^2 k_\perp \, d^{2} \zeta \, d^{2} \xi \, e^{- i {\un k} \cdot ({\un \zeta} - {\un \xi})} \left( \frac{{\un \zeta} + {\un \xi}}{2} \times {\un k}\right) \int\limits_0^{p_2^-} \frac{d k_1^-}{2\pi} \, \frac{P^+}{s}  \int\limits_{-\infty}^{\infty} d{y}^- d^2 y  \\ 
& \times  \,
\frac{(\ul{\zeta} - {\un y})^j}{|\ul{\zeta} - {\un y}|^2} \,  \mbox{Re} \left\langle  \mbox{T} \, \tr \left[ V_{\ul \zeta} \, V_{\un{y}} [ - \infty, y^-] \, \cev{D}^i (y^-, {\un y}) \, D^i  (y^-, {\un y}) \, V_{\un{y}} [ y^-, \infty]  \right] \right\rangle \, \frac{(\ul{\xi} - {\un y})^j}{|\ul{\xi} - {\un y}|^2}. \notag
\end{align}
Here $s = 2 P^+ k_1^-$ and the partial derivatives in $\cev{D}^i \, D^i$ act on the Wilson lines and on $\tfrac{(\ul{\zeta} - {\un y})^j}{|\ul{\zeta} - {\un y}|^2}$ and $\tfrac{(\ul{\xi} - {\un y})^j}{|\ul{\xi} - {\un y}|^2}$.

Following \cite{Cougoulic:2022gbk}, we rewrite 
\begin{align}\label{step_back}
\frac{\ul{\zeta} - {\un y}}{|\ul{\zeta} - {\un y}|^2} = \int \frac{d^2 k_1}{2 \pi i} e^{i {\un k}_1 \cdot (\ul{\zeta} - {\un y})} \, \frac{{\un k}_1}{{\un k}_1^2}, \ \ \ \frac{\ul{\xi} - {\un y}}{|\ul{\xi} - {\un y}|^2} = \int \frac{d^2 k_2}{2 \pi i} e^{i {\un k}_2 \cdot (\ul{\xi} - {\un y})} \, \frac{{\un k}_2}{{\un k}_2^2}.
\end{align}
Using these in \eq{LG2_1} yields
\begin{align}\label{LG2_2}
& - \frac{8 i P^+}{(2\pi)^5}    \int d^2 k_\perp \, d^{2} \zeta \, d^{2} \xi \, e^{- i {\un k} \cdot ({\un \zeta} - {\un \xi})} \left( \frac{{\un \zeta} + {\un \xi}}{2} \times {\un k}\right) \int\limits_0^{p_2^-} \frac{d k_1^-}{2\pi} \, \frac{P^+}{s}  \int\limits_{-\infty}^{\infty} d{y}^- d^2 y  \int \frac{d^2 k_1 \, d^2 k_2}{(2 \pi)^2} \, e^{i {\un k}_1 \cdot (\ul{\zeta} - {\un y}) + i {\un k}_2 \cdot (\ul{\xi} - {\un y})} \, \frac{{\un k}_1}{{\un k}_1^2} \cdot \frac{{\un k}_2}{{\un k}_2^2}  \\ 
& \times  \, \Bigg\{
\mbox{Re} \left\langle \mbox{T} \, \tr \left[ V_{\ul \zeta} \, V_{\un{y}} [ - \infty, y^-] \, \left[  \cev{D}^i (y^-, {\un y}) \, D^i  (y^-, {\un y}) - k_1^i \, k_2^i \right]  \, V_{\un{y}} [ y^-, \infty]  \right] \right\rangle  \notag \\
& + i\,\mbox{Im} \left\langle \mbox{T} \, \tr \left[ V_{\ul \zeta} \, V_{\un{y}} [ - \infty, y^-] \, \left[ i \, k_2^i \,  \cev{D}^i (y^-, {\un y}) + i \, k_1^i \, D^i  (y^-, {\un y})  \right]  \, V_{\un{y}} [ y^-, \infty]  \right] \right\rangle \Bigg\} . \notag
\end{align}

Again let us apply the passive PT transformation. We have
\begin{align}
V_{\ul \zeta} \ \xrightarrow[]{\textrm{PT}} \ V_{- \ul \zeta}^\dagger , \ \ \ V_{\un{y}} [ y^-, \infty]  \ \xrightarrow[]{\textrm{PT}} \ V_{-\un{y}} [ -y^-, - \infty ] , \ \ \ V_{\un{y}} [ -\infty , y^- ] \ \xrightarrow[]{\textrm{PT}} \ V_{-\un{y}} [ \infty , -y^-].
\end{align}
Note that it is not necessary to change the sign of the internal integration (position) variables, since their sign can always be changed back by a variable redefinition. Hence, we only change the signs of the infinite integration limits, along with $\tord \ \xrightarrow[]{\textrm{PT}} \ \atord$. \eq{LG2_2} transforms under PT to
\begin{align}\label{LG2_3}
& \notag - \frac{8 i P^+}{(2\pi)^5}    \int d^2 k_\perp \, d^{2} \zeta \, d^{2} \xi \, e^{- i {\un k} \cdot ({\un \zeta} - {\un \xi})} \left( \frac{{\un \zeta} + {\un \xi}}{2} \times {\un k}\right) \int\limits_0^{p_2^-} \frac{d k_1^-}{2\pi} \, \frac{P^+}{s}  \int\limits_{\infty}^{-\infty} d{y}^- d^2 y  \int \frac{d^2 k_1 \, d^2 k_2}{(2 \pi)^2} \, e^{i {\un k}_1 \cdot (\ul{\zeta} - {\un y}) + i {\un k}_2 \cdot (\ul{\xi} - {\un y})} \, \frac{{\un k}_1}{{\un k}_1^2} \cdot \frac{{\un k}_2}{{\un k}_2^2}  \\ 
& \times  \, \Bigg\{
\mbox{Re} \left\langle \atord \, \tr \left[ V_{\ul \zeta}^\dagger \, V_{\un{y}} [ \infty, y^-] \left[  \cev{D}^i (y^-, {\un y}) \, D^i  (y^-, {\un y}) - k_1^i \, k_2^i \right]  V_{\un{y}} [ y^-, - \infty]  \right] \right\rangle \\
& + i \, \mbox{Im} \left\langle \atord \, \tr \left[ V_{\ul \zeta}^\dagger \, V_{\un{y}} [ \infty, y^-] \left[ i \, k_2^i \,  \cev{D}^i (y^-, {\un y}) + i \, k_1^i \, D^i  (y^-, {\un y})  \right]  V_{\un{y}} [ y^-, - \infty]  \right] \right\rangle \Bigg\} . \notag
\end{align}
Note the limits of the $y^-$ integral.

To get back to the original expression form in \eq{LG2_2}, we can do a Hermitian conjugation of the matrix element under the Re sign, without affecting anything. Similarly, a Hermitian conjugation of the expression under the Im sign generates a minus sign. We get  
\begin{align}\label{LG2_4}
& + \frac{8 i P^+}{(2\pi)^5}    \int d^2 k_\perp \, d^{2} \zeta \, d^{2} \xi \, e^{- i {\un k} \cdot ({\un \zeta} - {\un \xi})} \left( \frac{{\un \zeta} + {\un \xi}}{2} \times {\un k}\right) \int\limits_0^{p_2^-} \frac{d k_1^-}{2\pi} \, \frac{P^+}{s}  \int\limits_{-\infty}^{\infty} d{y}^- d^2 y  \int \frac{d^2 k_1 \, d^2 k_2}{(2 \pi)^2} \, e^{i {\un k}_1 \cdot (\ul{\zeta} - {\un y}) + i {\un k}_2 \cdot (\ul{\xi} - {\un y})} \, \frac{{\un k}_1}{{\un k}_1^2} \cdot \frac{{\un k}_2}{{\un k}_2^2} \notag \\ 
& \times  \, \Bigg\{
\mbox{Re} \left\langle \tord \, \tr \left[ V_{\ul \zeta} \, V_{\un{y}} [- \infty, y^-] \left[  \cev{D}^i (y^-, {\un y}) \, D^i  (y^-, {\un y}) - k_1^i \, k_2^i \right]  V_{\un{y}} [ y^-, \infty]  \right] \right\rangle   \\
& + i\, \mbox{Im} \left\langle \tord \, \tr \left[ V_{\ul \zeta} \, V_{\un{y}} [ - \infty, y^-] \left[ i \, k_2^i \, D^i (y^-, {\un y}) + i \, k_1^i \,  \cev{D}^i  (y^-, {\un y})  \right]  V_{\un{y}} [ y^-, \infty]  \right] \right\rangle \Bigg\} . \notag
\end{align}
Comparing  \eq{LG2_4} to \eq{LG2_2} we see that the first term in the curly brackets is PT-odd: it cannot contribute to the quark OAM and should be discarded. 

To compare the second term in the curly brackets of Eqs.~\eqref{LG2_4} and \eqref{LG2_2} we notice that one can write
\begin{align}\label{DDdecomp}
i \, k_2^i \,  \cev{D}^i  + i \, k_1^i \, D^i  = \frac{i}{2} \, (k_1^i - k_2^i) \, (D^i  - \cev{D}^i) + \frac{i}{2} \, (k_1^i + k_2^i) \, (D^i  + \cev{D}^i) .
\end{align}
Using this in Eqs.~\eqref{LG2_4} and \eqref{LG2_2}  we see that the second term on the right of \eq{DDdecomp} also gives a PT-odd contribution in \eq{LG2_2} and should be discarded. Only the first (PT-even) term on the right of \eq{DDdecomp} survives in \eq{LG2_2}. We thus obtain 
\begin{align}\label{LG2_5}
& + \frac{8 P^+}{(2\pi)^5} \int d^2 k_\perp \, d^{2} \zeta \, d^{2} \xi \, e^{- i {\un k} \cdot ({\un \zeta} - {\un \xi})} \left( \frac{{\un \zeta} + {\un \xi}}{2} \times {\un k}\right) \int\limits_0^{p_2^-} \frac{d k_1^-}{2\pi} \,  \int\limits_{-\infty}^{\infty} d{y}^- d^2 y  \int \frac{d^2 k_1 \, d^2 k_2}{(2 \pi)^2} \, e^{i {\un k}_1 \cdot (\ul{\zeta} - {\un y}) + i {\un k}_2 \cdot (\ul{\xi} - {\un y})} \, \frac{{\un k}_1}{{\un k}_1^2} \cdot \frac{{\un k}_2}{{\un k}_2^2}  \\ 
& \times \,  (k_1^i - k_2^i) \,  \mbox{Re} \left\langle \frac{P^+}{2 s}  \,  \mbox{T} \, \tr \left[ V_{\ul \zeta} \, V_{\un{y}} [ - \infty, y^-] \, \left[ D^i  (y^-, {\un y})  -  \cev{D}^i (y^-, {\un y})  \right]  \, V_{\un{y}} [ y^-, \infty]  \right] \right\rangle . \notag
\end{align}

Defining \cite{Cougoulic:2022gbk,Kovchegov:2018znm}
\begin{align}\label{Vi}
V_{\un{z}}^{i \, \textrm{G} [2]} \equiv \frac{P^+}{2 s} \, \int\limits_{-\infty}^{\infty} d {z}^- \, V_{\un{z}} [ \infty, z^-] \, \left[ {D}^i (z^-, \un{z}) - \cev{D}^i (z^-, \un{z}) \right] \, V_{\un{z}} [ z^-, -\infty]  
\end{align}
we rewrite \eq{LG2_5} more compactly as
\begin{align}\label{LG2_6}
&  - \frac{8 P^+}{(2\pi)^5}   \int d^2 k_\perp \, d^{2} \zeta \, d^{2} \xi \, e^{- i {\un k} \cdot ({\un \zeta} - {\un \xi})} \left( \frac{{\un \zeta} + {\un \xi}}{2} \times {\un k}\right) \int\limits_0^{p_2^-} \frac{d k_1^-}{2\pi} \,  \int d^2 y  \int \frac{d^2 k_1 \, d^2 k_2}{(2 \pi)^2} \, e^{i {\un k}_1 \cdot (\ul{\zeta} - {\un y}) + i {\un k}_2 \cdot (\ul{\xi} - {\un y})} \, \frac{{\un k}_1}{{\un k}_1^2} \cdot \frac{{\un k}_2}{{\un k}_2^2}  \\ 
& \times \,  (k_1^i - k_2^i) \,  \mbox{Re} \left\langle \mbox{T} \, \tr \left[ V_{\ul \zeta} \, \left( V_{\un{y}}^{i \, \textrm{G} [2]} \right)^\dagger  \right]  + \mbox{T} \, \tr \left[ V_{\ul \zeta}^\dagger \, V_{\un{y}}^{i \, \textrm{G} [2] }  \right]  \right\rangle , \notag
\end{align}
where we have added the contribution of the second trace in \eq{LG2}, which is evaluated analogously. Equation \eqref{LG2_6} can be rewritten as
\begin{align}
    \label{LG2_7}
&  - \frac{8 N_c}{(2\pi)^6}  \int d^2 k_\perp \, d^{2} \zeta \, d^{2} \xi \, e^{- i {\un k} \cdot ({\un \zeta} - {\un \xi})} \left( \frac{{\un \zeta} + {\un \xi}}{2} \times {\un k}\right) \int\limits_\frac{\Lambda^2}{s}^{1} \frac{d z}{z} \,  \int d^2 y  \int \frac{d^2 k_1 \, d^2 k_2}{(2 \pi)^2} \, e^{i {\un k}_1 \cdot (\ul{\zeta} - {\un y}) + i {\un k}_2 \cdot (\ul{\xi} - {\un y})} \, \frac{{\un k}_1}{{\un k}_1^2} \cdot \frac{{\un k}_2}{{\un k}_2^2}  \\ 
& \times \,  (k_1^i - k_2^i) \, G^i_{{\un y}, {\un \zeta}} (zs) , \notag
\end{align}
where we have used $zs = 2 P^+ k^-_1$ to change the integration variable from $k^-_1$ to $z$ and defined the polarized dipole amplitude of the second kind \cite{Kovchegov:2018znm,Cougoulic:2022gbk}, 
\begin{align}
    \label{Gi_def}
    G^i_{\un y, \un \zeta}(zs) \equiv \frac{1}{2N_c} \mathrm{Re} \llangle \mathrm{T\,tr} \left[V_{\un \zeta}^\dagger V_{\un y}^{i\, G[2]} +  \left(V_{\un y}^{i\, G[2]}\right)^\dagger V_{\un \zeta}\right] \rrangle.
\end{align}
The scale $\Lambda$ is the infrared cutoff.

Let us define more convenient variables $\un{\tilde{\xi}} = \un \xi - \un y$ and $\un{\tilde{\zeta}} = \un \zeta - \un y$ so that \eq{LG2_7} becomes
\begin{align}
    \label{LG2_8}
&  - \frac{8 N_c}{(2\pi)^6}  \int d^2 k_\perp \, d^{2} \tilde{\zeta} \, d^{2} \tilde{\xi} \, d^2 y\, e^{- i {\un k} \cdot ({\un{\tilde{\zeta}}} - {\un{\tilde{\xi}}})} \left[ \left( \frac{{\un{\tilde{\zeta}}} + {\un{\tilde \xi}}}{2} + \un y \right)\times {\un k}\right] \int\limits_\frac{\Lambda^2}{s}^{1} \frac{d z}{z}  \int \frac{d^2 k_1 \, d^2 k_2}{(2 \pi)^2} \, e^{i {\un k}_1 \cdot \un{\tilde \zeta} + i {\un k}_2 \cdot  \un{\tilde \xi}} \, \frac{{\un k}_1}{{\un k}_1^2} \cdot \frac{{\un k}_2}{{\un k}_2^2} (k_1^i - k_2^i) \, G^i_{{\un y}, \un y + {\un \zeta}} (zs) 
\end{align}

The $\un{\tilde \xi}$ term in the square brackets can be written as $-i \un \nabla_{\un k}$ acting on $e^{i \un k \cdot \un{\tilde \xi}}$. After integrating by parts the $-i \un \nabla_{\un k}$ operator becomes $\un{\tilde \zeta}$. Hence we can replace $\un{\tilde \xi} \to \un{\tilde \zeta}$ in the square brackets, obtaining
\begin{align}
    \label{LG2_9}
& - \frac{8 N_c}{(2\pi)^6}  \int d^2 k_\perp \, d^{2} \tilde{\zeta} \, d^{2} \tilde{\xi} \, d^2 y\, e^{- i {\un k} \cdot (\un{\tilde\zeta} - \un{\tilde \xi})} \left[ \left( \un{\tilde \zeta} + \un y \right)\times {\un k}\right] \int\limits_\frac{\Lambda^2}{s}^{1} \frac{d z}{z}  \int \frac{d^2 k_1 \, d^2 k_2}{(2 \pi)^2} \, e^{i {\un k}_1 \cdot \un{\tilde \zeta} + i {\un k}_2 \cdot  \un{\tilde \xi}} \, \frac{{\un k}_1}{{\un k}_1^2} \cdot \frac{{\un k}_2}{{\un k}_2^2} (k_1^i - k_2^i) \, G^i_{{\un y}, \un y + {\un \zeta}} (zs) .
\end{align}
Now, we integrate over $\un{\tilde \xi}, \un k_2,$ and $\un k_1$. This gives 
\begin{align}
    \label{LG2_10}
&  + \frac{8 iN_c}{(2\pi)^5}  \int d^2 k_\perp \, d^{2} \tilde{\zeta} \, d^2 y\, e^{- i {\un k} \cdot \un{\tilde\zeta}} \left[ \left( \un{\tilde \zeta} + \un y \right)\times {\un k}\right] \int\limits_\frac{\Lambda^2}{s}^{1} \frac{d z}{z} \, 
\frac{k^i \, \un{\tilde \zeta}^2 \, ( \un k \cdot \un{\tilde \zeta} - i) + 2i \, \un k \cdot \un{\tilde{\zeta}} \, \tilde \zeta^i}{\un k^2 \, \un{\tilde \zeta}^4}
G^i_{{\un y}, \un y + {\un \zeta}} (zs) .
\end{align}

Next, let us simplify the second term in the curly brackets of \eq{OAM8} given in \eq{Qterm}. Evaluating it along the steps similar to the above we arrive at the same contribution as in \cite{Kovchegov:2019rrz} (see Eq.~(30) there, keeping in mind the sign difference mentioned in the main text),
\begin{align}
    \label{qOAMbint1}
    - \frac{8iN_c}{(2\pi)^6} \int d^2 k_\perp d^2 \zeta \, d^2 w \, e^{-i \un{k} \cdot \left( \un{\zeta} - \un{w} \right)} \int\limits^1_{\frac{\Lambda^2}{s}} \frac{dz}{z} \, \Bigg[
    2\pi i \frac{\un{\zeta} - \un{w}}{|\un{\zeta} - \un{w}|^2} \times \frac{\un{k}}{k_\perp^2} \left(
    \frac{\un{\zeta} + \un{w}}{2} \times \un{k} \right) - \pi \frac{\un{\zeta} - \un{w}}{|\un{\zeta} - \un{w}|^2} \cdot \frac{\un{k}}{k^2_\perp}
    \Bigg]
    Q_{\un{w},\un{\zeta}}(zs),
\end{align}
where we have integrated out $\un \xi$.

Adding Eqs.~\eqref{LG2_10} and \eqref{qOAMbint1} (while replacing $\un \zeta - \un w \to \un{\tilde \zeta}, \un w \to \un y$ in the latter), we find
\begin{align}
    \label{qOAMint7}
    &L_{q+\bar{q}}(x,Q^2) = -\frac{8 iN_c}{(2\pi)^5}  \int d^2 k_\perp \, d^{2} \tilde{\zeta} \, d^2 y\, e^{- i {\un k} \cdot \un{\tilde\zeta}} \int\limits_\frac{\Lambda^2}{s}^{1} \frac{d z}{z} \\ \notag
    & \times \left\{ 
    \left[i\,  \frac{\un{\tilde \zeta} \times \un k}{\un{\tilde \zeta}^2 k^2_\perp}\left(\frac{\un{\tilde \zeta} + 2 \un y}{2} \times \un k  \right) - \frac{1}{2} \frac{\un{\tilde \zeta}}{\un{\tilde \zeta}^2} \cdot \frac{\un k}{k^2_\perp} \right] Q_{\un y, \un y + \un{\tilde \zeta}}(zs)
   - \left[ \left( \un{\tilde \zeta} + \un y \right)\times {\un k}\right]
\frac{k^i \un{\tilde \zeta}^2 ( \un k \cdot \un{\tilde \zeta} - i) + 2i \, \un k \cdot \un{\tilde{\zeta}} \tilde \zeta^i}{k_\perp^2 \un{\tilde \zeta}^4}
G^i_{{\un y}, \un y + {\un \zeta}} (zs) 
\right\}.
\end{align}
Finally, replacing $\un{\tilde\zeta} \to - {\un x}_{10}$ and ${\un y} \to {\un x}_1$ in \eq{qOAMint7} we arrive at \eq{qOAMint77} in the main text.


\section{Ansätze for distribution functions and their ratios}
\label{ansatz_app}

Here we motivate the ansätze (\ref{approach1_ansatz}) and (\ref{ansatz2}) via the analytic solution to the helicity evolution equations \cite{Borden:2023ugd}. The general analytic solution of \eqs{helEvo_eqn1}, \eqref{Gamma_eq}, \eqref{G2_eq}, and \eqref{helEvo_eqn4} was constructed in \cite{Borden:2023ugd} using double Laplace transforms. The quark and gluon hPDFs from \eqs{s_DE} and (\ref{s_DG}) can be written in terms of the double Laplace transforms $G_{2\omega \gamma}, G^{(0)}_{2\omega \gamma}$ of the dipole amplitude $G_2 (x_{10}^2, zs)$ and of its initial condition $G_2^{(0)} (x_{10}^2, zs)$, respectively,  as (cf. Eqs.~(58) and (56) in \cite{Borden:2023ugd})
\begin{subequations}
    \label{analyticDFs}
\begin{align}   
    \Delta \Sigma(y, t) &= - \frac{N_f}{2\, \as \pi^2} \int \frac{d\omega}{2\pi i} \int \frac{d \gamma}{2\pi i} 
    \left( G_{2\omega \gamma} - G^{(0)}_{2\omega \gamma} \right) e^{\omega y + \gamma  t},  
    \\ 
    \Delta G(y, t) &= \frac{2 N_c}{\as \pi^2} \int \frac{d\omega}{2\pi i} \int \frac{d \gamma}{2\pi i} \,
     e^{\omega y + \gamma t}\, G_{2\omega \gamma},
\end{align}
\end{subequations}
where, for brevity, we have defined $y \equiv \sqrt{\bas} \, Y = \sqrt{\bas} \, \ln (1/x) , \, \, t \equiv \sqrt{\bas} \ln \frac{Q^2}{\Lambda^2}$, where $\bas = \frac{\as N_c}{2\pi}$, as defined above.\footnote{We omit the rescaling in Eq.~(52) of \cite{Borden:2023ugd}.} As usual, the $\omega$- and $\gamma$- contours run parallel to the imaginary axis to the right of the singularities of the integrand. Note that Re~$\omega >$~Re~$\gamma$ along these contours \cite{Borden:2023ugd}. The function $G^{(0)}_{2\omega \gamma}$ is determined by the initial conditions of the helicity evolution, while $G_{2\omega \gamma}$ is given by the solution found in \cite{Borden:2023ugd}, which, in turn, depends on $G^{(0)}_{2\omega \gamma}$ and $G^{(0)}_{\omega \gamma}$ (the latter is the double Laplace transform of the initial condition $G^{(0)} (x_{10}^2, zs)$ for the dipole amplitude $G (x_{10}^2, zs)$). For $G^{(0)}_2(s_{10}, \eta) = G^{(0)}(s_{10}, \eta)=1$ as used above, one can show that (see Eq.~(51) in  \cite{Borden:2023ugd})
\begin{subequations}
\label{ones_ICs}
\begin{align}
    G_{2\omega \gamma} &= \frac{1}{\omega (\gamma - \gamma^-_\omega)}\left(1+ \frac{2}{\gamma \gamma^+_\omega} \right),
    \\
    G^{(0)}_{2\omega \gamma} &= \frac{1}{\omega \gamma},
\end{align}
\end{subequations}
where 
\begin{align}
    \label{gammapmdef}
    \gamma^\pm_\omega = \frac{\omega}{2} \left[1 \pm \sqrt{1 - \frac{16}{\omega^2} \sqrt{1- \frac{4 }{\omega^2}}} \right].
\end{align}
Employing \eqs{ones_ICs} in \eqs{analyticDFs} and closing the $\gamma$ contours to the left (while keeping in mind that Re~$\omega >$~Re~$\gamma$ along the integration contours), we get
\begin{subequations}
    \label{ones_DFs}
\begin{align}   
    \label{qones}
    \Delta \Sigma(y, t) &=  - \frac{N_f}{2\, \as \pi^2} \int \frac{d\omega}{2\pi i}  \,
    e^{\omega y} 
     \left[
        e^{\gamma_\omega^- t} - 1 \right] \, \frac{1}{\omega}  \, \left(1+ \frac{2}{\gamma_\omega^- \gamma_\omega^+ } \right) , \\
    \label{gones}
    \Delta G(y, t) &= \frac{2 N_c}{\as \pi^2} \int \frac{d\omega}{2\pi i}  \,
     \frac{e^{\omega y}}{\omega} 
     \left[
        e^{\gamma_\omega^- t} \left(1 + \frac{2}{\gamma^+_\omega \gamma^-_\omega} \right) - \frac{2}{\gamma^+_\omega \gamma^-_\omega}
     \right].
\end{align}
\end{subequations}

\begin{figure}[t]
\centering
\includegraphics[width=0.4
\linewidth]{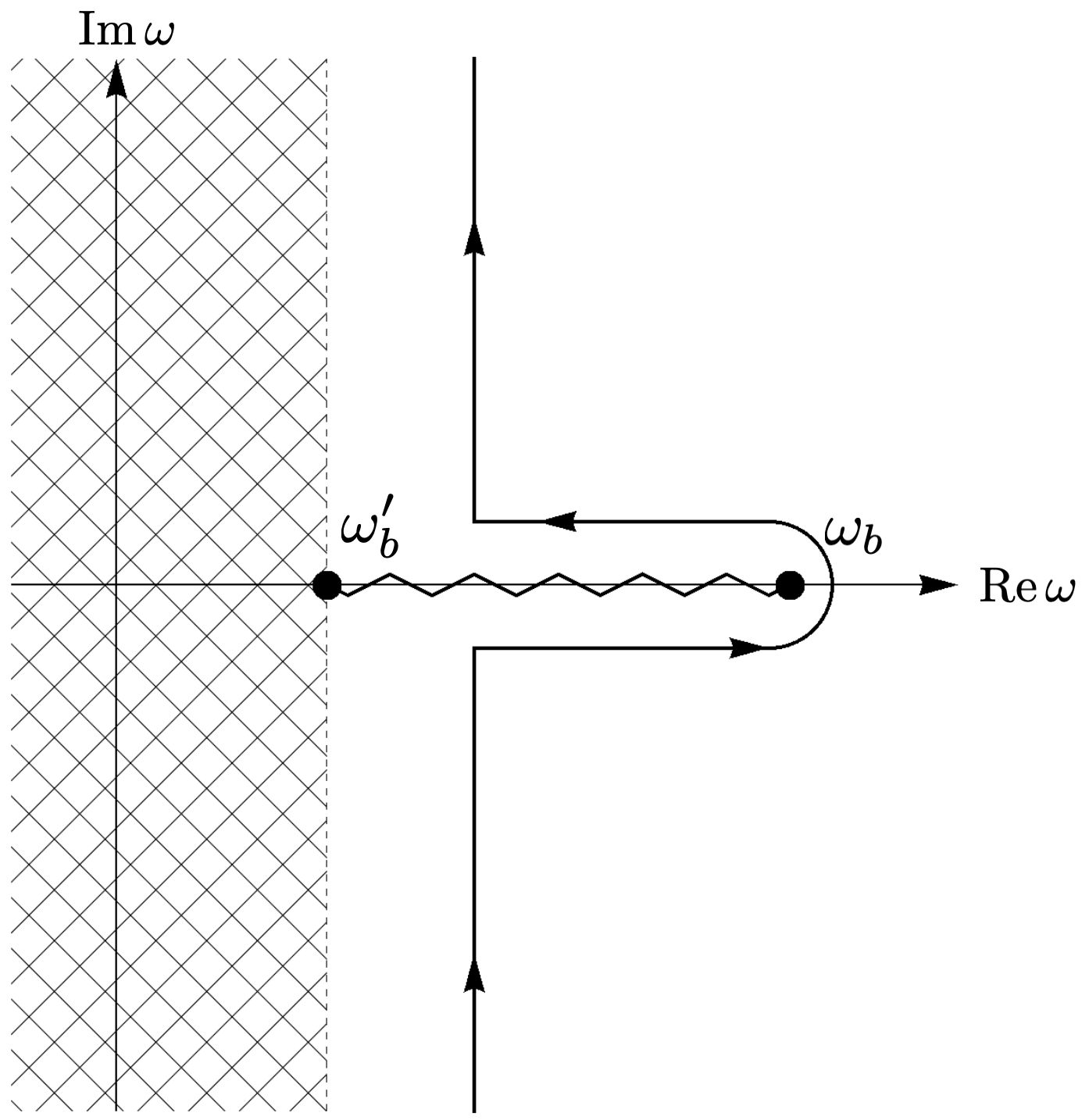} 
\caption{Analytic structure of $\gamma^-_\omega$ in the vicinity of its right-most branch point $\omega = \omega_b$, with the jagged line denoting the branch cut originating at $\omega_b$. The shaded region covers the rest of the complex plane, containing additional branch cuts of $\gamma^-_\omega$: those are irrelevant for our calculation. The integration contour, distorted to wrap around the branch cut, is denoted by the solid black line with arrows. }
\label{FIG:branch_cut}
\end{figure}

Now, to approximate the $\omega$ integrals in \eqs{ones_DFs} and extract the leading behavior at large $y$ (or $Y$), we can move the contours to the left until we hit the right-most singularity. As one can show \cite{Borden:2023ugd}, this singularity is given by the branch point at 
\begin{align}\label{bp}
    \omega = \omega_b \equiv \frac{4}{3^{1/3}} \sqrt{\mathrm{Re}\left[(-9 + i \sqrt{111})^{1/3} \right]}  \approx 3.66074, 
\end{align}
obtained by setting the large square root in \eq{gammapmdef} to $0$ and finding the root with the largest real part. At high $y$ the integrals in \eqs{ones_DFs} are dominated by the branch point \eqref{bp} and the branch cut originating at that branch point and going along the real axis for $\omega < \omega_b$ until the next branch point at $\omega = \omega'_b \approx 2.4$. Wrapping the integration contour around this branch cut, we approximate the integrals in \eqs{ones_DFs} by the integrals of the integrand discontinuities across the branch cut. This is illustrated in \fig{FIG:branch_cut}, where we do not show the entire branch cut structure of $\gamma_\omega^-$, concentrating on the dominant branch cut instead. The part of the contour wrapping around the branch in \fig{FIG:branch_cut} dominates the integral. Due to the $y$-dependent exponentials, we can discard the rest of the contribution from closing the contour to the left as sub-leading in $y$, and we can extend the upper limit of integration to infinity (effectively sending $\omega'_b \to - \infty$) {\sl after} we calculate the integrand discontinuity across the branch cut ending at $\omega_b$. If we denote the integrands (including the prefactors) in \eqs{qones} and (\ref{gones}) as $\Delta \Sigma_\omega$ and $\Delta G_\omega$, respectively, this approximation is represented via
\begin{subequations}\label{disc}
\begin{align}
    \Delta \Sigma(y,t) \approx \lim_{\epsilon \to 0^+}  \int\limits_0^\infty \frac{d\xi}{2\pi i} \big( \Delta \Sigma_{\omega_b - \xi + i \epsilon} - \Delta \Sigma_{\omega_b - \xi - i \epsilon} \big), 
    \\
    \Delta G(y,t) \approx \lim_{\epsilon \to 0^+} \int\limits_0^\infty \frac{d\xi}{2\pi i} \big( \Delta G_{\omega_b - \xi + i \epsilon} - \Delta G_{\omega_b - \xi - i \epsilon} \big),
\end{align}
\end{subequations}
where we have defined $\omega = \omega_b - \xi$.

Since \cite{Borden:2023ugd}
\begin{align}
    \gamma^+_\omega \gamma^-_\omega = 4 \, \sqrt{1- \frac{4}{\omega^2}}
\end{align}
we see that $\gamma^+_\omega \gamma^-_\omega$ has no branch cut discontinuity in the vicinity of $\omega_b$. Employing \eqs{disc} and \eqref{ones_DFs}, we then write
\begin{subequations}\label{DF2}
    \begin{align}
        \Delta \Sigma(y, t) & \approx  - \frac{N_f}{2\, \as \pi^2} \int\limits_0^\infty \frac{d\xi}{2\pi i}  \,
    e^{(\omega_b - \xi) y} 
     \left[
        e^{\gamma_{\omega_b - \xi + i \epsilon}^- t} - e^{\gamma_{\omega_b - \xi - i \epsilon}^- t} \right] \, \frac{1}{\omega_b - \xi }  \left(1+ \frac{2}{\gamma_{\omega_b - \xi}^- \gamma_{\omega_b - \xi}^+ } \right) , \\
    \label{gones2}
    \Delta G(y, t) & \approx \frac{2 N_c}{\as \pi^2} \int\limits_0^\infty \frac{d\xi}{2\pi i}  \,
     \frac{e^{(\omega_b - \xi) y}}{\omega_b - \xi} 
     \left[
        e^{\gamma_{\omega_b - \xi + i \epsilon}^- t} - e^{\gamma_{\omega_b - \xi - i \epsilon}^- t} \right] \,  \left(1+ \frac{2}{\gamma_{\omega_b - \xi}^- \gamma_{\omega_b - \xi}^+ } \right).
    \end{align}
\end{subequations}
Note again that in arriving at \eqs{DF2} we have extended the integrals over $\xi$ to infinity in the upper limit since the integrand is dominated by small values of $\xi$. The $e^{- \xi \, y}$ factor in both integrands ensures that $\xi \ll 1$ for large $y$. Therefore, the rest of the integrands can be expanded in the powers of $\xi$, with each additional power of $\xi \sim 1/y$ bringing in an extra power of $1/y$ suppression after the $\xi$-integration. 

To expand the integrands of \eqs{DF2} in $\xi$ we will need the following expansion of $\gamma^-_\omega$ near the branch point $\omega_b$, 
\begin{align}
    \label{gamma_approx}
    \gamma^-_{\omega_b - \xi \pm i \epsilon} 
& = \frac{\omega_b}{2} \mp  i \, \frac{8 \, \sqrt{2} \, \sqrt{\omega_b^2 -6}}{\omega_b^{5/2}} \, \xi^{1/2} - \frac{\xi}{2} \pm i \, \frac{512 \, \sqrt{2} \, (\omega_b^2 - 2)}{\omega_b^{15/2} \, \sqrt{\omega_b^2 - 6}} \, 
 \xi^{3/2} + \mathcal{O}(\xi^{\frac{5}{2}})
    \\ \notag
    &\equiv \frac{\omega_b}{2} \mp i a_1 \xi^{1/2} - \frac{\xi}{2} \pm i a_3 \xi^{3/2} + \mathcal{O}(\xi^{5/2}).
\end{align}
Using these in \eqs{DF2} while noticing that $\gamma_{\omega_b - \xi}^- \gamma_{\omega_b - \xi}^+ = \gamma^-_{\omega_b - \xi + i \epsilon} \, \gamma^-_{\omega_b - \xi - i \epsilon}$ we get
\begin{subequations}\label{approxDFs}
    \begin{align}
         \Delta \Sigma (y, t) &\approx \frac{N_f}{\as 2 \pi^2} \, t\, e^{\omega_b t/2} \int \displaylimits^\infty_0  \frac{d \xi}{2\pi}\, e^{(\wb - \xi) \, y} \Bigg\{ \frac{2 \, a_1 \, (\omega_b^2 + 8)}{\omega_b^3} \, \xi^{1/2} \\
         & \notag - \frac{1}{3 \, \omega_b^5} \Big[ 6 \, a_3 \, \omega_b^2 \, (\omega_b^2 + 8) + 3 \, a_1 \, \omega_b \, \left( t \, \omega_b \, (\omega_b^2 + 8) - 48 - 2 \, \omega_b^2\right) + a_1^3 \, \left(192 + 8 \, t^2 \, \omega_b^2 + t^2 \, \omega_b^4 \right) \Big] \, \xi^{3/2} + \mathcal{O}(\xi^{5/2}) \Bigg\} , \notag \\
         \Delta G(y,t) &\approx - \frac{2 N_c}{\as \pi^2} \, t\, e^{\omega_b t/2} \int \displaylimits^\infty_0  \frac{d \xi}{2\pi}\, e^{(\wb - \xi) \, y} \Bigg\{ \frac{2 \, a_1 \, (\omega_b^2 + 8)}{\omega_b^3} \, \xi^{1/2} \\
         & \notag - \frac{1}{3 \, \omega_b^5} \Big[ 6 \, a_3 \, \omega_b^2 \, (\omega_b^2 + 8) + 3 \, a_1 \, \omega_b \, \left( t \, \omega_b \, (\omega_b^2 + 8) - 48 - 2 \, \omega_b^2\right) + a_1^3 \, \left(192 + 8 \, t^2 \, \omega_b^2 + t^2 \, \omega_b^4 \right) \Big] \, \xi^{3/2} + \mathcal{O}(\xi^{5/2}) \Bigg\}. 
    \end{align}
\end{subequations}
Note that terms of $\mathcal{O}(\xi)$ do not contribute to \eqs{approxDFs}. Indeed, the $\mathcal{O}(\xi)$ term in \eq{gamma_approx} does not contain a discontinuity across the branch cut. From \eqs{approxDFs}, we see that in the small-$x$ asymptotic limit, $\Delta G(y,t) = -( 4N_c / N_f) \, \Delta \Sigma(y,t)$. This relation can also be seen directly from \eqs{analyticDFs}. For $N_c =3$ and $N_f=4$ we get $\Delta G(y,t) = - 3 \, \Delta \Sigma(y,t)$, which compares reasonably well with $\Delta G(y,t) \approx -2.29 \, \Delta \Sigma(y,t)$ found in \cite{Boussarie:2019icw} for the same $N_c$ and $N_f$, given that our version of this relation is derived in the large-$N_c$ limit while the work in \cite{Boussarie:2019icw} was done for any $N_c$ and $N_f$.

Performing the integration over $\xi$ in \eqs{approxDFs}, we get the following functional forms for the helicity PDFs
\begin{subequations}
    \label{func_forms}
\begin{align}
    \Delta \Sigma(y, t) &\approx \left[ \frac{c_{1, q}(t)}{y^{3/2}} + \frac{c_{2,q}(t)}{y^{5/2}} + \mathcal{O} \left( \frac{1}{y^{7/2}} \right) \right] \,  e^{\omega_b y}, \\ 
     \Delta G(y, t) &\approx \left[ \frac{c_{1,G}(t)}{y^{3/2}} + \frac{c_{2, G}(t)}{y^{5/2}} + \mathcal{O} \left( \frac{1}{y^{7/2}} \right) \right] \, e^{\omega_b  y},
\end{align}
\end{subequations}
where the coefficients are
\begin{subequations}\label{ones_coeff}
    \begin{align}
        c_{1,q}(t) &=  \frac{N_f}{4\pi^{5/2} \as} \, t \, e^{\wb t/2} \, \frac{a_1 \, (\omega_b^2 + 8)}{\omega_b^3} , \\ 
         c_{2,q}(t) &=  -\frac{3\,N_f}{16 \pi^{5/2} \as} \, t \, e^{\wb t/2} \, \frac{1}{3 \, \omega_b^5} \Big[ 6 \, a_3 \, \omega_b^2 \, (\omega_b^2 + 8) + 3 \, a_1 \, \omega_b \, \left( t \, \omega_b \, (\omega_b^2 + 8) - 48 - 2 \, \omega_b^2\right) \\ & + a_1^3 \, \left(192 + 8 \, t^2 \, \omega_b^2 + t^2 \, \omega_b^4 \right) \Big] ,  \notag \\
         c_{1,G}(t) &= - \frac{N_c}{\pi^{5/2} \as} \, t \, e^{\wb t/2} \, \frac{a_1 \, (\omega_b^2 + 8)}{\omega_b^3} , \\ 
         c_{2,G}(t) &=  \frac{3\,N_c}{4 \pi^{5/2} \as} \, t \, e^{\wb t/2} \, \frac{1}{3 \, \omega_b^5} \Big[ 6 \, a_3 \, \omega_b^2 \, (\omega_b^2 + 8) + 3 \, a_1 \, \omega_b \, \left( t \, \omega_b \, (\omega_b^2 + 8) - 48 - 2 \, \omega_b^2\right) \\ & + a_1^3 \, \left(192 + 8 \, t^2 \, \omega_b^2 + t^2 \, \omega_b^4 \right) \Big] . \notag
    \end{align}
\end{subequations}
The structure of the $y$-dependence in \eqs{func_forms} appears to be valid for any initial conditions. At the same time, the coefficients in \eqs{ones_coeff} are valid only for the specific choice of initial conditions $G^{(0)}_2(s_{10}, \eta) = G^{(0)}(s_{10}, \eta)=1$ we made here.  

Let us assume that the functional form in \eqs{func_forms} also applies to the OAM distributions $L_{q+\bar{q}}(Y,Q^2)$ and $L_G(Y,Q^2)$. Namely, we write 
\begin{subequations}
    \label{func_forms_L}
\begin{align}
    L_{q+\bar{q}} (y, t) &\approx \left[ \frac{{\bar c}_{1, q}(t)}{y^{3/2}} + \frac{{\bar c}_{2,q}(t)}{y^{5/2}} + \mathcal{O} \left( \frac{1}{y^{7/2}} \right) \right] \,  e^{\omega_b y}, \\ 
     L_G(y, t) &\approx \left[ \frac{{\bar c}_{1,G}(t)}{y^{3/2}} + \frac{{\bar c}_{2, G}(t)}{y^{5/2}} + \mathcal{O} \left( \frac{1}{y^{7/2}} \right) \right] \, e^{\omega_b  y},
\end{align}
\end{subequations}
This is indeed only an assumption, which nevertheless seems to be supported by our numerical analysis presented in the main text. 

For the OAM distributions and helicity PDFs obeying the functional form \eqref{func_forms}, \eqref{func_forms_L}, along with the OAM to hPDF ratios, changing the variables back to $Y = \ln(1/x)$ and $Q^2$, one can write the following ansätze (valid up to $\mathcal{O} (1/Y^2)$ terms), 
\begin{subequations}
    \label{ansatze_just} 
\begin{align}
    \ln |L_{q+\bar{q}}(Y,Q^2)| &= \omega_b \sqrt{\bas} Y + \bar{\beta}_{q}(Q^2) + \frac{\bar{\delta}_{q}(Q^2)}{Y} - \frac{3}{2} \ln Y , 
    \\ 
     \ln |\Delta \Sigma(Y,Q^2)| &=\omega_b \sqrt{\bas} Y + \beta_{q}(Q^2) + \frac{\delta_{q}(Q^2)}{Y} - \frac{3}{2} \ln Y, 
     \\ 
      \ln |L_G(Y,Q^2)| &= \omega_b \sqrt{\bas} Y + \bar{\beta}_{G}(Q^2) + \frac{\bar{\delta}_{G}(Q^2)}{Y} - \frac{3}{2} \ln Y ,
      \\
       \ln |\Delta G(Y,Q^2)| &= \omega_b \sqrt{\bas} Y + \beta_{G}(Q^2) + \frac{\delta_{G}(Q^2)}{Y} - \frac{3}{2} \ln Y  ,
       \\ 
       \frac{L_{q+\bar{q}} (Y,Q^2)}{\Delta \Sigma(Y, Q^2)} &= A_q(Q^2) + \frac{B_q(Q^2)}{Y} ,
       \\ 
       \frac{L_{G} (Y,Q^2)}{\Delta G(Y, Q^2)} &= A_G(Q^2) + \frac{B_G(Q^2)}{Y} ,
\end{align}
\end{subequations}
where the parameters $\beta, \delta, A, B$ are related to the coefficients $c_1, c_2, {\bar c}_1, {\bar c}_2$ via 
\begin{subequations}
    \begin{align}
        \beta &= \ln \frac{|c_1|}{\bas^{3/4}}, \label{beta_c} \\ 
        \delta &= \frac{c_2}{c_1 \bas^{1/2}}, \label{delta_c} \\ 
        A &= \frac{\bar{c}_1}{c_1}, \\ 
        B &= \frac{{\bar c}_1}{\bas \, c_1} \left( \frac{{\bar c}_2}{{\bar c}_1} - \frac{c_2}{c_1} \right) .
    \end{align}
\end{subequations}
Here we suppress the quark ($q$) and gluon ($G$) subscripts for brevity.

Equations~\eqref{ansatze_just} now motivate the ansätze in \eqs{approach1_ansatz} and (\ref{ansatz2}). We see that in the numerical analysis of Section~\ref{method1sec} above, the coefficients in front of the $\ln Y$ terms are consistent with $-3/2$ from \eqs{ansatze_just} and the coefficients in front of the linear terms in $Y$ are consistent with $\wb \sqrt{\bas} \approx 1.2648$ (for $\as = 0.25$ and $N_c =3$); both of these results have been derived analytically in this Appendix (for hPDFs), per \eqs{func_forms}. For $\as = 0.25, \, N_c = N_f =3, \, Q^2 = 10\, \mathrm{GeV}^2, \, \Lambda = 0.938 \, \mathrm{GeV}$, \eqs{ones_coeff}, when used in Eqs.~\eqref{beta_c} and \eqref{delta_c}, give us  
\begin{subequations}
\begin{align}
    \beta_q(Q^2 = 10 \, \mathrm{GeV}^2) &\approx 0.545,
    \\
    \delta_q(Q^2 = 10 \, \mathrm{GeV}^2) &\approx -1.835, 
    \\ 
    \beta_G(Q^2 = 10 \, \mathrm{GeV}^2) &\approx 1.933,
    \\
    \delta_G(Q^2 = 10 \, \mathrm{GeV}^2) &\approx -1.835,
\end{align}
\end{subequations}
in reasonable agreement with the results for $\ln |\Delta\Sigma|$ and $\ln |\Delta G|$ in Table \ref{tab:params_CL}. The analogous coefficients for the OAM distributions, $\bar{\beta}_{q,G}, \bar{\delta}_{q,G}$ can be computed with the analytic solution of \eqs{all_oam_eqns}, which we leave for future work \cite{Manley:2023}.


\providecommand{\href}[2]{#2}\begingroup\raggedright\endgroup

\end{document}